\documentclass[amssymb,twocolumn,floats,amsmath,superscriptaddress]{revtex4-2}
\usepackage[english]{babel}
\usepackage[utf8]{inputenc}
\usepackage[colorinlistoftodos, color=green!40, prependcaption]{todonotes}
\usepackage{amsthm}
\usepackage{mathtools}
\usepackage{physics}
\usepackage{xcolor}
\usepackage{graphicx}
\usepackage[left=23mm,right=13mm,top=35mm,columnsep=15pt]{geometry} 
\usepackage{adjustbox}
\usepackage{placeins}
\usepackage[T1]{fontenc}
\usepackage{csquotes}
\def \AC{II. Institute of Physics B and JARA-FIT, RWTH-Aachen University, 52074 Aachen, Germany}
\def \ACa{2nd Institute of Physics and JARA-FIT, RWTH-Aachen University, 52074 Aachen, Germany}
\def \Haifa{Schulich Faculty of Chemistry, Solid State Institute, Russell Berrie Nanotechnology Institute and Helen Diller Quantum Center, Technion – Israel Institute of Technology, Haifa 3200003, Israel}
\def \HaifaMat{Department of Materials Science and Engineering, Technion – Israel Institute of Technology, Haifa 3200003, Israel}
\def \FZJ{Forschungszentrum Jülich,
Peter Grünberg Institute (PGI-6),
52428 Jülich, Germany}
\def \FZJtheo{Forschungszentrum Jülich,
Peter Grünberg Institute (PGI-1),
52428 Jülich, Germany}
\def \Dui{Faculty of Physics, University of Duisburg-Essen, 47057 Duisburg, Germany}

\providecommand{\keywords}[1]
{
  \small	
  \textbf{\textit{keywords---}} #1
}

\begin{document}
\title{Identifying band structure changes of FePS$_3$ across the antiferromagnetic phase transition}
\author{Benjamin Pestka}\affiliation{\AC}
\author{Jeff Strasdas}\affiliation{\AC}
\author{Gustav Bihlmayer}\affiliation{\FZJtheo}
\author{Adam K. Budniak}\affiliation{\Haifa}
\author{Marcus Liebmann}\affiliation{\AC}
\author{Niklas Leuth}\affiliation{\AC}
\author{Honey Boban}\affiliation{\FZJ}
\author{Vitaliy Feyer}\affiliation{\FZJ}
\author{Iulia Cojocariu}\affiliation{\FZJ}
\author{Daniel Baranowski}\affiliation{\FZJ}
\author{Simone Mearini}\affiliation{\FZJ}
\author{Yaron Amouyal}\affiliation{\HaifaMat}
\author{Lutz Waldecker}\affiliation{\ACa}
\author{Bernd Beschoten}\affiliation{\ACa}
\author{Christoph Stampfer}\affiliation{\ACa}
\author{Lukasz Plucinski}\affiliation{\FZJ}
\author{Efrat Lifshitz}\affiliation{\Haifa} 
\author{Peter Kratzer}\affiliation{\Dui}
\author{Markus Morgenstern}\affiliation{\AC}
\date{\today} 

\begin{abstract}
\vspace{\baselineskip}
\section*{Abstract}
\vspace{-\baselineskip}
Magnetic 2D materials enable novel tuning options of magnetism. As an example, the van der Waals material FePS$_3$, a zigzag-type intralayer antiferromagnet, exhibits very strong magnetoelastic coupling due to the different bond lengths along different ferromagnetic and antiferromagnetic coupling directions enabling elastic tuning of magnetic properties. The likely cause of the length change is the intricate competition between direct exchange of the Fe atoms and superexchange via the S and P atoms. To elucidate this interplay,
we study the band structure of exfoliated FePS$_3$ by  $\mu$m scale ARPES (Angular Resolved Photoelectron Spectroscopy), both, above and, for the first time, below the Néel temperature $T_{\rm N}$. We find three characteristic changes across $T_{\rm N}$. They involve S 3p-type bands, Fe 3d-type bands and P 3p-type bands, respectively, as attributed by comparison with density functional theory calculations (DFT+U). This highlights the involvement of all the atoms in the magnetic phase transition providing independent evidence for the intricate exchange paths. 
\end{abstract}

\keywords{Magnetic 2D materials, $\mu$-ARPES, density functional theory}

\maketitle

\noindent {{Corresponding author: } 
M.~Morgenstern, email: mmorgens@physik.rwth-aachen.de } \\
\noindent {\\ {\bf{Keywords: }} 
magnetic 2D materials, angular resolved photoelectron spectroscopy, layered magnetism, density functional theory.}

\section{Introduction}
Two-dimensional (2D) van der Waals (vdW) ferromagnets (FMs) and antiferromagnets (AFMs)  \cite{Huang2017,Sun2019b,Gong2017,Wang2018d,Deng2018,Fei2018} 
enable novel types of spintronics devices, magnetic sensors, and magneto-optical devices \cite{Li2019,Zhang2019,Burch2018,Gong2019, Huang2020}. They are mostly based on the versatile tuning options of the magnetism 
via magnetic field, electrostatic gating, strain, light, ion intercalation, proximity effects and moiré lattices \cite{Burch2018,Li2019,Mi2022, Ahn2024}.
The 2D magnets, moreover, reveal rather direct access to proximity-induced interactions between different types of quantum materials and to pronounced magneto-electric, magneto-optic, and  magneto-elastic couplings  
\cite{Burch2018,Huang2020,Gong2019,Wang2022,Vaclavkova2021, Ressouche2010, Kang2020,Hwangbo2021,Dirnberger2022,Ahn2024}. 

\begin{figure*}[!htb]
\centering
\includegraphics[width=\textwidth]{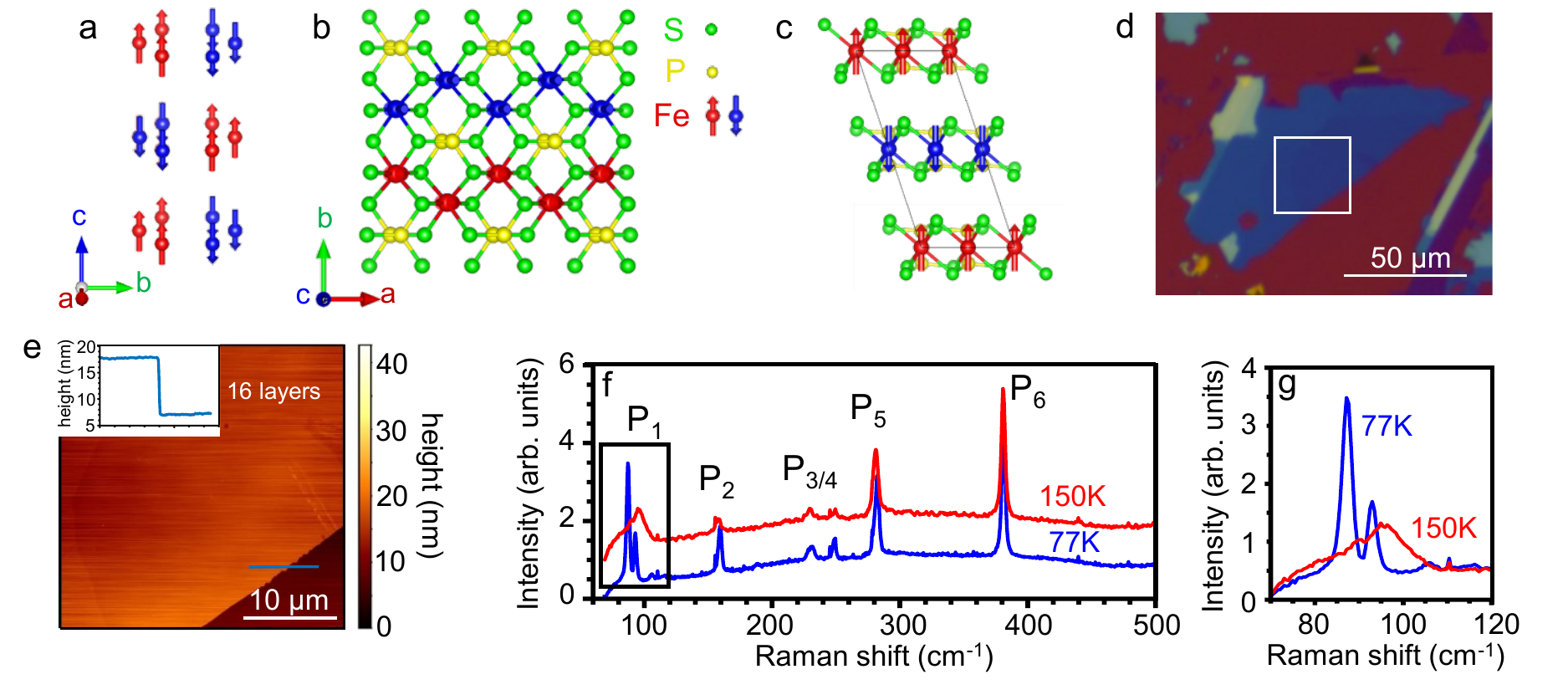}
\vspace{-0.4cm}
\caption{{\bf Magnetic structure, exfoliation and Raman measurements}. (a)  Magnetic arrangement of the Fe atoms in a tilted side view. Arrows are the magnetic moments of Fe with different directions highlighted by the color. (b) Top view of the atomic arrangement in FePS$_3$ using the same color code for the Fe spin directions as in a. (c) Side view along a zigzag direction with identical orientation of all Fe spins (Graphics are produced with the VESTA program \cite{VESTA}). (d) Optical microscope image of the FePS$_3$ flake used for  ARPES. It is exfoliated on Au/Ti/SiO$_2$/Si. (e) Tapping-mode atomic force microscopy image of the highlighted area in d recorded after the ARPES measurements, inset: height profile along the blue line resulting in a thickness of 16 layers. (f) Raman spectra of the probed flake after the ARPES measurements recorded above and below the Néel temperature $T_{\rm N}=117$\,K, spectra are vertically offset for clarity, excitation: $532$\,nm, $100$\,µW. P1-P6 mark known peaks of FePS$_3$ \cite{Lee2016b}. (g) Zoom into the black box of f around P1 exhibiting the known splittings of the broad P1 feature below $T_{\rm N}$ \cite{Lee2016b}. The lowest energy mode is a backfolded phononic excitation \cite{Lee2016b}.} \label{Fig_1}
\end{figure*}

An interesting class of 2D magnets are transition metal phosphorus trisulfides (TmPS$_3$) as a rare example of intralayer AFMs \cite{Joy1992,Wang2018,chittari2016,Kim2019b,Autieri2022}. 
They are  semiconductors with a layered honeycomb structure of the Tm atoms, each surrounded by covalently bonded (P$_2$S$_6$)$^{4-}$ bipyramids (Figure.~\ref{Fig_1}b--c). Their magnetism is governed by competing direct exchange between neighboring Tm atoms and indirect super-exchange interaction via S and P atoms \cite{Autieri2022,Yan2023,Rybak2024}. Depending on the relative strength of the competing interactions, zigzag, Néel, and stripy AFM order with different types of anisotropy have been found \cite{Joy1992,Sivadas2015}. Hence, the materials provide a rather unique platform to study distinct AFM arrangements that can additionally be tuned via
alloying of the transition metals \cite{Basnet2024,Basnet2022,Khan2024}.

FePS$_3$ is particularly interesting as it exhibits the zigzag type   
AFM structure with Ising-type out-of-plane spins (Néel temperature $T_{\rm N}=117$\,K). Each Fe atom is coupled ferromagnetically to two Fe neighbors and antiferromagnetically to one Fe neighbor (Figure~\ref{Fig_1}a) \cite{Jernberg1984}. Only one of the ferromagnetic bonds is longer by 0.14\,\AA\ according to density functional theory (DFT) calculations \cite{Amirabbasi2023} and neutron diffraction \cite{Lanon2016}. This surprising behaviour is attributed to a rotational ordering of the singly occupied Fe 3d$_{x^2-y^2}$ levels that optimizes the phase-matched overlap with adjacent in-plane S 3p levels along that bond \cite{Amirabbasi2023}. 
The resulting interrelation of structural and magnetic properties enables, e.\,g., to change $T_{\rm N}$ by strain with 100\,K/\% \cite{Siskins2020}, leads to a magnon-phonon anticrossing with a large cooperativity of 29 \cite{Vaclavkova2021,zhang2021b,Liu2021}, induces a 50\% frequency change for a suspended FePS$_3$ membrane across $T_{\rm N}$ \cite{Houmes2023,Siskins2023,Siskins2020}, and exhibits intertwined shear oscillations consisting of magnons and phonons after a laser pulse induced demagnetization \cite{Zhou2022b,Zong2023}.  
Additionally, the third nearest neighbor exchange constant, that propagates along the S and the P atoms, is as large as the next-nearest neighbor one \cite{Lanon2016,Kim2021,Amirabbasi2023}. 

To elucidate the complex interplay of orbitals in the magnetic exchange interaction, we performed $\mu$-ARPES \cite{Wiemann2011} on FePS$_3$ flakes above and below $T_{\rm N}$, using the approach that we recently developed for the isostructural MnPS$_3$ \cite{Strasdas2023}. We employed DFT+U calculations to attribute the orbital contributions to the measured bands and found three distinct changes across $T_{\rm N}$. Firstly, a low energy, nearly parabolic P 3p$_z$/S 3p$_z$ band shifts downward above $T_{\rm N}$ by about $80$\,meV with respect to the Fermi energy $E_{\rm F}$ and with respect to a pronounced, flat Fe 3d band. Secondly, Fe 3d bands that form a rhombus in the band structure by hybridizing with S 3p bands also shift downwards by a similar amount. Thirdly, a bunch of relatively flat S 3p bands mixed with Fe 3d levels and located close to the valence band maximum reorder, i.\,e. they form two separate bunches of bands above $T_{\rm N}$, but only a single broad bunch below $T_{\rm N}$. These three groups of bands also changed substantially in the DFT+U calculations between zigzag and Néel AFM configuration corroborating their susceptibility to the magnetic order. 
We conjecture that the involvement of orbitals from all elements of FePS$_3$  
is directly related to the multiple exchange paths of similar strengths in the material as outlined below.

\begin{figure*}[!htb]
\centering
\includegraphics[width=\textwidth]{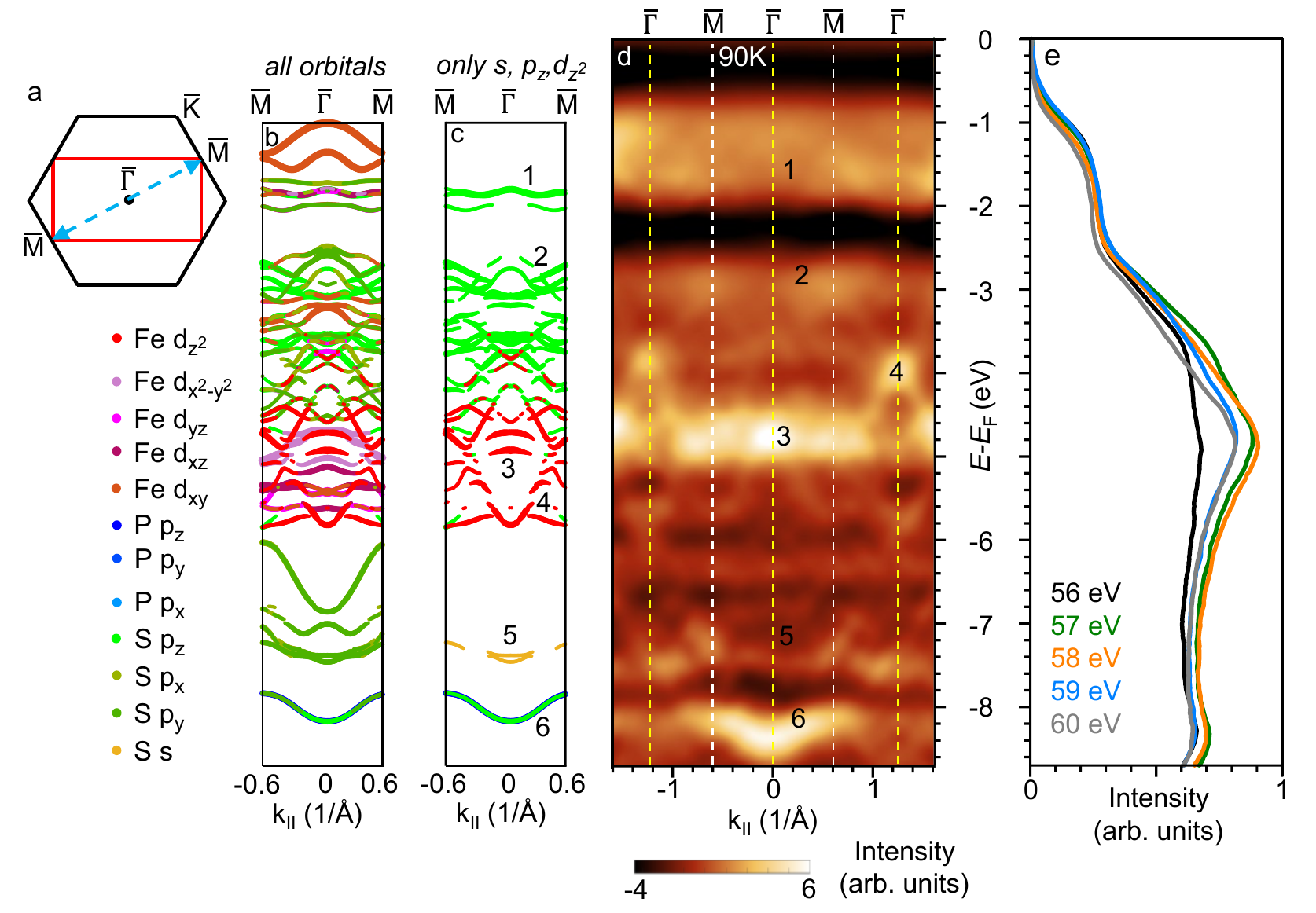}
\vspace{-0.4cm}
\caption{{\bf Electronic band structure of FePS$_3$ below $\bf{T_{\rm N}}$}. 
(a) Surface projected Brillouin zone (BZ) of the atomic arrangement (black) and of the AFM zigzag arrangement (red) of FePS$_3$ with marked high symmetry points and $k_\parallel$ direction of the plots in b--c (dashed blue arrow).
(b) Band structure obtained by DFT+U-J calculation after unfolding \cite{Popescu2012}, $U=1.4$\,eV, $J=0.2$\,eV, $k_z =-0.07/$\AA, all orbital contributions larger than $10$\,\% are displayed for each state, 
the line width of the colored bands is proportional to the orbital contribution, 100\,\% corresponds to the diameter of the symbol in the legend (left).
The energy axis is rigidly shifted to adapt to the experimental data in d revealing $E_{\rm VBM}-E_{\rm F}\approx 1.0$\,eV. (c) Same as b, but only displaying s- , p$_z$ and d$_{z^2}$-type orbital contributions (larger $10$\,\%) as required by the simplified selection rules \cite{Moser2017}. Numbers are for comparison with d (see text).
(d) ARPES curvature 
obtained after smoothing the raw data prior and after curvature determination (Suppl. Section 1H), $h\nu=58$\,eV, $T=90$\,K
(raw data in Suppl. Figure~S5). 
(e) Raw ARPES intensity  $I(E-E_{\rm F})$ at the central $\overline{\Gamma}$ for different photon energies $h\nu$ as color-coded, a resonant enhancement of the Fe 3d bands  appears at $E-E_{\rm F}\approx -5$\,eV, ARPES intensity is normalized to the photocurrent measured at the entrance  
of the analyzer.}
\label{Fig_2}
\end{figure*}

\section{Methods}
For preparing the FePS$_3$, we exfoliated thin flakes onto a conductive Au/Ti film deposited on Si/SiO$_2$, itself necessary for identifying the flakes optically (Suppl. Section S1B). The conductive Au/Ti layer prevents the charging of the sample due to photoelectrons, that usually emerges for semiconductors at low temperatures \cite{Bianchi2023,DeVita2022, Nitschke2023,Koitzsch2023, Voloshina2023}. The method has meanwhile also been applied successfully for CrSBr \cite{watson2024}.

Figures~\ref{Fig_1}d-e show an optical and an atomic force microscopy image of the FePS$_3$ flake area that is probed by ARPES. 
The flake is 16 layers thick (inset, Fig.~\ref{Fig_1}e) with rms surface roughness of $0.19$\,nm only. Raman spectroscopy was employed after ARPES to verify that the investigated flake indeed shows the expected magnetic phase transition. Previous Raman data revealed a strong change of the low energy Raman peaks rationalized by backfolding of phononic bands from the Brillouin zone edge to $\Gamma$ due to the enlarged magnetic unit cell  \cite{Lee2016b, Scagliotti1985, Scagliotti1987}. We observe the same signatures, i.\,e. two strong and a few weaker Raman peaks at P$_1$ (out-of-plane Fe oscillations) and a significant increase of Raman intensity for P$_2$ (in-plane Fe oscillations) below $T_{\rm N}$  (Figure~\ref{Fig_1}f-g). This corroborates that the FePS$_3$ flake exhibits the usual AFM phase transition into the zigzag phase.  

\section{Results}
Figure~\ref{Fig_2} compares the band structure from DFT+U calculations with the ARPES data. The surface projections of the geometric and the magnetic Brillouin zone (BZ) are shown in Figure~\ref{Fig_2}a with hexagonal and rectangular appearance, respectively. In the ARPES data, we do not observe any patterns related to the rectangular BZ as usual (Suppl. Section S1F), since the changes in the wave functions due to the AFM order are relatively weak \cite{Hüfner2013}.  Consequently, we display the calculated band structure of the AFM zigzag phase after unfolding the states to the hexagonal non-magnetic BZ via wave function projection \cite{Popescu2012} (Figure~\ref{Fig_2}b, Suppl. Section S2A/B). By detailed comparison with the ARPES data, we adapted the effective onsite energy at the Fe atom $U_{\rm eff}=U-J=1.2$\,eV ($J=0.2$\,eV: Hund's rule exchange coupling) and the out-of-plane wave vector $k_z=-0.07/$\AA\  (Suppl. Section S2D/F) \cite{Strasdas2023}.
A simplified selection rule, that projects the initial orbitals to final state plane waves via the matrix elements corresponding to our photon beam geometry \cite{Moser2017}, reveals that only states with the orbital character of s-, p$_z$ or d$_{z^2}$ are visible (Suppl. Section S2C) \cite{Strasdas2023}. We, hence display only these contributions, if larger than 10\,\%, in Figure~\ref{Fig_2}c. This separates energy regions with different orbitals, from top to bottom, namely S 3p$_z$ (feature 1,2), Fe 3d$_{z^2}$ (feature 3, 4), S 3s (feature 5) and a P 3p$_z$ band with some S 3p$_z$ contributions (feature 6). Notice that all these bands have additional contributions from other not visible orbitals (Suppl. Figure~S18). 

Figure~\ref{Fig_2}d shows ARPES data along $\overline{\rm M}\overline{\Gamma}\overline{\rm M}$ displayed as curvature  $C(E,k_\parallel)$ ($E$: energy, $k_\parallel$: in-plane wave vector) for  better visibility. The photoelectrons are recorded from a small area (diameter: $\sim 5$\,$\mu$m) of the 16 layer flake (Figure~\ref{Fig_1}d-e). The band energies are related to the Fermi level $E_{\rm F}$ measured on the Au substrate nearby.
The data are similar to previously published ARPES data of FePS$_3$ recorded at room temperature \cite{Yan2023,Nitschke2023, Koitzsch2023}. 
Figure~\ref{Fig_2}d displays three BZs, since some 
 features exhibit different visibility in different BZs. For example, a rhombus at $-4$\,eV to $-6$\,eV similar to the DFT band structure appears only in the second BZ (feature 4, Figure~\ref{Fig_2}c-d).
 Such differences can be rationalized by final state effects \cite{Hüfner2013}  beyond the projection 
 to final plane waves
 \cite{Moser2017}. 
 
 The calculated band structures have been rigidly shifted to match the measured ones as well as possible using the marked features 1--6.
 
\begin{figure*}[htb]
\centering
\includegraphics[width=\textwidth]{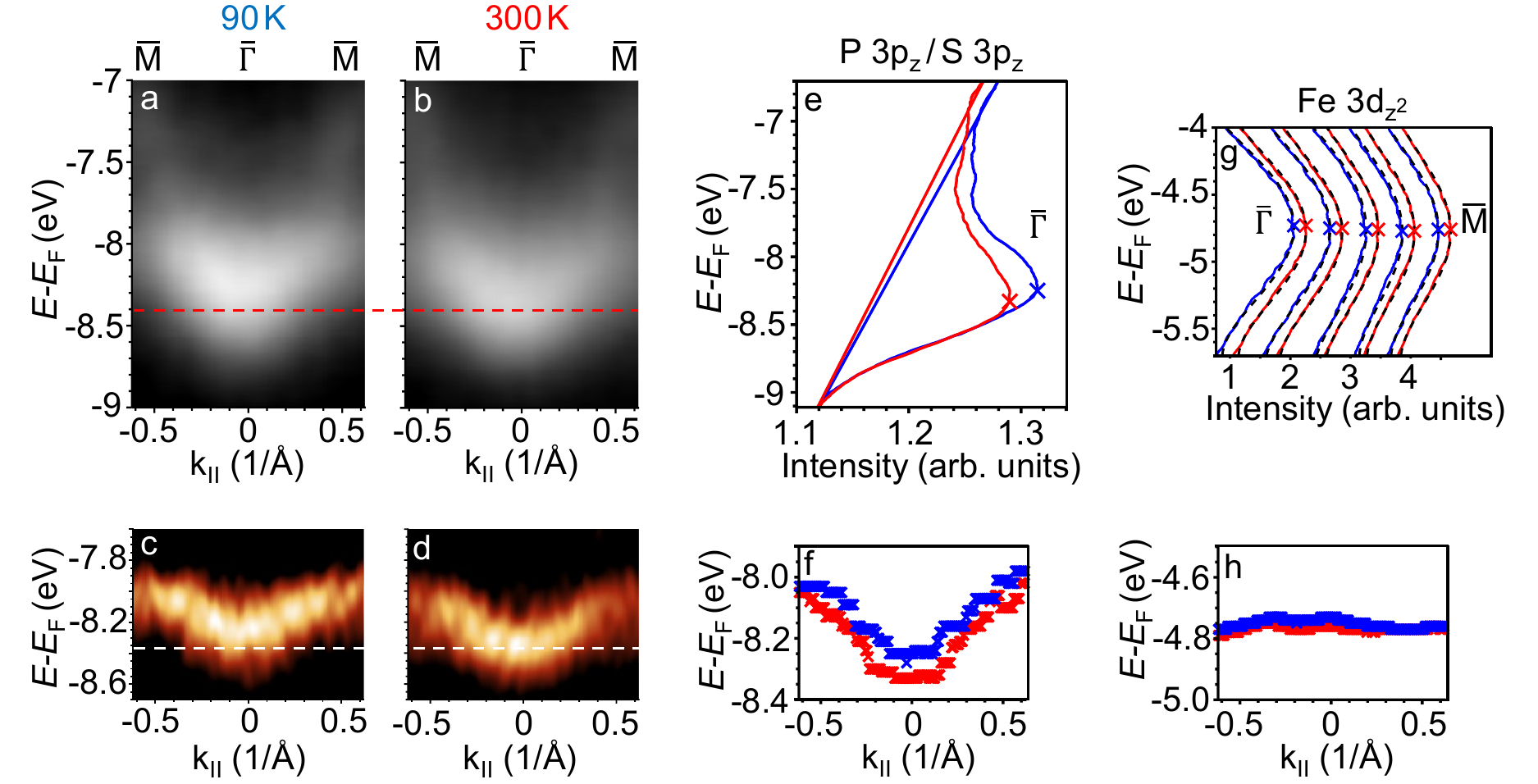}
\vspace{-0.4cm}
\caption{{\bf Energy shift of P 3p$_z$/S 3p$_z$ band across  $\bf{T_{\rm N}}$, $h\nu=58$\,eV}. 
(a), (b) ARPES intensity $I(E,k_\parallel)$ after linear background subtraction 
below and above $T_{\rm N}=117$\,K, dashed line: guide to the eye. (c), (d) Curvature $C(E,k_\parallel)$ deduced from a, b, respectively, with the same dashed line. (e) Raw data $I(E-E_{\rm F})$ at $\overline{\Gamma}$ for $T=90$\,K (blue), $T=300$\,K (red) with adapted linear background lines (same color) and marked maxima deduced after background subtraction (crosses).  
(f) Intensity maxima obtained as in e for various $k_\parallel$ along $\overline{\rm M}$$\overline{\Gamma}$$\overline{\rm M}$ (data from a,b), blue: $90$\,K, red: $300$\,K. 
(g) Raw data $I(E-E_{\rm F})$ at different $k_\parallel$ in the energy region of the flat Fe 3d$_{z^2}$ band (feature 3, Figure~\ref{Fig_2}d) at $90$\,K (blue) and $300$\,K (red), left curve at $\overline{\Gamma}$, right curve at $\overline{\rm M}$, curves in between at intermediate, equidistant $k_\parallel$, constant horizontal offset between red and blue curves. The marked maxima (crosses) are deduced from a fourth-order polynomial fit (dashed lines). 
(h) Intensity maxima obtained as in panel g for various $k_\parallel$ along $\overline{\rm M}$$\overline{\Gamma}$$\overline{\rm M}$, blue: $90$\,K, red: $300$\,K.}
\label{Fig_3}
\end{figure*}
The lowest energy band at $E-E_{\rm F} \approx -8.2$\,eV (feature 6) is most easily attributed to the mixed P 3p$_z$/S 3p$_z$ band in the calculations. The S 3s band of the calculations (feature 5) is faintly visible in the 1$^{\rm st}$ and the 2$^{\rm nd}$ BZ. The rhombus with dominating Fe 3d$_{z^2}$ character, spreading from $E-E_{\rm F}=-5.8$\,eV to $-4$\,eV (feature 4), is visible in the 2$^{\rm nd}$ BZ, but barely in the 1$^{\rm st}$ one. In contrast, the flat Fe 3d$_{z^2}$ bands in the center of the rhombus (feature 3) are more pronounced in the 1$^{\rm st}$ BZ. The strong dispersion of the Fe bands within the rhombus is likely caused by the formation of bonding and anti-bonding linear combinations with S 3p orbitals (Suppl. Figure~S18), most pronounced along the FM chains \cite{Amirabbasi2023}. The bunch of S 3p$_z$ bands at slightly higher energy (feature 2) exhibits a rather blurred structure in the experimental data similar to the flat bands (feature 1) appearing between $-2$\,eV and $-1$\,eV in the ARPES data. The highest occupied bands in the calculations correspond to  mixed Fe 3d$_{xy}$ / S 3p$_{x,y}$ orbitals that are not visible for our ARPES geometry due to the selection rules \cite{Moser2017}. Unfortunately, these bands carry the single 3d electron in the minority Fe spin channel and are, according to DFT calculations, centrally involved in the long-range magnetic interaction \cite{Amirabbasi2023}.

Importantly, we can attribute most features of the experimental data to bands of the calculation. However, the calculated bands appear to be compressed in energy with respect to the measured ones by $\sim 7$\,\% (Suppl. Figure~S9). This could not be cured by changing $U_{\rm eff}$ or $k_z$ (Suppl. Figure~S8, S10) and agrees roughly with the 17\,\% compression found earlier above $T_{\rm N}$ for a slightly larger $U_{\rm eff}$ \cite{Nitschke2023}. Hence, DFT+U is not sufficient to capture the measured band structure by ARPES quantitatively. 
A possible explanation is the dynamical, energy-dependent screening of the photohole that is generated during ARPES \cite{Hüfner2013}. It would require a calculation of the electronic self-energy which is beyond the scope of this work.
We note in passing that we observe an even stronger compression of the DFT+U data with respect to the ARPES data for NiPS$_3$ 
(to be published), but not for MnPS$_3$ \cite{Strasdas2023}.

To confirm that the measured features 3 and 4 are indeed caused by Fe 3d orbitals, we employed resonant excitations from occupied Fe 3p to empty Fe 3d levels that are known to enhance the photoelectron emission from occupied Fe 3d levels \cite{Choi1994,Jin2022}. The measured Fe 3p core level at $E-E_{\rm F}=-56$\,eV (not shown) implies a resonance at photon energies $h\nu=57-60$\,eV \cite{Choi1994,Jin2022}. For these $h\nu$, an enhancement of the photoelectron intensity is indeed observed in the energy region of the occupied Fe 3d$_{z^2}$ bands (Figure~\ref{Fig_2}e).

\begin{figure*}[!htb]
\centering
\includegraphics[width=\textwidth]{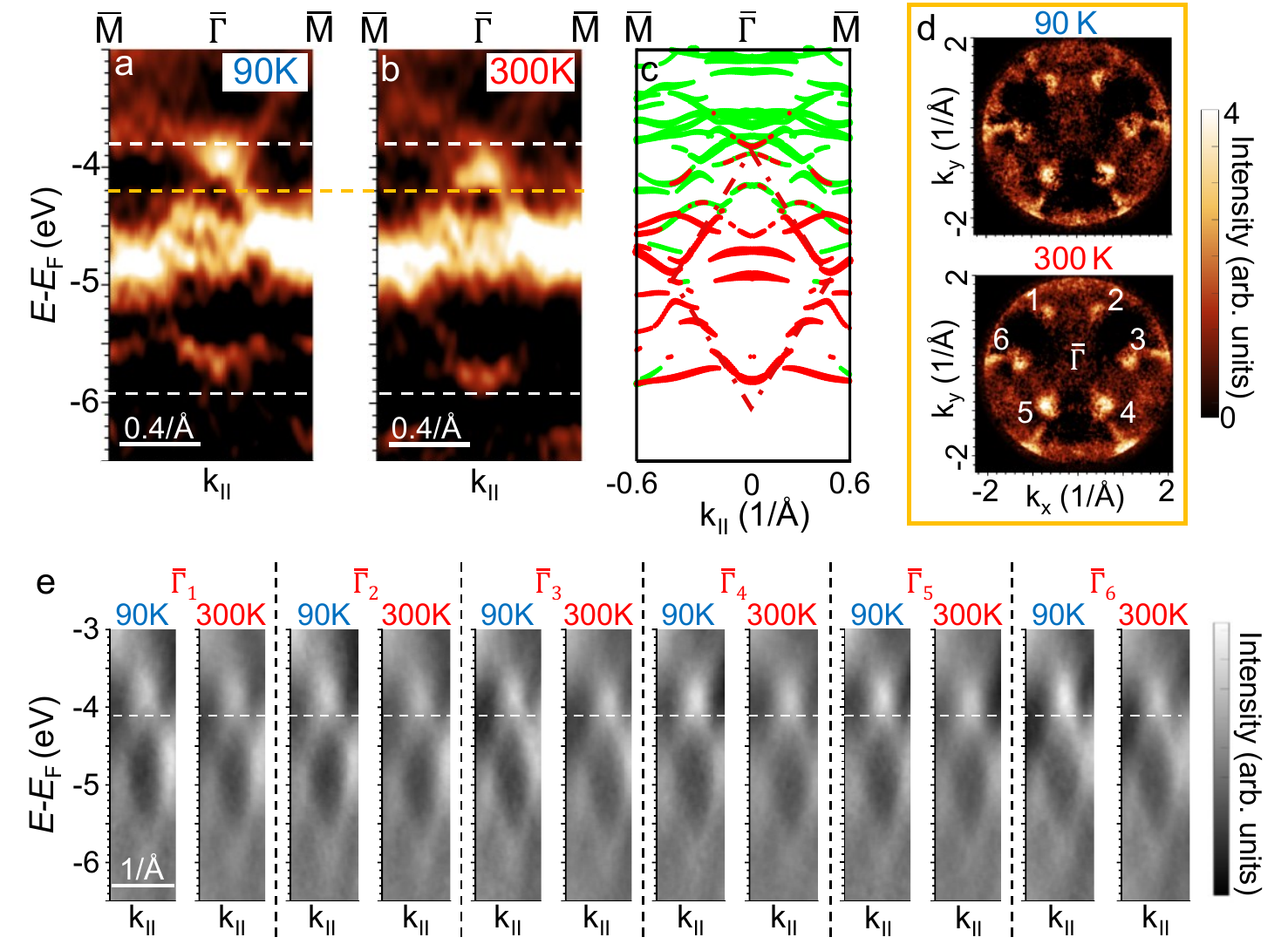}
\vspace{-0.3cm}
\caption{{\bf Energy shift of Fe 3d$_{z^2}$ dominated rhombus, $h\nu=58$\,eV}. 
(a), (b) Curvature plots $C(E,k_\parallel)$ centered at $\overline{\Gamma}_6$ as marked in d by 6, below and above $T_{\rm N}=117$\,K,  red dashed-dotted lines mark the rhombus, white dashed line: guide to the eye, orange dashed line: energy of $C(k_x,k_y)$ plots in d. (c) DFT+U band structure (selected orbitals) of the same area as in a, b (zoom into Figure~\ref{Fig_2}c) with highlighted rhombus (dashed-dotted red line).
(d) Curvature $C(k_x,k_y)$ integrated for $E-E_{\rm F}= -4.18$\,eV to $-4.2$\,eV below and above $T_{\rm N}$, $\overline{\Gamma}_n$ of different BZs are labeled in the bottom image, a, b and d share the same color scale.
(e) Raw ARPES data $I(E,k_\parallel)$ after background subtraction 
(Suppl. Section S1G) centered at the various $\overline{\Gamma}_n$ points as labeled in d, $\overline{\rm M \Gamma \rm M}$ direction, $T$ on top, white line marks the energy used in d. 
}
\label{Fig_5}
\end{figure*}

Having identified the orbital band characters, we next describe the observed band changes across $T_{\rm N}$. A comparison of the complete experimental band structures is displayed in Suppl. Figure~S13. 
The first change is an energy shift of the rather parabolic P 3p$_z$/S 3p$_z$ band at low energy (feature 6, Figure~\ref{Fig_2}d). Figures~\ref{Fig_3}a-d show the raw data and the curvature data of this band above and below $T_{\rm N}$. A slight shift downwards above $T_{\rm N}$ is apparent, more striking in the curvature data (Figure~\ref{Fig_3}c-d). To quantify the shift, we use the raw $I(E)$ data for each $k_\parallel$, shown for $\overline{\Gamma}$ in Figure~\ref{Fig_3}e, subtract a linear background (lines in Figure~\ref{Fig_3}e) and determine the resulting maximum (crosses in Figure~\ref{Fig_3}e). The maxima are displayed for both temperatures as a function of $k_\parallel$ in Figure~\ref{Fig_3}f exhibiting a rather rigid downwards shift by $\sim 80$\,meV  without changing the band dispersion significantly. Besides using $E_{\rm F}$ as a reference, that might change, e.g., by a change of band gap, we use the strong and flat Fe 3d band at $E-E_{\rm F}\approx -4.8$\,eV as an additional reference. Here, we use a polynomial fit to determine the maximum of $I(E)$ at various $k_\parallel$ along $\overline{\rm M}$$\overline{\Gamma}$$\overline{\rm M}$ (Fig.~\ref{Fig_3}g). The Fe band barely shifts across $T_{\rm N}$ on the $10$\,meV scale (Fig.~\ref{Fig_3}h). 
Interestingly, the optical band gap exhibits a blue shift of about $40$\,meV \cite{Budniak2022} below $T_{\rm N}$ exactly opposite to the trend of the parabolic P 3p$_z$/S 3p$_z$  band with respect to $E_{\rm F}$.

 \begin{figure*}[!htb]
\centering
\includegraphics[width=\textwidth]{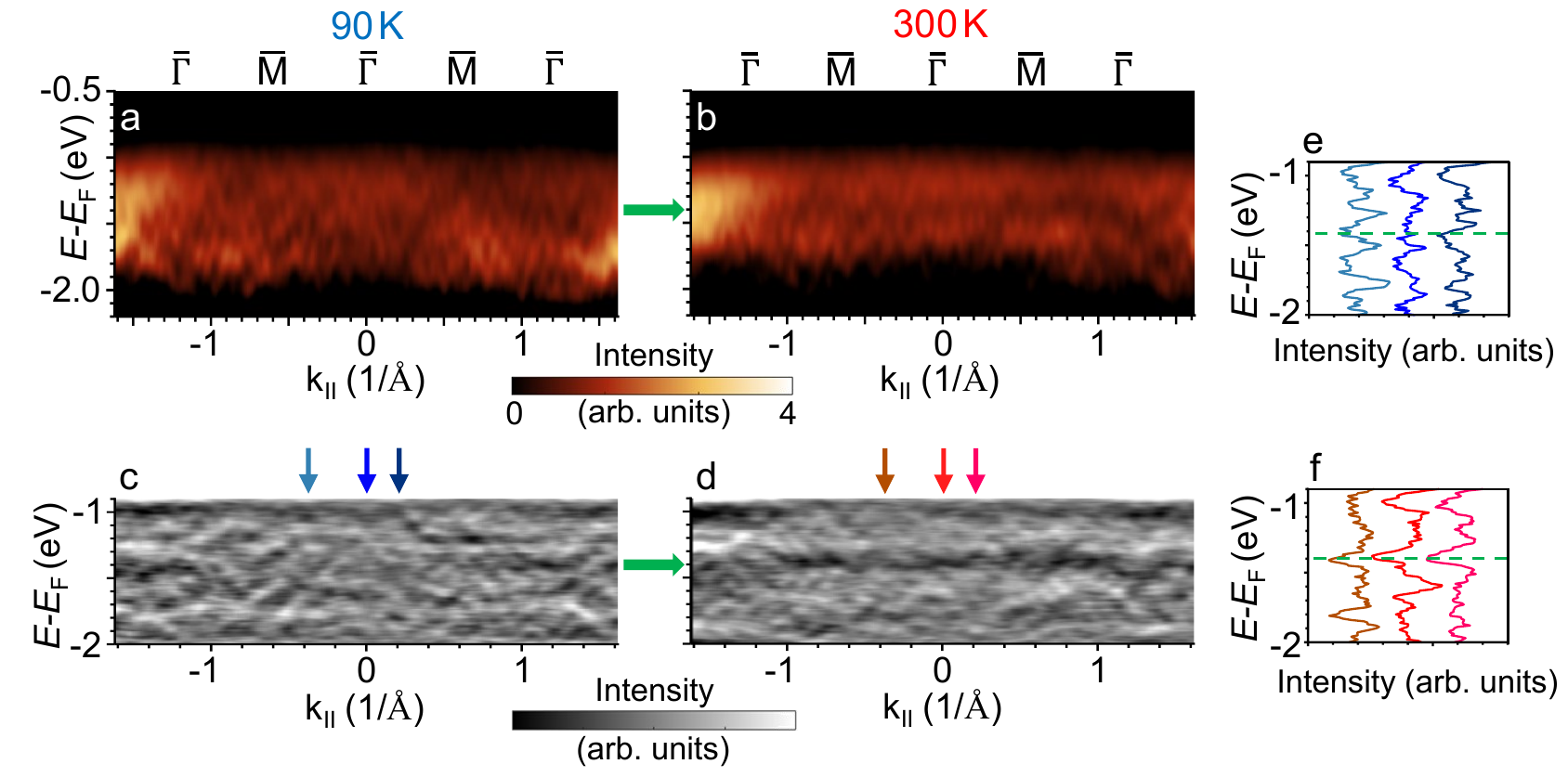}
\vspace{-0.8cm}
\caption{{\bf Band rearrangements in S 3p$_z$ valence band, $h\nu=58$\,eV}  (a), (b) Curvature plots $C(E, k_\parallel)$ at $T$ as given on top, 
green arrow marks the energy of reduced intensity in b. (c), (d) Raw ARPES data $I(E,k_\parallel)$ after fourth order polynomial background subtraction with green arrow at the same energy as in b, (c) $T=90$\,K, (d) $T=300$\,K, blue and red arrows: $k_\parallel$ positions of line cuts in e, f.  
(e), (f) Line cuts $I(E-E_{\rm F})$ at the $k_\parallel$ marked in c, d, respectively. Green dashed line: same energy as arrows in b, d, where a dip is consistently observed in e.}
\label{Fig_4}
\end{figure*}
A similar downward energy shift with respect to $E_{\rm F}$ is found for the rhombus feature visible in the 2$^{\rm nd}$ BZ and attributed to a dominating Fe 3d$_{z^2}$ character. Comparing the curvature plots (Figure~\ref{Fig_5}a-b) reveals that the top of the rhombus and the adjacent side arms are shifted downwards above $T_{\rm N}$. 
Figure~\ref{Fig_5}d shows the curvature $C(k_x,k_y)$ 
above and below $T_{\rm N}$, both at the same energy slightly below the top of the rhombus (orange line, Figure~\ref{Fig_5}a--b). Below $T_{\rm N}$, mostly a ring is observed around the various $\overline{\Gamma}_n$ of the 2$^{\rm nd}$ BZs that evolves into a disk above $T_{\rm N}$. This can be rationalized straightforwardly by the downward shift of the rhombus above $T_{\rm N}$.
To corroborate the shift within the raw data, Figure~\ref{Fig_5}e shows them  after background subtraction in the energy direction.
For this, we averaged the $I(E-E_{\rm F})$ curves in the displayed $k_\parallel$ range of each of the subfigures and subtracted the average 
from $I(E-E_{\rm F})$ of the individual $k_\parallel$ (Suppl. Section S1G).  Thus, we get rid of the otherwise dominating flat bands and most of the background originating from secondary electrons. 
All six $2^{\rm nd}$ BZs are displayed above and below $T_{\rm N}$. In the upper part of the rhombus, a downwards energy shift of the intensity maximum appears for all 2$^{\rm nd}$ BZs  above $T_{\rm N}$.
In the lower part, the transition from dark areas to bright areas shifts visibly downwards except for $\overline{\Gamma}_4$. 
The average energy shift is deduced from the intensity maxima at $\overline{\Gamma}$ and amounts to $\sim 75$\,meV 
(Suppl. Section S3C). Possible shape changes of the rhombus are difficult to extract 
due to variations between the different 2$^{\rm nd}$ BZs likely caused by the remaining noise in the data.

Finally, we compare the high energy valence band (feature 1, Figure~\ref{Fig_2}d), visible via its S 3p$_z$ character (orbital selection rules), but exhibiting contributions from the other S 3p levels and all Fe 3d levels except 3d$_{\rm z^2}$ (Suppl. Section S3E).
The curvature plots (Figure~\ref{Fig_4}a-b) reveal that the bands occupy a more localized energy region above $T_{\rm N}$ and exhibit a central stripe of reduced intensity (green arrow in Figure~\ref{Fig_4}b). This gap-like feature is corroborated by the background-subtracted raw data (Figure~\ref{Fig_4}c-d), where a reduced intensity appears at the same energy, and the $I(E-E_{\rm F})$ line cuts, where a dip is consistently observed above $T_{\rm N}$ (Figure~\ref{Fig_4}e-f). 

The energy shifts of the nearly parabolic S 3p$_z$/P 3p$_z$ band and the Fe 3d$_{z^2}$-type rhombus as well as the dip in the S 3p$_z$ bands at higher energy are nearly identically observed in the $\overline{{\rm K}\Gamma\rm{K}}$  direction (Suppl. Section S3D) implying little in-plane anisotropy. Nevertheless, the broken threefold rotational symmetry  (Figure~\ref{Fig_2}a), caused by the AFM zigzag phase \cite{Jernberg1984} and the structural distortion below $T_{\rm N}$ \cite{Lanon2016}, leads to non-equivalent $\overline{\rm M \Gamma \rm M}$ directions of the hexagonal, atomic BZ (Figure~\ref{Fig_2}a). Indeed, the calculated band structure shows related differences 
(Suppl. Figure~S17). However, they are local in $\boldsymbol{k}_\parallel$ space and relatively small in energy ($<100$\,meV) such that   
we could not spot them in the ARPES data (Suppl. Section S3D) 
likely due to the finite energy resolution ($50$\,meV, Suppl. Section S1C). 

\section{Conclusions}
To rationalize the observed band structure changes across $T_{\rm N}$, we performed additional DFT+U calculations for various magnetic configurations (Suppl. Section 3F). 
This emulates the variety of magnetic environments of the Fe  moments in a paramagnet. 
While in the zigzag ground state (Figure~\ref{Fig_1}a) each Fe atom has two aligned and one anti-aligned Fe neighbor, in a Néel AFM all neighboring magnetic moments are antiparallel.
The two calculations indeed exhibit pronounced variations for the bands that experimentally change across $T_{\rm N}$.  The P 3p$_z$/S 3p$_z$ band and the Fe 3d$_{z^2}$ rhombus shift upwards by 100-200\,meV for the Néel phase, if compared to the zigzag configuration. The upper S 3p$_z$ band also shifts upwards and, in addition, gets more structured  (Suppl. Figure~S19).  
This corroborates that the experimentally identified bands are indeed most strongly influenced by the relative magnetic orientation of the Fe atoms. 
Finally, we comment on the relation of the band changes to the magnetic exchange paths. The dominant part of the magnetic exchange in FePS$_3$ involves next neighbor superexchange of the Fe 3d levels via neighboring S 3p orbitals  (Figure~\ref{Fig_1}b) \cite{Lanon2016,Amirabbasi2023,Kim2021,Yang2024}. The upper S 3p bands are indeed  strongly hybridized with neighboring Fe 3d$_{xy}$, 3d$_{xz}$ and 3d$_{yz}$ orbitals (Suppl. Figure~S18). This straightforwardly explains why they change across $T_{\rm N}$ via the changed superexchange. The Fe 3d$_{z^2}$ rhombus has significant contributions from all S 3p orbitals as well (Suppl. Figure~S18), hence is also involved in the next-neighbor superexchange. Its energetic shift could be caused by a charge rearrangement within the Fe multiplets. Interestingly, the Fe 3d contributions and the S 3p ones of the rhombus separate in energy for the Néel AFM (Suppl. Figure~S19). This implies that their hybridization is crucial for the FM exchange only present in the zigzag configuration as expected naively from the close to 90$^\circ$ Fe-S-Fe bond configuration. Besides the nearest-neighbor superexchange,
there is also a strong AFM-type interaction between third-nearest Fe neighbors \cite{Lanon2016, Amirabbasi2023,Kim2021}. It proceeds along a Fe - S - P - S - Fe path (Figure~\ref{Fig_1}b) exploiting the attractive potential of the P dumbbell to overlap neighboring S 3p orbitals \cite{Yang2024}.
The resulting additional charge at the dumbbell might explain the upward shift of the P 3p$_z$/S 3p$_z$ band.
Hence, all three chemical elements are involved in the magnetic transition as reflected in the change of related bands from all elements.

In summary, we have probed the band structure of exfoliated FePS$_3$ flakes above and below $T_{\rm N}$
pinpointing three changes. A nearly parabolic band of mixed P 3p$_z$ and S 3p$_z$ character shifts up in energy by 80\,meV below $T_{\rm N}$. A similar energy shift is found for a rhombus-like feature around $\overline{\Gamma}$ that has a dominating Fe 3d$_{z^2}$ character with minor S 3p contributions and is mostly visible in the 2$^{\rm nd}$ BZ. Finally, upper valence bands of mixed S 3p and Fe 3d (except 3d$_{z^2}$) characters get more distributed in energy below $T_{\rm N}$ closing a gap-like feature at its central energy range. The fact that all elements are involved in the changing bands, partly without any direct orbital contributions from Fe, underlines the complex magnetic interactions in the material involving strong superexchange-type mechanisms.

\section{Experimental Section}

Details on sample growth, sample preparation, characterization, Raman, ARPES and DFT+U calculations are provided in the Supporting Information.

\section*{Acknowledgement}
We are grateful to M. Amirabbasi for supporting us with the DFT+U calculations and thank M. Birowska and M. Rybak for helpful discussions. 
We acknowledge financial support from the German Research Foundation (DFG) via the project Mo 858/19-1, the German Ministry of Education and Research (Project 05K2022 -ioARPES) and Germany’s Excellence Strategy — Cluster of Excellence: Matter and Light for Quantum Computing(ML4Q) EXC 2004/1 (project Nr. 390534769). 
We acknowledge access to the NanoESCA beamline of Elettra, the Italian synchrotron facility. E.L. is supported by the Israel Science Foundation (project no. 2528/19) and by Deutsch - Israelisches Programm (project no. NA1223/2-1).
A.K.B and E.L. were supported by the European Commission via the Marie-Sklodowska Curie action Phonsi (H2020-MSCA-ITN-642656).

\section*{Conflict of Interest}
The authors declare no conflict of interest.



\begin{thebibliography}{60}
\expandafter\ifx\csname natexlab\endcsname\relax\def\natexlab#1{#1}\fi
\expandafter\ifx\csname bibnamefont\endcsname\relax
  \def\bibnamefont#1{#1}\fi
\expandafter\ifx\csname bibfnamefont\endcsname\relax
  \def\bibfnamefont#1{#1}\fi
\expandafter\ifx\csname citenamefont\endcsname\relax
  \def\citenamefont#1{#1}\fi
\expandafter\ifx\csname url\endcsname\relax
  \def\url#1{\texttt{#1}}\fi
\expandafter\ifx\csname urlprefix\endcsname\relax\def\urlprefix{URL }\fi
\providecommand{\bibinfo}[2]{#2}
\providecommand{\eprint}[2][]{\url{#2}}

\bibitem[{\citenamefont{Huang et~al.}(2017)\citenamefont{Huang, Clark, Navarro-Moratalla, Klein, Cheng, Seyler, Zhong, Schmidgall, McGuire, Cobden et~al.}}]{Huang2017}
\bibinfo{author}{\bibfnamefont{B.}~\bibnamefont{Huang}}, \bibinfo{author}{\bibfnamefont{G.}~\bibnamefont{Clark}}, \bibinfo{author}{\bibfnamefont{E.}~\bibnamefont{Navarro-Moratalla}}, \bibinfo{author}{\bibfnamefont{D.~R.} \bibnamefont{Klein}}, \bibinfo{author}{\bibfnamefont{R.}~\bibnamefont{Cheng}}, \bibinfo{author}{\bibfnamefont{K.~L.} \bibnamefont{Seyler}}, \bibinfo{author}{\bibfnamefont{D.}~\bibnamefont{Zhong}}, \bibinfo{author}{\bibfnamefont{E.}~\bibnamefont{Schmidgall}}, \bibinfo{author}{\bibfnamefont{M.~A.} \bibnamefont{McGuire}}, \bibinfo{author}{\bibfnamefont{D.~H.} \bibnamefont{Cobden}}, \bibinfo{author}{\bibfnamefont{W.}~\bibnamefont{Yao}}, \bibinfo{author}{\bibfnamefont{D.}~\bibnamefont{Xiao}}, \bibinfo{author}{\bibfnamefont{P.}~\bibnamefont{Jarillo-Herrero}}, \bibnamefont{and} \bibinfo{author}{\bibfnamefont{X.}~\bibnamefont{Xu}}, \emph{\bibinfo{title}{Layer-dependent ferromagnetism in a van der Waals crystal down to the monolayer limit}}, \bibinfo{journal}{Nature} \textbf{\bibinfo{volume}{546}},
  \bibinfo{pages}{270} (\bibinfo{year}{2017}), \urlprefix\url{https://doi.org/10.1038/nature22391}.

\bibitem[{\citenamefont{Sun et~al.}(2019)\citenamefont{Sun, Yi, Song, Clark, Huang, Shan, Wu, Huang, Gao, Chen et~al.}}]{Sun2019b}
\bibinfo{author}{\bibfnamefont{Z.}~\bibnamefont{Sun}}, \bibinfo{author}{\bibfnamefont{Y.}~\bibnamefont{Yi}}, \bibinfo{author}{\bibfnamefont{T.}~\bibnamefont{Song}}, \bibinfo{author}{\bibfnamefont{G.}~\bibnamefont{Clark}}, \bibinfo{author}{\bibfnamefont{B.}~\bibnamefont{Huang}}, \bibinfo{author}{\bibfnamefont{Y.}~\bibnamefont{Shan}}, \bibinfo{author}{\bibfnamefont{S.}~\bibnamefont{Wu}}, \bibinfo{author}{\bibfnamefont{D.}~\bibnamefont{Huang}}, \bibinfo{author}{\bibfnamefont{C.}~\bibnamefont{Gao}}, \bibinfo{author}{\bibfnamefont{Z.}~\bibnamefont{Chen}}, \bibinfo{author}{\bibfnamefont{M.}~\bibnamefont{McGuire}}, \bibinfo{author}{\bibfnamefont{T.}~\bibnamefont{Cao}}, \bibinfo{author}{\bibfnamefont{D.}~\bibnamefont{Xiao}}, \bibinfo{author}{\bibfnamefont{W.-T.} \bibnamefont{Liu}}, \bibinfo{author}{\bibfnamefont{W.}~\bibnamefont{Yao}}, \bibinfo{author}{\bibfnamefont{X.}~\bibnamefont{Xu}}, \bibnamefont{and} \bibinfo{author}{\bibfnamefont{S.}~\bibnamefont{Wu}}, \emph{\bibinfo{title}{Giant nonreciprocal second-harmonic
  generation from antiferromagnetic bilayer CrI3}}, \bibinfo{journal}{Nature} \textbf{\bibinfo{volume}{572}}, \bibinfo{pages}{497} (\bibinfo{year}{2019}), ISSN \bibinfo{issn}{1476-4687}, \urlprefix\url{http://dx.doi.org/10.1038/s41586-019-1445-3}.

\bibitem[{\citenamefont{Gong et~al.}(2017)\citenamefont{Gong, Li, Li, Ji, Stern, Xia, Cao, Bao, Wang, Wang et~al.}}]{Gong2017}
\bibinfo{author}{\bibfnamefont{C.}~\bibnamefont{Gong}}, \bibinfo{author}{\bibfnamefont{L.}~\bibnamefont{Li}}, \bibinfo{author}{\bibfnamefont{Z.}~\bibnamefont{Li}}, \bibinfo{author}{\bibfnamefont{H.}~\bibnamefont{Ji}}, \bibinfo{author}{\bibfnamefont{A.}~\bibnamefont{Stern}}, \bibinfo{author}{\bibfnamefont{Y.}~\bibnamefont{Xia}}, \bibinfo{author}{\bibfnamefont{T.}~\bibnamefont{Cao}}, \bibinfo{author}{\bibfnamefont{W.}~\bibnamefont{Bao}}, \bibinfo{author}{\bibfnamefont{C.}~\bibnamefont{Wang}}, \bibinfo{author}{\bibfnamefont{Y.}~\bibnamefont{Wang}}, \bibinfo{author}{\bibfnamefont{Z.~Q.} \bibnamefont{Qiu}}, \bibinfo{author}{\bibfnamefont{R.~J.} \bibnamefont{Cava}}, \bibinfo{author}{\bibfnamefont{S.~G.} \bibnamefont{Louie}}, \bibinfo{author}{\bibfnamefont{J.}~\bibnamefont{Xia}}, \bibnamefont{and} \bibinfo{author}{\bibfnamefont{X.}~\bibnamefont{Zhang}}, \emph{\bibinfo{title}{Discovery of intrinsic ferromagnetism in two-dimensional van der Waals crystals}}, \bibinfo{journal}{Nature} \textbf{\bibinfo{volume}{546}},
  \bibinfo{pages}{265} (\bibinfo{year}{2017}), \urlprefix\url{https://doi.org/10.1038/nature22060}.

\bibitem[{\citenamefont{Wang et~al.}(2018{\natexlab{a}})\citenamefont{Wang, Zhang, Ding, Dong, Li, Chen, Li, Huang, Wang, Zhao et~al.}}]{Wang2018d}
\bibinfo{author}{\bibfnamefont{Z.}~\bibnamefont{Wang}}, \bibinfo{author}{\bibfnamefont{T.}~\bibnamefont{Zhang}}, \bibinfo{author}{\bibfnamefont{M.}~\bibnamefont{Ding}}, \bibinfo{author}{\bibfnamefont{B.}~\bibnamefont{Dong}}, \bibinfo{author}{\bibfnamefont{Y.}~\bibnamefont{Li}}, \bibinfo{author}{\bibfnamefont{M.}~\bibnamefont{Chen}}, \bibinfo{author}{\bibfnamefont{X.}~\bibnamefont{Li}}, \bibinfo{author}{\bibfnamefont{J.}~\bibnamefont{Huang}}, \bibinfo{author}{\bibfnamefont{H.}~\bibnamefont{Wang}}, \bibinfo{author}{\bibfnamefont{X.}~\bibnamefont{Zhao}}, \bibinfo{author}{\bibfnamefont{Y.}~\bibnamefont{Li}}, \bibinfo{author}{\bibfnamefont{D.}~\bibnamefont{Li}}, \bibinfo{author}{\bibfnamefont{C.}~\bibnamefont{Jia}}, \bibinfo{author}{\bibfnamefont{L.}~\bibnamefont{Sun}}, \bibinfo{author}{\bibfnamefont{H.}~\bibnamefont{Guo}}, \bibinfo{author}{\bibfnamefont{Y.}~\bibnamefont{Ye}}, \bibinfo{author}{\bibfnamefont{D.}~\bibnamefont{Sun}}, \bibinfo{author}{\bibfnamefont{Y.}~\bibnamefont{Chen}},
  \bibinfo{author}{\bibfnamefont{T.}~\bibnamefont{Yang}}, \bibinfo{author}{\bibfnamefont{J.}~\bibnamefont{Zhang}}, \bibinfo{author}{\bibfnamefont{S.}~\bibnamefont{Ono}}, \bibinfo{author}{\bibfnamefont{Z.}~\bibnamefont{Han}}, \bibnamefont{and} \bibinfo{author}{\bibfnamefont{Z.}~\bibnamefont{Zhang}}, \emph{\bibinfo{title}{Electric-field control of magnetism in a few-layered van der Waals ferromagnetic semiconductor}}, \bibinfo{journal}{Nat. Nanotechnol.} \textbf{\bibinfo{volume}{13}}, \bibinfo{pages}{554} (\bibinfo{year}{2018}{\natexlab{a}}), ISSN \bibinfo{issn}{1748-3395}, \urlprefix\url{http://dx.doi.org/10.1038/s41565-018-0186-z}.

\bibitem[{\citenamefont{Deng et~al.}(2018)\citenamefont{Deng, Yu, Song, Zhang, Wang, Sun, Yi, Wu, Wu, Zhu et~al.}}]{Deng2018}
\bibinfo{author}{\bibfnamefont{Y.}~\bibnamefont{Deng}}, \bibinfo{author}{\bibfnamefont{Y.}~\bibnamefont{Yu}}, \bibinfo{author}{\bibfnamefont{Y.}~\bibnamefont{Song}}, \bibinfo{author}{\bibfnamefont{J.}~\bibnamefont{Zhang}}, \bibinfo{author}{\bibfnamefont{N.~Z.} \bibnamefont{Wang}}, \bibinfo{author}{\bibfnamefont{Z.}~\bibnamefont{Sun}}, \bibinfo{author}{\bibfnamefont{Y.}~\bibnamefont{Yi}}, \bibinfo{author}{\bibfnamefont{Y.~Z.} \bibnamefont{Wu}}, \bibinfo{author}{\bibfnamefont{S.}~\bibnamefont{Wu}}, \bibinfo{author}{\bibfnamefont{J.}~\bibnamefont{Zhu}}, \bibinfo{author}{\bibfnamefont{J.}~\bibnamefont{Wang}}, \bibinfo{author}{\bibfnamefont{X.~H.} \bibnamefont{Chen}}, \bibnamefont{and} \bibinfo{author}{\bibfnamefont{Y.}~\bibnamefont{Zhang}}, \emph{\bibinfo{title}{Gate-tunable room-temperature ferromagnetism in two-dimensional Fe3GeTe2}}, \bibinfo{journal}{Nature} \textbf{\bibinfo{volume}{563}}, \bibinfo{pages}{94} (\bibinfo{year}{2018}), \urlprefix\url{https://doi.org/10.1038/s41586-018-0626-9}.

\bibitem[{\citenamefont{Fei et~al.}(2018)\citenamefont{Fei, Huang, Malinowski, Wang, Song, Sanchez, Yao, Xiao, Zhu, May et~al.}}]{Fei2018}
\bibinfo{author}{\bibfnamefont{Z.}~\bibnamefont{Fei}}, \bibinfo{author}{\bibfnamefont{B.}~\bibnamefont{Huang}}, \bibinfo{author}{\bibfnamefont{P.}~\bibnamefont{Malinowski}}, \bibinfo{author}{\bibfnamefont{W.}~\bibnamefont{Wang}}, \bibinfo{author}{\bibfnamefont{T.}~\bibnamefont{Song}}, \bibinfo{author}{\bibfnamefont{J.}~\bibnamefont{Sanchez}}, \bibinfo{author}{\bibfnamefont{W.}~\bibnamefont{Yao}}, \bibinfo{author}{\bibfnamefont{D.}~\bibnamefont{Xiao}}, \bibinfo{author}{\bibfnamefont{X.}~\bibnamefont{Zhu}}, \bibinfo{author}{\bibfnamefont{A.~F.} \bibnamefont{May}}, \bibinfo{author}{\bibfnamefont{W.}~\bibnamefont{Wu}}, \bibinfo{author}{\bibfnamefont{D.~H.} \bibnamefont{Cobden}}, \bibinfo{author}{\bibfnamefont{J.-H.} \bibnamefont{Chu}}, \bibnamefont{and} \bibinfo{author}{\bibfnamefont{X.}~\bibnamefont{Xu}}, \emph{\bibinfo{title}{Two-dimensional itinerant ferromagnetism in atomically thin Fe3GeTe2}}, \bibinfo{journal}{Nat. Mater.} \textbf{\bibinfo{volume}{17}}, \bibinfo{pages}{778} (\bibinfo{year}{2018}),
  \urlprefix\url{https://doi.org/10.1038/s41563-018-0149-7}.

\bibitem[{\citenamefont{Li et~al.}(2019)\citenamefont{Li, Ruan, and Zeng}}]{Li2019}
\bibinfo{author}{\bibfnamefont{H.}~\bibnamefont{Li}}, \bibinfo{author}{\bibfnamefont{S.}~\bibnamefont{Ruan}}, \bibnamefont{and} \bibinfo{author}{\bibfnamefont{Y.}~\bibnamefont{Zeng}}, \emph{\bibinfo{title}{Intrinsic Van Der Waals Magnetic Materials from Bulk to the 2D Limit: New Frontiers of Spintronics}}, \bibinfo{journal}{Adv. Mater.} \textbf{\bibinfo{volume}{31}}, \bibinfo{pages}{1900065} (\bibinfo{year}{2019}), ISSN \bibinfo{issn}{1521-4095}, \urlprefix\url{http://dx.doi.org/10.1002/adma.201900065}.

\bibitem[{\citenamefont{Zhang et~al.}(2019)\citenamefont{Zhang, Wong, Zhu, and Wee}}]{Zhang2019}
\bibinfo{author}{\bibfnamefont{W.}~\bibnamefont{Zhang}}, \bibinfo{author}{\bibfnamefont{P.~K.~J.} \bibnamefont{Wong}}, \bibinfo{author}{\bibfnamefont{R.}~\bibnamefont{Zhu}}, \bibnamefont{and} \bibinfo{author}{\bibfnamefont{A.~T.~S.} \bibnamefont{Wee}}, \emph{\bibinfo{title}{Van der Waals magnets: Wonder building blocks for two‐dimensional spintronics?}}, \bibinfo{journal}{InfoMat} \textbf{\bibinfo{volume}{1}}, \bibinfo{pages}{479} (\bibinfo{year}{2019}), ISSN \bibinfo{issn}{2567-3165}, \urlprefix\url{http://dx.doi.org/10.1002/inf2.12048}.

\bibitem[{\citenamefont{Burch et~al.}(2018)\citenamefont{Burch, Mandrus, and Park}}]{Burch2018}
\bibinfo{author}{\bibfnamefont{K.~S.} \bibnamefont{Burch}}, \bibinfo{author}{\bibfnamefont{D.}~\bibnamefont{Mandrus}}, \bibnamefont{and} \bibinfo{author}{\bibfnamefont{J.-G.} \bibnamefont{Park}}, \emph{\bibinfo{title}{Magnetism in two-dimensional van der Waals materials}}, \bibinfo{journal}{Nature} \textbf{\bibinfo{volume}{563}}, \bibinfo{pages}{47} (\bibinfo{year}{2018}), \urlprefix\url{https://doi.org/10.1038/s41586-018-0631-z}.

\bibitem[{\citenamefont{Gong and Zhang}(2019)}]{Gong2019}
\bibinfo{author}{\bibfnamefont{C.}~\bibnamefont{Gong}} \bibnamefont{and} \bibinfo{author}{\bibfnamefont{X.}~\bibnamefont{Zhang}}, \emph{\bibinfo{title}{Two-dimensional magnetic crystals and emergent heterostructure devices}}, \bibinfo{journal}{Science} \textbf{\bibinfo{volume}{363}}, \bibinfo{pages}{706} (\bibinfo{year}{2019}), \urlprefix\url{https://doi.org/10.1126/science.aav4450}.

\bibitem[{\citenamefont{Huang et~al.}(2020)\citenamefont{Huang, McGuire, May, Xiao, Jarillo-Herrero, and Xu}}]{Huang2020}
\bibinfo{author}{\bibfnamefont{B.}~\bibnamefont{Huang}}, \bibinfo{author}{\bibfnamefont{M.~A.} \bibnamefont{McGuire}}, \bibinfo{author}{\bibfnamefont{A.~F.} \bibnamefont{May}}, \bibinfo{author}{\bibfnamefont{D.}~\bibnamefont{Xiao}}, \bibinfo{author}{\bibfnamefont{P.}~\bibnamefont{Jarillo-Herrero}}, \bibnamefont{and} \bibinfo{author}{\bibfnamefont{X.}~\bibnamefont{Xu}}, \emph{\bibinfo{title}{Emergent phenomena and proximity effects in two-dimensional magnets and heterostructures}}, \bibinfo{journal}{Nat. Mater.} \textbf{\bibinfo{volume}{19}}, \bibinfo{pages}{1276} (\bibinfo{year}{2020}), \urlprefix\url{https://doi.org/10.1038/s41563-020-0791-8}.

\bibitem[{\citenamefont{Mi et~al.}(2022)\citenamefont{Mi, Zheng, Wang, Zhou, Yu, Xiao, Song, Shen, Li, Bai et~al.}}]{Mi2022}
\bibinfo{author}{\bibfnamefont{M.}~\bibnamefont{Mi}}, \bibinfo{author}{\bibfnamefont{X.}~\bibnamefont{Zheng}}, \bibinfo{author}{\bibfnamefont{S.}~\bibnamefont{Wang}}, \bibinfo{author}{\bibfnamefont{Y.}~\bibnamefont{Zhou}}, \bibinfo{author}{\bibfnamefont{L.}~\bibnamefont{Yu}}, \bibinfo{author}{\bibfnamefont{H.}~\bibnamefont{Xiao}}, \bibinfo{author}{\bibfnamefont{H.}~\bibnamefont{Song}}, \bibinfo{author}{\bibfnamefont{B.}~\bibnamefont{Shen}}, \bibinfo{author}{\bibfnamefont{F.}~\bibnamefont{Li}}, \bibinfo{author}{\bibfnamefont{L.}~\bibnamefont{Bai}}, \bibinfo{author}{\bibfnamefont{Y.}~\bibnamefont{Chen}}, \bibinfo{author}{\bibfnamefont{S.}~\bibnamefont{Wang}}, \bibinfo{author}{\bibfnamefont{X.}~\bibnamefont{Liu}}, \bibnamefont{and} \bibinfo{author}{\bibfnamefont{Y.}~\bibnamefont{Wang}}, \emph{\bibinfo{title}{Variation between Antiferromagnetism and Ferrimagnetism in NiPS3 by Electron Doping}}, \bibinfo{journal}{Adv. Funct. Mater.} \textbf{\bibinfo{volume}{32}}, \bibinfo{pages}{2112750} (\bibinfo{year}{2022}),
  ISSN \bibinfo{issn}{1616-3028}, \urlprefix\url{http://dx.doi.org/10.1002/adfm.202112750}.

\bibitem[{\citenamefont{Ahn et~al.}(2024)\citenamefont{Ahn, Guo, Son, Sun, and Zhao}}]{Ahn2024}
\bibinfo{author}{\bibfnamefont{Y.}~\bibnamefont{Ahn}}, \bibinfo{author}{\bibfnamefont{X.}~\bibnamefont{Guo}}, \bibinfo{author}{\bibfnamefont{S.}~\bibnamefont{Son}}, \bibinfo{author}{\bibfnamefont{Z.}~\bibnamefont{Sun}}, \bibnamefont{and} \bibinfo{author}{\bibfnamefont{L.}~\bibnamefont{Zhao}}, \emph{\bibinfo{title}{Progress and prospects in two-dimensional magnetism of van der Waals materials}}, \bibinfo{journal}{Progr. Quant. Electr.} \textbf{\bibinfo{volume}{93}}, \bibinfo{pages}{100498} (\bibinfo{year}{2024}), ISSN \bibinfo{issn}{0079-6727}, \urlprefix\url{http://dx.doi.org/10.1016/j.pquantelec.2024.100498}.

\bibitem[{\citenamefont{Wang et~al.}(2022)\citenamefont{Wang, Cao, Li, Lu, Cohen, Haldar, Kitadai, Tan, Burch, Smirnov et~al.}}]{Wang2022}
\bibinfo{author}{\bibfnamefont{X.}~\bibnamefont{Wang}}, \bibinfo{author}{\bibfnamefont{J.}~\bibnamefont{Cao}}, \bibinfo{author}{\bibfnamefont{H.}~\bibnamefont{Li}}, \bibinfo{author}{\bibfnamefont{Z.}~\bibnamefont{Lu}}, \bibinfo{author}{\bibfnamefont{A.}~\bibnamefont{Cohen}}, \bibinfo{author}{\bibfnamefont{A.}~\bibnamefont{Haldar}}, \bibinfo{author}{\bibfnamefont{H.}~\bibnamefont{Kitadai}}, \bibinfo{author}{\bibfnamefont{Q.}~\bibnamefont{Tan}}, \bibinfo{author}{\bibfnamefont{K.~S.} \bibnamefont{Burch}}, \bibinfo{author}{\bibfnamefont{D.}~\bibnamefont{Smirnov}}, \bibinfo{author}{\bibfnamefont{W.}~\bibnamefont{Xu}}, \bibinfo{author}{\bibfnamefont{S.}~\bibnamefont{Sharifzadeh}}, \bibinfo{author}{\bibfnamefont{L.}~\bibnamefont{Liang}}, \bibnamefont{and} \bibinfo{author}{\bibfnamefont{X.}~\bibnamefont{Ling}}, \emph{\bibinfo{title}{Electronic Raman scattering in the 2D antiferromagnet {NiPS}$_3$}}, \bibinfo{journal}{Sci. Adv.} \textbf{\bibinfo{volume}{8}}, \bibinfo{pages}{eabl7707} (\bibinfo{year}{2022}),
  \urlprefix\url{https://doi.org/10.1126/sciadv.abl7707}.

\bibitem[{\citenamefont{Vaclavkova et~al.}(2021)\citenamefont{Vaclavkova, Palit, Wyzula, Ghosh, Delhomme, Maity, Kapuscinski, Ghosh, Veis, Grzeszczyk et~al.}}]{Vaclavkova2021}
\bibinfo{author}{\bibfnamefont{D.}~\bibnamefont{Vaclavkova}}, \bibinfo{author}{\bibfnamefont{M.}~\bibnamefont{Palit}}, \bibinfo{author}{\bibfnamefont{J.}~\bibnamefont{Wyzula}}, \bibinfo{author}{\bibfnamefont{S.}~\bibnamefont{Ghosh}}, \bibinfo{author}{\bibfnamefont{A.}~\bibnamefont{Delhomme}}, \bibinfo{author}{\bibfnamefont{S.}~\bibnamefont{Maity}}, \bibinfo{author}{\bibfnamefont{P.}~\bibnamefont{Kapuscinski}}, \bibinfo{author}{\bibfnamefont{A.}~\bibnamefont{Ghosh}}, \bibinfo{author}{\bibfnamefont{M.}~\bibnamefont{Veis}}, \bibinfo{author}{\bibfnamefont{M.}~\bibnamefont{Grzeszczyk}}, \bibinfo{author}{\bibfnamefont{C.}~\bibnamefont{Faugeras}}, \bibinfo{author}{\bibfnamefont{M.}~\bibnamefont{Orlita}}, \bibinfo{author}{\bibfnamefont{S.}~\bibnamefont{Datta}}, \bibnamefont{and} \bibinfo{author}{\bibfnamefont{M.}~\bibnamefont{Potemski}}, \emph{\bibinfo{title}{Magnon polarons in the van der Waals antiferromagnet {FePS}3}}, \bibinfo{journal}{Phys. Rev. B} \textbf{\bibinfo{volume}{104}}, \bibinfo{pages}{134437}
  (\bibinfo{year}{2021}), \urlprefix\url{https://doi.org/10.1103/physrevb.104.134437}.

\bibitem[{\citenamefont{Ressouche et~al.}(2010)\citenamefont{Ressouche, Loire, Simonet, Ballou, Stunault, and Wildes}}]{Ressouche2010}
\bibinfo{author}{\bibfnamefont{E.}~\bibnamefont{Ressouche}}, \bibinfo{author}{\bibfnamefont{M.}~\bibnamefont{Loire}}, \bibinfo{author}{\bibfnamefont{V.}~\bibnamefont{Simonet}}, \bibinfo{author}{\bibfnamefont{R.}~\bibnamefont{Ballou}}, \bibinfo{author}{\bibfnamefont{A.}~\bibnamefont{Stunault}}, \bibnamefont{and} \bibinfo{author}{\bibfnamefont{A.}~\bibnamefont{Wildes}}, \emph{\bibinfo{title}{Magnetoelectric ${\text{MnPS}}_{3}$ as a candidate for ferrotoroidicity}}, \bibinfo{journal}{Phys. Rev. B} \textbf{\bibinfo{volume}{82}}, \bibinfo{pages}{100408} (\bibinfo{year}{2010}), \urlprefix\url{https://link.aps.org/doi/10.1103/PhysRevB.82.100408}.

\bibitem[{\citenamefont{Kang et~al.}(2020)\citenamefont{Kang, Kim, Kim, Kim, Sim, Lee, Lee, Park, Yun, Kim et~al.}}]{Kang2020}
\bibinfo{author}{\bibfnamefont{S.}~\bibnamefont{Kang}}, \bibinfo{author}{\bibfnamefont{K.}~\bibnamefont{Kim}}, \bibinfo{author}{\bibfnamefont{B.~H.} \bibnamefont{Kim}}, \bibinfo{author}{\bibfnamefont{J.}~\bibnamefont{Kim}}, \bibinfo{author}{\bibfnamefont{K.~I.} \bibnamefont{Sim}}, \bibinfo{author}{\bibfnamefont{J.-U.} \bibnamefont{Lee}}, \bibinfo{author}{\bibfnamefont{S.}~\bibnamefont{Lee}}, \bibinfo{author}{\bibfnamefont{K.}~\bibnamefont{Park}}, \bibinfo{author}{\bibfnamefont{S.}~\bibnamefont{Yun}}, \bibinfo{author}{\bibfnamefont{T.}~\bibnamefont{Kim}}, \bibinfo{author}{\bibfnamefont{A.}~\bibnamefont{Nag}}, \bibinfo{author}{\bibfnamefont{A.}~\bibnamefont{Walters}}, \bibinfo{author}{\bibfnamefont{M.}~\bibnamefont{Garcia-Fernandez}}, \bibinfo{author}{\bibfnamefont{J.}~\bibnamefont{Li}}, \bibinfo{author}{\bibfnamefont{L.}~\bibnamefont{Chapon}}, \bibinfo{author}{\bibfnamefont{K.-J.} \bibnamefont{Zhou}}, \bibinfo{author}{\bibfnamefont{Y.-W.} \bibnamefont{Son}}, \bibinfo{author}{\bibfnamefont{J.~H.}
  \bibnamefont{Kim}}, \bibinfo{author}{\bibfnamefont{H.}~\bibnamefont{Cheong}}, \bibnamefont{and} \bibinfo{author}{\bibfnamefont{J.-G.} \bibnamefont{Park}}, \emph{\bibinfo{title}{Coherent many-body exciton in van der Waals antiferromagnet {NiPS}$_3$}}, \bibinfo{journal}{Nature} \textbf{\bibinfo{volume}{583}}, \bibinfo{pages}{785} (\bibinfo{year}{2020}), \urlprefix\url{https://doi.org/10.1038/s41586-020-2520-5}.

\bibitem[{\citenamefont{Hwangbo et~al.}(2021)\citenamefont{Hwangbo, Zhang, Jiang, Wang, Fonseca, Wang, Diederich, Gamelin, Xiao, Chu et~al.}}]{Hwangbo2021}
\bibinfo{author}{\bibfnamefont{K.}~\bibnamefont{Hwangbo}}, \bibinfo{author}{\bibfnamefont{Q.}~\bibnamefont{Zhang}}, \bibinfo{author}{\bibfnamefont{Q.}~\bibnamefont{Jiang}}, \bibinfo{author}{\bibfnamefont{Y.}~\bibnamefont{Wang}}, \bibinfo{author}{\bibfnamefont{J.}~\bibnamefont{Fonseca}}, \bibinfo{author}{\bibfnamefont{C.}~\bibnamefont{Wang}}, \bibinfo{author}{\bibfnamefont{G.~M.} \bibnamefont{Diederich}}, \bibinfo{author}{\bibfnamefont{D.~R.} \bibnamefont{Gamelin}}, \bibinfo{author}{\bibfnamefont{D.}~\bibnamefont{Xiao}}, \bibinfo{author}{\bibfnamefont{J.-H.} \bibnamefont{Chu}}, \bibinfo{author}{\bibfnamefont{W.}~\bibnamefont{Yao}}, \bibnamefont{and} \bibinfo{author}{\bibfnamefont{X.}~\bibnamefont{Xu}}, \emph{\bibinfo{title}{Highly anisotropic excitons and multiple phonon bound states in a van der Waals antiferromagnetic insulator}}, \bibinfo{journal}{Nat. Nanotechnol.} \textbf{\bibinfo{volume}{16}}, \bibinfo{pages}{655} (\bibinfo{year}{2021}), \urlprefix\url{https://doi.org/10.1038/s41565-021-00873-9}.

\bibitem[{\citenamefont{Dirnberger et~al.}(2022)\citenamefont{Dirnberger, Bushati, Datta, Kumar, MacDonald, Baldini, and Menon}}]{Dirnberger2022}
\bibinfo{author}{\bibfnamefont{F.}~\bibnamefont{Dirnberger}}, \bibinfo{author}{\bibfnamefont{R.}~\bibnamefont{Bushati}}, \bibinfo{author}{\bibfnamefont{B.}~\bibnamefont{Datta}}, \bibinfo{author}{\bibfnamefont{A.}~\bibnamefont{Kumar}}, \bibinfo{author}{\bibfnamefont{A.~H.} \bibnamefont{MacDonald}}, \bibinfo{author}{\bibfnamefont{E.}~\bibnamefont{Baldini}}, \bibnamefont{and} \bibinfo{author}{\bibfnamefont{V.~M.} \bibnamefont{Menon}}, \emph{\bibinfo{title}{Spin-correlated exciton{\textendash}polaritons in a van der Waals magnet}}, \bibinfo{journal}{Nat. Nanotechnol.} \textbf{\bibinfo{volume}{17}}, \bibinfo{pages}{1060} (\bibinfo{year}{2022}), \urlprefix\url{https://doi.org/10.1038/s41565-022-01204-2}.

\bibitem[{\citenamefont{Joy and Vasudevan}(1992)}]{Joy1992}
\bibinfo{author}{\bibfnamefont{P.~A.} \bibnamefont{Joy}} \bibnamefont{and} \bibinfo{author}{\bibfnamefont{S.}~\bibnamefont{Vasudevan}}, \emph{\bibinfo{title}{Magnetism in the layered transition-metal {thiophosphates MPS}$_3$ (M=Mn, Fe, and Ni)}}, \bibinfo{journal}{Phys. Rev. B} \textbf{\bibinfo{volume}{46}}, \bibinfo{pages}{5425} (\bibinfo{year}{1992}), \urlprefix\url{https://doi.org/10.1103/physrevb.46.5425}.

\bibitem[{\citenamefont{Wang et~al.}(2018{\natexlab{b}})\citenamefont{Wang, Shifa, Yu, He, Liu, Wang, Wang, Zhan, Lou, Xia et~al.}}]{Wang2018}
\bibinfo{author}{\bibfnamefont{F.}~\bibnamefont{Wang}}, \bibinfo{author}{\bibfnamefont{T.~A.} \bibnamefont{Shifa}}, \bibinfo{author}{\bibfnamefont{P.}~\bibnamefont{Yu}}, \bibinfo{author}{\bibfnamefont{P.}~\bibnamefont{He}}, \bibinfo{author}{\bibfnamefont{Y.}~\bibnamefont{Liu}}, \bibinfo{author}{\bibfnamefont{F.}~\bibnamefont{Wang}}, \bibinfo{author}{\bibfnamefont{Z.}~\bibnamefont{Wang}}, \bibinfo{author}{\bibfnamefont{X.}~\bibnamefont{Zhan}}, \bibinfo{author}{\bibfnamefont{X.}~\bibnamefont{Lou}}, \bibinfo{author}{\bibfnamefont{F.}~\bibnamefont{Xia}}, \bibnamefont{and} \bibinfo{author}{\bibfnamefont{J.}~\bibnamefont{He}}, \emph{\bibinfo{title}{New Frontiers on van der Waals Layered Metal Phosphorous Trichalcogenides}}, \bibinfo{journal}{Adv. Funct. Mater.} \textbf{\bibinfo{volume}{28}}, \bibinfo{pages}{1802151} (\bibinfo{year}{2018}{\natexlab{b}}), \urlprefix\url{https://doi.org/10.1002/adfm.201802151}.

\bibitem[{\citenamefont{Chittari et~al.}(2016)\citenamefont{Chittari, Park, Lee, Han, MacDonald, Hwang, and Jung}}]{chittari2016}
\bibinfo{author}{\bibfnamefont{B.~L.} \bibnamefont{Chittari}}, \bibinfo{author}{\bibfnamefont{Y.}~\bibnamefont{Park}}, \bibinfo{author}{\bibfnamefont{D.}~\bibnamefont{Lee}}, \bibinfo{author}{\bibfnamefont{M.}~\bibnamefont{Han}}, \bibinfo{author}{\bibfnamefont{A.~H.} \bibnamefont{MacDonald}}, \bibinfo{author}{\bibfnamefont{E.}~\bibnamefont{Hwang}}, \bibnamefont{and} \bibinfo{author}{\bibfnamefont{J.}~\bibnamefont{Jung}}, \emph{\bibinfo{title}{Electronic and magnetic properties of single-layer $M\mathrm{P}{X}_{3}$ metal phosphorous trichalcogenides}}, \bibinfo{journal}{Phys. Rev. B} \textbf{\bibinfo{volume}{94}}, \bibinfo{pages}{184428} (\bibinfo{year}{2016}).

\bibitem[{\citenamefont{Kim et~al.}(2019)\citenamefont{Kim, Lim, Lee, Lee, Kim, Park, Jeon, Park, Park, and Cheong}}]{Kim2019b}
\bibinfo{author}{\bibfnamefont{K.}~\bibnamefont{Kim}}, \bibinfo{author}{\bibfnamefont{S.~Y.} \bibnamefont{Lim}}, \bibinfo{author}{\bibfnamefont{J.-U.} \bibnamefont{Lee}}, \bibinfo{author}{\bibfnamefont{S.}~\bibnamefont{Lee}}, \bibinfo{author}{\bibfnamefont{T.~Y.} \bibnamefont{Kim}}, \bibinfo{author}{\bibfnamefont{K.}~\bibnamefont{Park}}, \bibinfo{author}{\bibfnamefont{G.~S.} \bibnamefont{Jeon}}, \bibinfo{author}{\bibfnamefont{C.-H.} \bibnamefont{Park}}, \bibinfo{author}{\bibfnamefont{J.-G.} \bibnamefont{Park}}, \bibnamefont{and} \bibinfo{author}{\bibfnamefont{H.}~\bibnamefont{Cheong}}, \emph{\bibinfo{title}{Suppression of magnetic ordering in {XXZ}-type antiferromagnetic monolayer {NiPS}$_3$}}, \bibinfo{journal}{Nat. Commun.} \textbf{\bibinfo{volume}{10}}, \bibinfo{pages}{345} (\bibinfo{year}{2019}), \urlprefix\url{https://doi.org/10.1038/s41467-018-08284-6}.

\bibitem[{\citenamefont{Autieri et~al.}(2022)\citenamefont{Autieri, Cuono, Noce, Rybak, Kotur, Agrapidis, Wohlfeld, and Birowska}}]{Autieri2022}
\bibinfo{author}{\bibfnamefont{C.}~\bibnamefont{Autieri}}, \bibinfo{author}{\bibfnamefont{G.}~\bibnamefont{Cuono}}, \bibinfo{author}{\bibfnamefont{C.}~\bibnamefont{Noce}}, \bibinfo{author}{\bibfnamefont{M.}~\bibnamefont{Rybak}}, \bibinfo{author}{\bibfnamefont{K.~M.} \bibnamefont{Kotur}}, \bibinfo{author}{\bibfnamefont{C.~E.} \bibnamefont{Agrapidis}}, \bibinfo{author}{\bibfnamefont{K.}~\bibnamefont{Wohlfeld}}, \bibnamefont{and} \bibinfo{author}{\bibfnamefont{M.}~\bibnamefont{Birowska}}, \emph{\bibinfo{title}{Limited Ferromagnetic Interactions in Monolayers of MPS$_3$ (M = Mn and Ni)}}, \bibinfo{journal}{J. Phys. Chem. C} \textbf{\bibinfo{volume}{126}}, \bibinfo{pages}{6791} (\bibinfo{year}{2022}), ISSN \bibinfo{issn}{1932-7447}, \urlprefix\url{https://doi.org/10.1021/acs.jpcc.2c00646}.

\bibitem[{\citenamefont{Yan et~al.}(2023)\citenamefont{Yan, Du, Zhang, Wan, and Wang}}]{Yan2023}
\bibinfo{author}{\bibfnamefont{S.}~\bibnamefont{Yan}}, \bibinfo{author}{\bibfnamefont{Y.}~\bibnamefont{Du}}, \bibinfo{author}{\bibfnamefont{X.}~\bibnamefont{Zhang}}, \bibinfo{author}{\bibfnamefont{X.}~\bibnamefont{Wan}}, \bibnamefont{and} \bibinfo{author}{\bibfnamefont{D.}~\bibnamefont{Wang}}, \emph{\bibinfo{title}{First-principles study of magnetic interactions and excitations in antiferromagnetic van der Waals material MPX3 (M=Mn, Fe, Co, Ni; X=S, Se)}}, \bibinfo{journal}{J. Phys. Condens. Matter} \textbf{\bibinfo{volume}{36}}, \bibinfo{pages}{065502} (\bibinfo{year}{2023}), ISSN \bibinfo{issn}{1361-648X}, \urlprefix\url{http://dx.doi.org/10.1088/1361-648X/ad06ef}.

\bibitem[{\citenamefont{Rybak et~al.}(2024)\citenamefont{Rybak, Faria~Junior, Woźniak, Scharoch, Fabian, and Birowska}}]{Rybak2024}
\bibinfo{author}{\bibfnamefont{M.}~\bibnamefont{Rybak}}, \bibinfo{author}{\bibfnamefont{P.~E.} \bibnamefont{Faria~Junior}}, \bibinfo{author}{\bibfnamefont{T.}~\bibnamefont{Woźniak}}, \bibinfo{author}{\bibfnamefont{P.}~\bibnamefont{Scharoch}}, \bibinfo{author}{\bibfnamefont{J.}~\bibnamefont{Fabian}}, \bibnamefont{and} \bibinfo{author}{\bibfnamefont{M.}~\bibnamefont{Birowska}}, \emph{\bibinfo{title}{Magneto-optical anisotropies of two-dimensional antiferromagnetic MPX$_3$ from first principles}}, \bibinfo{journal}{Phys. Rev. B} \textbf{\bibinfo{volume}{109}}, \bibinfo{pages}{054426} (\bibinfo{year}{2024}), ISSN \bibinfo{issn}{2469-9969}, \urlprefix\url{http://dx.doi.org/10.1103/PhysRevB.109.054426}.

\bibitem[{\citenamefont{Sivadas et~al.}(2015)\citenamefont{Sivadas, Daniels, Swendsen, Okamoto, and Xiao}}]{Sivadas2015}
\bibinfo{author}{\bibfnamefont{N.}~\bibnamefont{Sivadas}}, \bibinfo{author}{\bibfnamefont{M.~W.} \bibnamefont{Daniels}}, \bibinfo{author}{\bibfnamefont{R.~H.} \bibnamefont{Swendsen}}, \bibinfo{author}{\bibfnamefont{S.}~\bibnamefont{Okamoto}}, \bibnamefont{and} \bibinfo{author}{\bibfnamefont{D.}~\bibnamefont{Xiao}}, \emph{\bibinfo{title}{Magnetic ground state of semiconducting transition-metal trichalcogenide monolayers}}, \bibinfo{journal}{Phys. Rev. B} \textbf{\bibinfo{volume}{91}}, \bibinfo{pages}{235425} (\bibinfo{year}{2015}), \urlprefix\url{https://doi.org/10.1103/physrevb.91.235425}.

\bibitem[{\citenamefont{Basnet et~al.}(2024)\citenamefont{Basnet, Patel, Wang, Upreti, Chhetri, Acharya, Nabi, Sakon, and Hu}}]{Basnet2024}
\bibinfo{author}{\bibfnamefont{R.}~\bibnamefont{Basnet}}, \bibinfo{author}{\bibfnamefont{T.}~\bibnamefont{Patel}}, \bibinfo{author}{\bibfnamefont{J.}~\bibnamefont{Wang}}, \bibinfo{author}{\bibfnamefont{D.}~\bibnamefont{Upreti}}, \bibinfo{author}{\bibfnamefont{S.~K.} \bibnamefont{Chhetri}}, \bibinfo{author}{\bibfnamefont{G.}~\bibnamefont{Acharya}}, \bibinfo{author}{\bibfnamefont{M.~R.~U.} \bibnamefont{Nabi}}, \bibinfo{author}{\bibfnamefont{J.}~\bibnamefont{Sakon}}, \bibnamefont{and} \bibinfo{author}{\bibfnamefont{J.}~\bibnamefont{Hu}}, \emph{\bibinfo{title}{Understanding and Tuning Magnetism in Layered Ising‐Type Antiferromagnet FePSe3 for Potential 2D Magnet}}, \bibinfo{journal}{Adv. Electr. Mater.} p. \bibinfo{pages}{2300738} (\bibinfo{year}{2024}), ISSN \bibinfo{issn}{2199-160X}, \urlprefix\url{http://dx.doi.org/10.1002/aelm.202300738}.

\bibitem[{\citenamefont{Basnet et~al.}(2022)\citenamefont{Basnet, Kotur, Rybak, Stephenson, Bishop, Autieri, Birowska, and Hu}}]{Basnet2022}
\bibinfo{author}{\bibfnamefont{R.}~\bibnamefont{Basnet}}, \bibinfo{author}{\bibfnamefont{K.~M.} \bibnamefont{Kotur}}, \bibinfo{author}{\bibfnamefont{M.}~\bibnamefont{Rybak}}, \bibinfo{author}{\bibfnamefont{C.}~\bibnamefont{Stephenson}}, \bibinfo{author}{\bibfnamefont{S.}~\bibnamefont{Bishop}}, \bibinfo{author}{\bibfnamefont{C.}~\bibnamefont{Autieri}}, \bibinfo{author}{\bibfnamefont{M.}~\bibnamefont{Birowska}}, \bibnamefont{and} \bibinfo{author}{\bibfnamefont{J.}~\bibnamefont{Hu}}, \emph{\bibinfo{title}{Controlling magnetic exchange and anisotropy by nonmagnetic ligand substitution in layered $M\mathrm{P}{X}_{3}$ ($M=\mathrm{Ni}$, Mn; $X=\mathrm{S}$, Se)}}, \bibinfo{journal}{Phys. Rev. Research} \textbf{\bibinfo{volume}{4}}, \bibinfo{pages}{023256} (\bibinfo{year}{2022}), \urlprefix\url{https://link.aps.org/doi/10.1103/PhysRevResearch.4.023256}.

\bibitem[{\citenamefont{Khan et~al.}(2024)\citenamefont{Khan, Kumar, Kumar, Shemerliuk, Selter, B\"{u}chner, Pal, Aswartham, and Kumar}}]{Khan2024}
\bibinfo{author}{\bibfnamefont{N.}~\bibnamefont{Khan}}, \bibinfo{author}{\bibfnamefont{D.}~\bibnamefont{Kumar}}, \bibinfo{author}{\bibfnamefont{V.}~\bibnamefont{Kumar}}, \bibinfo{author}{\bibfnamefont{Y.}~\bibnamefont{Shemerliuk}}, \bibinfo{author}{\bibfnamefont{S.}~\bibnamefont{Selter}}, \bibinfo{author}{\bibfnamefont{B.}~\bibnamefont{B\"{u}chner}}, \bibinfo{author}{\bibfnamefont{K.}~\bibnamefont{Pal}}, \bibinfo{author}{\bibfnamefont{S.}~\bibnamefont{Aswartham}}, \bibnamefont{and} \bibinfo{author}{\bibfnamefont{P.}~\bibnamefont{Kumar}}, \emph{\bibinfo{title}{Interplay of topology and antiferromagnetic order in two-dimensional van der Waals crystals of (NixFe1-x)2P2S6}}, \bibinfo{journal}{2D Mater.} pp. \bibinfo{pages}{DOI 10.1088/2053--1583/ad3e0a} (\bibinfo{year}{2024}), ISSN \bibinfo{issn}{2053-1583}, \urlprefix\url{http://dx.doi.org/10.1088/2053-1583/ad3e0a}.

\bibitem[{\citenamefont{Momma and Izumi}(2011)}]{VESTA}
\bibinfo{author}{\bibfnamefont{K.}~\bibnamefont{Momma}} \bibnamefont{and} \bibinfo{author}{\bibfnamefont{F.}~\bibnamefont{Izumi}}, \emph{\bibinfo{title}{{{\it VESTA3} for three-dimensional visualization of crystal, volumetric and morphology data}}}, \bibinfo{journal}{J. Appl. Crystal.} \textbf{\bibinfo{volume}{44}}, \bibinfo{pages}{1272} (\bibinfo{year}{2011}), \urlprefix\url{https://doi.org/10.1107/S0021889811038970}.

\bibitem[{\citenamefont{Lee et~al.}(2016)\citenamefont{Lee, Lee, Ryoo, Kang, Kim, Kim, Park, Park, and Cheong}}]{Lee2016b}
\bibinfo{author}{\bibfnamefont{J.-U.} \bibnamefont{Lee}}, \bibinfo{author}{\bibfnamefont{S.}~\bibnamefont{Lee}}, \bibinfo{author}{\bibfnamefont{J.~H.} \bibnamefont{Ryoo}}, \bibinfo{author}{\bibfnamefont{S.}~\bibnamefont{Kang}}, \bibinfo{author}{\bibfnamefont{T.~Y.} \bibnamefont{Kim}}, \bibinfo{author}{\bibfnamefont{P.}~\bibnamefont{Kim}}, \bibinfo{author}{\bibfnamefont{C.-H.} \bibnamefont{Park}}, \bibinfo{author}{\bibfnamefont{J.-G.} \bibnamefont{Park}}, \bibnamefont{and} \bibinfo{author}{\bibfnamefont{H.}~\bibnamefont{Cheong}}, \emph{\bibinfo{title}{Ising-Type Magnetic Ordering in Atomically Thin {FePS}3}}, \bibinfo{journal}{Nano Lett.} \textbf{\bibinfo{volume}{16}}, \bibinfo{pages}{7433} (\bibinfo{year}{2016}), \urlprefix\url{https://doi.org/10.1021/acs.nanolett.6b03052}.

\bibitem[{\citenamefont{Jernberg et~al.}(1984)\citenamefont{Jernberg, Bjarman, and W\"{a}ppling}}]{Jernberg1984}
\bibinfo{author}{\bibfnamefont{P.}~\bibnamefont{Jernberg}}, \bibinfo{author}{\bibfnamefont{S.}~\bibnamefont{Bjarman}}, \bibnamefont{and} \bibinfo{author}{\bibfnamefont{R.}~\bibnamefont{W\"{a}ppling}}, \emph{\bibinfo{title}{{FePS}$_3$: A first-order phase transition in a 2D Ising antiferromagnet}}, \bibinfo{journal}{J. Magn. Magn. Mater.} \textbf{\bibinfo{volume}{46}}, \bibinfo{pages}{178} (\bibinfo{year}{1984}), \urlprefix\url{https://doi.org/10.1016/0304-8853(84)90355-x}.

\bibitem[{\citenamefont{Amirabbasi and Kratzer}(2023)}]{Amirabbasi2023}
\bibinfo{author}{\bibfnamefont{M.}~\bibnamefont{Amirabbasi}} \bibnamefont{and} \bibinfo{author}{\bibfnamefont{P.}~\bibnamefont{Kratzer}}, \emph{\bibinfo{title}{Orbital and magnetic ordering in single-layer FePS$_3$: A DFT+U study}}, \bibinfo{journal}{Phys. Rev. B} \textbf{\bibinfo{volume}{107}}, \bibinfo{pages}{024401} (\bibinfo{year}{2023}), \urlprefix\url{https://doi.org/10.1103/physrevb.107.024401}.

\bibitem[{\citenamefont{Lancon et~al.}(2016)\citenamefont{Lancon, Walker, Ressouche, Ouladdiaf, Rule, McIntyre, Hicks, R{\o}nnow, and Wildes}}]{Lanon2016}
\bibinfo{author}{\bibfnamefont{D.}~\bibnamefont{Lancon}}, \bibinfo{author}{\bibfnamefont{H.~C.} \bibnamefont{Walker}}, \bibinfo{author}{\bibfnamefont{E.}~\bibnamefont{Ressouche}}, \bibinfo{author}{\bibfnamefont{B.}~\bibnamefont{Ouladdiaf}}, \bibinfo{author}{\bibfnamefont{K.~C.} \bibnamefont{Rule}}, \bibinfo{author}{\bibfnamefont{G.~J.} \bibnamefont{McIntyre}}, \bibinfo{author}{\bibfnamefont{T.~J.} \bibnamefont{Hicks}}, \bibinfo{author}{\bibfnamefont{H.~M.} \bibnamefont{R{\o}nnow}}, \bibnamefont{and} \bibinfo{author}{\bibfnamefont{A.~R.} \bibnamefont{Wildes}}, \emph{\bibinfo{title}{Magnetic structure and magnon dynamics of the quasi-two-dimensional antiferromagnet {FePS}$_3$}}, \bibinfo{journal}{Phys. Rev. B} \textbf{\bibinfo{volume}{94}}, \bibinfo{pages}{214407} (\bibinfo{year}{2016}), \urlprefix\url{https://doi.org/10.1103/physrevb.94.214407}.

\bibitem[{\citenamefont{{\v{S}}i{\v{s}}kins et~al.}(2020)\citenamefont{{\v{S}}i{\v{s}}kins, Lee, Ma{\~{n}}as-Valero, Coronado, Blanter, van~der Zant, and Steeneken}}]{Siskins2020}
\bibinfo{author}{\bibfnamefont{M.}~\bibnamefont{{\v{S}}i{\v{s}}kins}}, \bibinfo{author}{\bibfnamefont{M.}~\bibnamefont{Lee}}, \bibinfo{author}{\bibfnamefont{S.}~\bibnamefont{Ma{\~{n}}as-Valero}}, \bibinfo{author}{\bibfnamefont{E.}~\bibnamefont{Coronado}}, \bibinfo{author}{\bibfnamefont{Y.~M.} \bibnamefont{Blanter}}, \bibinfo{author}{\bibfnamefont{H.~S.~J.} \bibnamefont{van~der Zant}}, \bibnamefont{and} \bibinfo{author}{\bibfnamefont{P.~G.} \bibnamefont{Steeneken}}, \emph{\bibinfo{title}{Magnetic and electronic phase transitions probed by nanomechanical resonators}}, \bibinfo{journal}{Nat. Commun.} \textbf{\bibinfo{volume}{11}}, \bibinfo{pages}{2698} (\bibinfo{year}{2020}), \urlprefix\url{https://doi.org/10.1038/s41467-020-16430-2}.

\bibitem[{\citenamefont{Zhang et~al.}(2021)\citenamefont{Zhang, Ozerov, Boström, Cui, Suri, Jiang, Wang, Wu, Hwangbo, Chu et~al.}}]{zhang2021b}
\bibinfo{author}{\bibfnamefont{Q.}~\bibnamefont{Zhang}}, \bibinfo{author}{\bibfnamefont{M.}~\bibnamefont{Ozerov}}, \bibinfo{author}{\bibfnamefont{E.~V.} \bibnamefont{Boström}}, \bibinfo{author}{\bibfnamefont{J.}~\bibnamefont{Cui}}, \bibinfo{author}{\bibfnamefont{N.}~\bibnamefont{Suri}}, \bibinfo{author}{\bibfnamefont{Q.}~\bibnamefont{Jiang}}, \bibinfo{author}{\bibfnamefont{C.}~\bibnamefont{Wang}}, \bibinfo{author}{\bibfnamefont{F.}~\bibnamefont{Wu}}, \bibinfo{author}{\bibfnamefont{K.}~\bibnamefont{Hwangbo}}, \bibinfo{author}{\bibfnamefont{J.-H.} \bibnamefont{Chu}}, \bibinfo{author}{\bibfnamefont{D.}~\bibnamefont{Xiao}}, \bibinfo{author}{\bibfnamefont{A.}~\bibnamefont{Rubio}}, \bibnamefont{and} \bibinfo{author}{\bibfnamefont{X.}~\bibnamefont{Xu}}, \emph{\bibinfo{title}{Coherent strong-coupling of terahertz magnons and phonons in a Van der Waals antiferromagnetic insulator}}, \bibinfo{journal}{arXiv:} p. \bibinfo{pages}{2108.11619} (\bibinfo{year}{2021}).

\bibitem[{\citenamefont{Liu et~al.}(2021)\citenamefont{Liu, Granados~del \'Aguila, Bhowmick, Gan, Thu Ha~Do, Prosnikov, Sedmidubsk\'y, Sofer, Christianen, Sengupta et~al.}}]{Liu2021}
\bibinfo{author}{\bibfnamefont{S.}~\bibnamefont{Liu}}, \bibinfo{author}{\bibfnamefont{A.}~\bibnamefont{Granados~del \'Aguila}}, \bibinfo{author}{\bibfnamefont{D.}~\bibnamefont{Bhowmick}}, \bibinfo{author}{\bibfnamefont{C.~K.} \bibnamefont{Gan}}, \bibinfo{author}{\bibfnamefont{T.}~\bibnamefont{Thu Ha~Do}}, \bibinfo{author}{\bibfnamefont{M.~A.} \bibnamefont{Prosnikov}}, \bibinfo{author}{\bibfnamefont{D.}~\bibnamefont{Sedmidubsk\'y}}, \bibinfo{author}{\bibfnamefont{Z.}~\bibnamefont{Sofer}}, \bibinfo{author}{\bibfnamefont{P.~C.~M.} \bibnamefont{Christianen}}, \bibinfo{author}{\bibfnamefont{P.}~\bibnamefont{Sengupta}}, \bibnamefont{and} \bibinfo{author}{\bibfnamefont{Q.}~\bibnamefont{Xiong}}, \emph{\bibinfo{title}{Direct Observation of Magnon-Phonon Strong Coupling in Two-Dimensional Antiferromagnet at High Magnetic Fields}}, \bibinfo{journal}{Phys. Rev. Lett.} \textbf{\bibinfo{volume}{127}}, \bibinfo{pages}{097401} (\bibinfo{year}{2021}), \urlprefix\url{https://link.aps.org/doi/10.1103/PhysRevLett.127.097401}.

\bibitem[{\citenamefont{Houmes et~al.}(2023)\citenamefont{Houmes, Baglioni, Šiškins, Lee, Esteras, Ruiz, Mañas-Valero, Boix-Constant, Baldoví, Coronado et~al.}}]{Houmes2023}
\bibinfo{author}{\bibfnamefont{M.~J.~A.} \bibnamefont{Houmes}}, \bibinfo{author}{\bibfnamefont{G.}~\bibnamefont{Baglioni}}, \bibinfo{author}{\bibfnamefont{M.}~\bibnamefont{Šiškins}}, \bibinfo{author}{\bibfnamefont{M.}~\bibnamefont{Lee}}, \bibinfo{author}{\bibfnamefont{D.~L.} \bibnamefont{Esteras}}, \bibinfo{author}{\bibfnamefont{A.~M.} \bibnamefont{Ruiz}}, \bibinfo{author}{\bibfnamefont{S.}~\bibnamefont{Mañas-Valero}}, \bibinfo{author}{\bibfnamefont{C.}~\bibnamefont{Boix-Constant}}, \bibinfo{author}{\bibfnamefont{J.~J.} \bibnamefont{Baldoví}}, \bibinfo{author}{\bibfnamefont{E.}~\bibnamefont{Coronado}}, \bibinfo{author}{\bibfnamefont{Y.~M.} \bibnamefont{Blanter}}, \bibinfo{author}{\bibfnamefont{P.~G.} \bibnamefont{Steeneken}}, \bibnamefont{and} \bibinfo{author}{\bibfnamefont{H.~S.~J.} \bibnamefont{van~der Zant}}, \emph{\bibinfo{title}{Magnetic order in 2D antiferromagnets revealed by spontaneous anisotropic magnetostriction}}, \bibinfo{journal}{Nat. Commun.} \textbf{\bibinfo{volume}{14}},
  \bibinfo{pages}{8503} (\bibinfo{year}{2023}), ISSN \bibinfo{issn}{2041-1723}, \urlprefix\url{http://dx.doi.org/10.1038/s41467-023-44180-4}.

\bibitem[{\citenamefont{Šiškins et~al.}(2023)\citenamefont{Šiškins, Keşkekler, Houmes, Mañas-Valero, Coronado, Blanter, van~der Zant, Steeneken, and Alijani}}]{Siskins2023}
\bibinfo{author}{\bibfnamefont{M.}~\bibnamefont{Šiškins}}, \bibinfo{author}{\bibfnamefont{A.}~\bibnamefont{Keşkekler}}, \bibinfo{author}{\bibfnamefont{M.~J.~A.} \bibnamefont{Houmes}}, \bibinfo{author}{\bibfnamefont{S.}~\bibnamefont{Mañas-Valero}}, \bibinfo{author}{\bibfnamefont{E.}~\bibnamefont{Coronado}}, \bibinfo{author}{\bibfnamefont{Y.~M.} \bibnamefont{Blanter}}, \bibinfo{author}{\bibfnamefont{H.~S.~J.} \bibnamefont{van~der Zant}}, \bibinfo{author}{\bibfnamefont{P.~G.} \bibnamefont{Steeneken}}, \bibnamefont{and} \bibinfo{author}{\bibfnamefont{F.}~\bibnamefont{Alijani}}, \emph{\bibinfo{title}{Nonlinear dynamics and magneto-elasticity of nanodrums near the phase transition}}, \bibinfo{journal}{arXiv:} p. \bibinfo{pages}{2309.09672} (\bibinfo{year}{2023}), \eprint{2309.09672}.

\bibitem[{\citenamefont{Zhou et~al.}(2022)\citenamefont{Zhou, Hwangbo, Zhang, Wang, Shen, Zhang, Jiang, Zong, Su, Zajac et~al.}}]{Zhou2022b}
\bibinfo{author}{\bibfnamefont{F.}~\bibnamefont{Zhou}}, \bibinfo{author}{\bibfnamefont{K.}~\bibnamefont{Hwangbo}}, \bibinfo{author}{\bibfnamefont{Q.}~\bibnamefont{Zhang}}, \bibinfo{author}{\bibfnamefont{C.}~\bibnamefont{Wang}}, \bibinfo{author}{\bibfnamefont{L.}~\bibnamefont{Shen}}, \bibinfo{author}{\bibfnamefont{J.}~\bibnamefont{Zhang}}, \bibinfo{author}{\bibfnamefont{Q.}~\bibnamefont{Jiang}}, \bibinfo{author}{\bibfnamefont{A.}~\bibnamefont{Zong}}, \bibinfo{author}{\bibfnamefont{Y.}~\bibnamefont{Su}}, \bibinfo{author}{\bibfnamefont{M.}~\bibnamefont{Zajac}}, \bibinfo{author}{\bibfnamefont{Y.}~\bibnamefont{Ahn}}, \bibinfo{author}{\bibfnamefont{D.~A.} \bibnamefont{Walko}}, \bibinfo{author}{\bibfnamefont{R.~D.} \bibnamefont{Schaller}}, \bibinfo{author}{\bibfnamefont{J.-H.} \bibnamefont{Chu}}, \bibinfo{author}{\bibfnamefont{N.}~\bibnamefont{Gedik}}, \bibinfo{author}{\bibfnamefont{X.}~\bibnamefont{Xu}}, \bibinfo{author}{\bibfnamefont{D.}~\bibnamefont{Xiao}}, \bibnamefont{and}
  \bibinfo{author}{\bibfnamefont{H.}~\bibnamefont{Wen}}, \emph{\bibinfo{title}{Dynamical criticality of spin-shear coupling in van der Waals antiferromagnets}}, \bibinfo{journal}{Nat. Commun.} \textbf{\bibinfo{volume}{13}}, \bibinfo{pages}{6598} (\bibinfo{year}{2022}), \urlprefix\url{https://doi.org/10.1038/s41467-022-34376-5}.

\bibitem[{\citenamefont{Zong et~al.}(2023)\citenamefont{Zong, Zhang, Zhou, Su, Hwangbo, Shen, Jiang, Liu, Gage, Walko et~al.}}]{Zong2023}
\bibinfo{author}{\bibfnamefont{A.}~\bibnamefont{Zong}}, \bibinfo{author}{\bibfnamefont{Q.}~\bibnamefont{Zhang}}, \bibinfo{author}{\bibfnamefont{F.}~\bibnamefont{Zhou}}, \bibinfo{author}{\bibfnamefont{Y.}~\bibnamefont{Su}}, \bibinfo{author}{\bibfnamefont{K.}~\bibnamefont{Hwangbo}}, \bibinfo{author}{\bibfnamefont{X.}~\bibnamefont{Shen}}, \bibinfo{author}{\bibfnamefont{Q.}~\bibnamefont{Jiang}}, \bibinfo{author}{\bibfnamefont{H.}~\bibnamefont{Liu}}, \bibinfo{author}{\bibfnamefont{T.~E.} \bibnamefont{Gage}}, \bibinfo{author}{\bibfnamefont{D.~A.} \bibnamefont{Walko}}, \bibinfo{author}{\bibfnamefont{M.~E.} \bibnamefont{Kozina}}, \bibinfo{author}{\bibfnamefont{D.}~\bibnamefont{Luo}}, \bibinfo{author}{\bibfnamefont{A.~H.} \bibnamefont{Reid}}, \bibinfo{author}{\bibfnamefont{J.}~\bibnamefont{Yang}}, \bibinfo{author}{\bibfnamefont{S.}~\bibnamefont{Park}}, \bibinfo{author}{\bibfnamefont{S.~H.} \bibnamefont{Lapidus}}, \bibinfo{author}{\bibfnamefont{J.-H.} \bibnamefont{Chu}},
  \bibinfo{author}{\bibfnamefont{I.}~\bibnamefont{Arslan}}, \bibinfo{author}{\bibfnamefont{X.}~\bibnamefont{Wang}}, \bibinfo{author}{\bibfnamefont{D.}~\bibnamefont{Xiao}}, \bibinfo{author}{\bibfnamefont{X.}~\bibnamefont{Xu}}, \bibinfo{author}{\bibfnamefont{N.}~\bibnamefont{Gedik}}, \bibnamefont{and} \bibinfo{author}{\bibfnamefont{H.}~\bibnamefont{Wen}}, \emph{\bibinfo{title}{Spin-mediated shear oscillators in a van der Waals antiferromagnet}}, \bibinfo{journal}{Nature} \textbf{\bibinfo{volume}{620}}, \bibinfo{pages}{988} (\bibinfo{year}{2023}), ISSN \bibinfo{issn}{1476-4687}, \urlprefix\url{http://dx.doi.org/10.1038/s41586-023-06279-y}.

\bibitem[{\citenamefont{Kim and Park}(2021)}]{Kim2021}
\bibinfo{author}{\bibfnamefont{T.~Y.} \bibnamefont{Kim}} \bibnamefont{and} \bibinfo{author}{\bibfnamefont{C.-H.} \bibnamefont{Park}}, \emph{\bibinfo{title}{Magnetic Anisotropy and Magnetic Ordering of Transition-Metal Phosphorus Trisulfides}}, \bibinfo{journal}{Nano Lett.} \textbf{\bibinfo{volume}{21}}, \bibinfo{pages}{10114} (\bibinfo{year}{2021}), \urlprefix\url{https://doi.org/10.1021/acs.nanolett.1c03992}.

\bibitem[{\citenamefont{Wiemann et~al.}(2011)\citenamefont{Wiemann, Patt, Krug, Weber, Escher, Merkel, and Schneider}}]{Wiemann2011}
\bibinfo{author}{\bibfnamefont{C.}~\bibnamefont{Wiemann}}, \bibinfo{author}{\bibfnamefont{M.}~\bibnamefont{Patt}}, \bibinfo{author}{\bibfnamefont{I.~P.} \bibnamefont{Krug}}, \bibinfo{author}{\bibfnamefont{N.~B.} \bibnamefont{Weber}}, \bibinfo{author}{\bibfnamefont{M.}~\bibnamefont{Escher}}, \bibinfo{author}{\bibfnamefont{M.}~\bibnamefont{Merkel}}, \bibnamefont{and} \bibinfo{author}{\bibfnamefont{C.~M.} \bibnamefont{Schneider}}, \emph{\bibinfo{title}{A New Nanospectroscopy Tool with Synchrotron Radiation: NanoESCA@Elettra}}, \bibinfo{journal}{e-J. Surf. Sci. Nanotechnol.} \textbf{\bibinfo{volume}{9}}, \bibinfo{pages}{395} (\bibinfo{year}{2011}).

\bibitem[{\citenamefont{Strasdas et~al.}(2023)\citenamefont{Strasdas, Pestka, Rybak, Budniak, Leuth, Boban, Feyer, Cojocariu, Baranowski, Avila et~al.}}]{Strasdas2023}
\bibinfo{author}{\bibfnamefont{J.}~\bibnamefont{Strasdas}}, \bibinfo{author}{\bibfnamefont{B.}~\bibnamefont{Pestka}}, \bibinfo{author}{\bibfnamefont{M.}~\bibnamefont{Rybak}}, \bibinfo{author}{\bibfnamefont{A.~K.} \bibnamefont{Budniak}}, \bibinfo{author}{\bibfnamefont{N.}~\bibnamefont{Leuth}}, \bibinfo{author}{\bibfnamefont{H.}~\bibnamefont{Boban}}, \bibinfo{author}{\bibfnamefont{V.}~\bibnamefont{Feyer}}, \bibinfo{author}{\bibfnamefont{I.}~\bibnamefont{Cojocariu}}, \bibinfo{author}{\bibfnamefont{D.}~\bibnamefont{Baranowski}}, \bibinfo{author}{\bibfnamefont{J.}~\bibnamefont{Avila}}, \bibinfo{author}{\bibfnamefont{P.}~\bibnamefont{Dudin}}, \bibinfo{author}{\bibfnamefont{A.}~\bibnamefont{Bostwick}}, \bibinfo{author}{\bibfnamefont{C.}~\bibnamefont{Jozwiak}}, \bibinfo{author}{\bibfnamefont{E.}~\bibnamefont{Rotenberg}}, \bibinfo{author}{\bibfnamefont{C.}~\bibnamefont{Autieri}}, \bibinfo{author}{\bibfnamefont{Y.}~\bibnamefont{Amouyal}}, \bibinfo{author}{\bibfnamefont{L.}~\bibnamefont{Plucinski}},
  \bibinfo{author}{\bibfnamefont{E.}~\bibnamefont{Lifshitz}}, \bibinfo{author}{\bibfnamefont{M.}~\bibnamefont{Birowska}}, \bibnamefont{and} \bibinfo{author}{\bibfnamefont{M.}~\bibnamefont{Morgenstern}}, \emph{\bibinfo{title}{Electronic Band Structure Changes across the Antiferromagnetic Phase Transition of Exfoliated MnPS$_3$ Flakes Probed by $\mu$-ARPES}}, \bibinfo{journal}{Nano Lett.} \textbf{\bibinfo{volume}{23}}, \bibinfo{pages}{10342} (\bibinfo{year}{2023}), ISSN \bibinfo{issn}{1530-6992}, \urlprefix\url{http://dx.doi.org/10.1021/acs.nanolett.3c02906}.

\bibitem[{\citenamefont{Popescu and Zunger}(2012)}]{Popescu2012}
\bibinfo{author}{\bibfnamefont{V.}~\bibnamefont{Popescu}} \bibnamefont{and} \bibinfo{author}{\bibfnamefont{A.}~\bibnamefont{Zunger}}, \emph{\bibinfo{title}{Extracting $E$ versus ${k}$ effective band structure from supercell calculations on alloys and impurities}}, \bibinfo{journal}{Phys. Rev. B} \textbf{\bibinfo{volume}{85}}, \bibinfo{pages}{085201} (\bibinfo{year}{2012}), \urlprefix\url{https://link.aps.org/doi/10.1103/PhysRevB.85.085201}.

\bibitem[{\citenamefont{Moser}(2017)}]{Moser2017}
\bibinfo{author}{\bibfnamefont{S.}~\bibnamefont{Moser}}, \emph{\bibinfo{title}{An experimentalist{\textquotesingle}s guide to the matrix element in angle resolved photoemission}}, \bibinfo{journal}{J. Electr. Spectr. Rel. Phen.} \textbf{\bibinfo{volume}{214}}, \bibinfo{pages}{29} (\bibinfo{year}{2017}), \urlprefix\url{https://doi.org/10.1016/j.elspec.2016.11.007}.

\bibitem[{\citenamefont{Bianchi et~al.}(2023)\citenamefont{Bianchi, Acharya, Dirnberger, Klein, Pashov, Mosina, Sofer, Rudenko, Katsnelson, van Schilfgaarde et~al.}}]{Bianchi2023}
\bibinfo{author}{\bibfnamefont{M.}~\bibnamefont{Bianchi}}, \bibinfo{author}{\bibfnamefont{S.}~\bibnamefont{Acharya}}, \bibinfo{author}{\bibfnamefont{F.}~\bibnamefont{Dirnberger}}, \bibinfo{author}{\bibfnamefont{J.}~\bibnamefont{Klein}}, \bibinfo{author}{\bibfnamefont{D.}~\bibnamefont{Pashov}}, \bibinfo{author}{\bibfnamefont{K.}~\bibnamefont{Mosina}}, \bibinfo{author}{\bibfnamefont{Z.}~\bibnamefont{Sofer}}, \bibinfo{author}{\bibfnamefont{A.~N.} \bibnamefont{Rudenko}}, \bibinfo{author}{\bibfnamefont{M.~I.} \bibnamefont{Katsnelson}}, \bibinfo{author}{\bibfnamefont{M.}~\bibnamefont{van Schilfgaarde}}, \bibinfo{author}{\bibfnamefont{M.}~\bibnamefont{Rösner}}, \bibnamefont{and} \bibinfo{author}{\bibfnamefont{P.}~\bibnamefont{Hofmann}}, \emph{\bibinfo{title}{Paramagnetic Electronic Structure of CrSBr: Comparison between Ab Initio GW Theory and Angle-Resolved Photoemission Spectroscopy}}, \bibinfo{journal}{arXiv:} p. \bibinfo{pages}{2303.01292} (\bibinfo{year}{2023}),
  \urlprefix\url{https://arxiv.org/abs/2303.01292}.

\bibitem[{\citenamefont{De~Vita et~al.}(2022)\citenamefont{De~Vita, Nguyen, Sant, Pierantozzi, Amoroso, Bigi, Polewczyk, Vinai, Nguyen, Kong et~al.}}]{DeVita2022}
\bibinfo{author}{\bibfnamefont{A.}~\bibnamefont{De~Vita}}, \bibinfo{author}{\bibfnamefont{T.~T.~P.} \bibnamefont{Nguyen}}, \bibinfo{author}{\bibfnamefont{R.}~\bibnamefont{Sant}}, \bibinfo{author}{\bibfnamefont{G.~M.} \bibnamefont{Pierantozzi}}, \bibinfo{author}{\bibfnamefont{D.}~\bibnamefont{Amoroso}}, \bibinfo{author}{\bibfnamefont{C.}~\bibnamefont{Bigi}}, \bibinfo{author}{\bibfnamefont{V.}~\bibnamefont{Polewczyk}}, \bibinfo{author}{\bibfnamefont{G.}~\bibnamefont{Vinai}}, \bibinfo{author}{\bibfnamefont{L.~T.} \bibnamefont{Nguyen}}, \bibinfo{author}{\bibfnamefont{T.}~\bibnamefont{Kong}}, \bibinfo{author}{\bibfnamefont{J.}~\bibnamefont{Fujii}}, \bibinfo{author}{\bibfnamefont{I.}~\bibnamefont{Vobornik}}, \bibinfo{author}{\bibfnamefont{N.~B.} \bibnamefont{Brookes}}, \bibinfo{author}{\bibfnamefont{G.}~\bibnamefont{Rossi}}, \bibinfo{author}{\bibfnamefont{R.~J.} \bibnamefont{Cava}}, \bibinfo{author}{\bibfnamefont{F.}~\bibnamefont{Mazzola}}, \bibinfo{author}{\bibfnamefont{K.}~\bibnamefont{Yamauchi}},
  \bibinfo{author}{\bibfnamefont{S.}~\bibnamefont{Picozzi}}, \bibnamefont{and} \bibinfo{author}{\bibfnamefont{G.}~\bibnamefont{Panaccione}}, \emph{\bibinfo{title}{Influence of Orbital Character on the Ground State Electronic Properties in the van Der Waals Transition Metal Iodides VI3 and CrI3}}, \bibinfo{journal}{Nano Lett.} \textbf{\bibinfo{volume}{22}}, \bibinfo{pages}{7034–} (\bibinfo{year}{2022}), ISSN \bibinfo{issn}{1530-6992}, \urlprefix\url{http://dx.doi.org/10.1021/acs.nanolett.2c01922}.

\bibitem[{\citenamefont{Nitschke et~al.}(2023)\citenamefont{Nitschke, Esteras, Gutnikov, Schiller, Ma{\~n}as-Valero, Coronado, Stupar, Zamborlini, Ponzoni, Baldov\'i et~al.}}]{Nitschke2023}
\bibinfo{author}{\bibfnamefont{J.~E.} \bibnamefont{Nitschke}}, \bibinfo{author}{\bibfnamefont{D.~L.} \bibnamefont{Esteras}}, \bibinfo{author}{\bibfnamefont{M.}~\bibnamefont{Gutnikov}}, \bibinfo{author}{\bibfnamefont{K.}~\bibnamefont{Schiller}}, \bibinfo{author}{\bibfnamefont{S.}~\bibnamefont{Ma{\~n}as-Valero}}, \bibinfo{author}{\bibfnamefont{E.}~\bibnamefont{Coronado}}, \bibinfo{author}{\bibfnamefont{M.}~\bibnamefont{Stupar}}, \bibinfo{author}{\bibfnamefont{G.}~\bibnamefont{Zamborlini}}, \bibinfo{author}{\bibfnamefont{S.}~\bibnamefont{Ponzoni}}, \bibinfo{author}{\bibfnamefont{J.~J.} \bibnamefont{Baldov\'i}}, \bibnamefont{and} \bibinfo{author}{\bibfnamefont{M.}~\bibnamefont{Cinchetti}}, \emph{\bibinfo{title}{Valence band electronic structure of the van der Waals antiferromagnet FePS3}}, \bibinfo{journal}{Materials Today Electronics} \textbf{\bibinfo{volume}{6}}, \bibinfo{pages}{100061} (\bibinfo{year}{2023}), ISSN \bibinfo{issn}{2772-9494},
  \urlprefix\url{https://www.sciencedirect.com/science/article/pii/S2772949423000372}.

\bibitem[{\citenamefont{Koitzsch et~al.}(2023)\citenamefont{Koitzsch, Klaproth, Selter, Shemerliuk, Aswartham, Janson, B\"{u}chner, and Knupfer}}]{Koitzsch2023}
\bibinfo{author}{\bibfnamefont{A.}~\bibnamefont{Koitzsch}}, \bibinfo{author}{\bibfnamefont{T.}~\bibnamefont{Klaproth}}, \bibinfo{author}{\bibfnamefont{S.}~\bibnamefont{Selter}}, \bibinfo{author}{\bibfnamefont{Y.}~\bibnamefont{Shemerliuk}}, \bibinfo{author}{\bibfnamefont{S.}~\bibnamefont{Aswartham}}, \bibinfo{author}{\bibfnamefont{O.}~\bibnamefont{Janson}}, \bibinfo{author}{\bibfnamefont{B.}~\bibnamefont{B\"{u}chner}}, \bibnamefont{and} \bibinfo{author}{\bibfnamefont{M.}~\bibnamefont{Knupfer}}, \emph{\bibinfo{title}{Intertwined electronic and magnetic structure of the van-der-Waals antiferromagnet Fe2P2S6}}, \bibinfo{journal}{npj Quantum Mater.} \textbf{\bibinfo{volume}{8}}, \bibinfo{pages}{27} (\bibinfo{year}{2023}), ISSN \bibinfo{issn}{2397-4648}, \urlprefix\url{http://dx.doi.org/10.1038/s41535-023-00560-z}.

\bibitem[{\citenamefont{Voloshina et~al.}(2023)\citenamefont{Voloshina, Jin, and Dedkov}}]{Voloshina2023}
\bibinfo{author}{\bibfnamefont{E.}~\bibnamefont{Voloshina}}, \bibinfo{author}{\bibfnamefont{Y.}~\bibnamefont{Jin}}, \bibnamefont{and} \bibinfo{author}{\bibfnamefont{Y.}~\bibnamefont{Dedkov}}, \emph{\bibinfo{title}{{ARPES} studies of the ground state electronic properties of the van der Waals transition metal trichalcogenide {CoPS}$_3$}}, \bibinfo{journal}{Chem. Phys. Lett.} \textbf{\bibinfo{volume}{823}}, \bibinfo{pages}{140511} (\bibinfo{year}{2023}), \urlprefix\url{https://doi.org/10.1016/j.cplett.2023.140511}.

\bibitem[{\citenamefont{Watson et~al.}(2024)\citenamefont{Watson, Acharya, Nunn, Nagireddy, Pashov, R\"{o}sner, van Schilfgaarde, Wilson, and Cacho}}]{watson2024}
\bibinfo{author}{\bibfnamefont{M.~D.} \bibnamefont{Watson}}, \bibinfo{author}{\bibfnamefont{S.}~\bibnamefont{Acharya}}, \bibinfo{author}{\bibfnamefont{J.~E.} \bibnamefont{Nunn}}, \bibinfo{author}{\bibfnamefont{L.}~\bibnamefont{Nagireddy}}, \bibinfo{author}{\bibfnamefont{D.}~\bibnamefont{Pashov}}, \bibinfo{author}{\bibfnamefont{M.}~\bibnamefont{R\"{o}sner}}, \bibinfo{author}{\bibfnamefont{M.}~\bibnamefont{van Schilfgaarde}}, \bibinfo{author}{\bibfnamefont{N.~R.} \bibnamefont{Wilson}}, \bibnamefont{and} \bibinfo{author}{\bibfnamefont{C.}~\bibnamefont{Cacho}}, \emph{\bibinfo{title}{Giant exchange splitting in the electronic structure of A-type 2D antiferromagnet CrSBr}}, \bibinfo{journal}{npj 2D Mater. Applic.} \textbf{\bibinfo{volume}{8}}, \bibinfo{pages}{54} (\bibinfo{year}{2024}), ISSN \bibinfo{issn}{2397-7132}, \urlprefix\url{http://dx.doi.org/10.1038/s41699-024-00492-7}.

\bibitem[{\citenamefont{Scagliotti et~al.}(1985)\citenamefont{Scagliotti, Jouanne, Balkanski, and Ouvrard}}]{Scagliotti1985}
\bibinfo{author}{\bibfnamefont{M.}~\bibnamefont{Scagliotti}}, \bibinfo{author}{\bibfnamefont{M.}~\bibnamefont{Jouanne}}, \bibinfo{author}{\bibfnamefont{M.}~\bibnamefont{Balkanski}}, \bibnamefont{and} \bibinfo{author}{\bibfnamefont{G.}~\bibnamefont{Ouvrard}}, \emph{\bibinfo{title}{Spin dependent phonon Raman scattering in antiferromagnetic FePS3 layer-type compound}}, \bibinfo{journal}{Solid State Communications} \textbf{\bibinfo{volume}{54}}, \bibinfo{pages}{291} (\bibinfo{year}{1985}), ISSN \bibinfo{issn}{0038-1098}, \urlprefix\url{https://www.sciencedirect.com/science/article/pii/0038109885910877}.

\bibitem[{\citenamefont{Scagliotti et~al.}(1987)\citenamefont{Scagliotti, Jouanne, Balkanski, Ouvrard, and Benedek}}]{Scagliotti1987}
\bibinfo{author}{\bibfnamefont{M.}~\bibnamefont{Scagliotti}}, \bibinfo{author}{\bibfnamefont{M.}~\bibnamefont{Jouanne}}, \bibinfo{author}{\bibfnamefont{M.}~\bibnamefont{Balkanski}}, \bibinfo{author}{\bibfnamefont{G.}~\bibnamefont{Ouvrard}}, \bibnamefont{and} \bibinfo{author}{\bibfnamefont{G.}~\bibnamefont{Benedek}}, \emph{\bibinfo{title}{Raman scattering in antiferromagnetic ${\mathrm{FePS}}_{3}$ and ${\mathrm{FePSe}}_{3}$ crystals}}, \bibinfo{journal}{Phys. Rev. B} \textbf{\bibinfo{volume}{35}}, \bibinfo{pages}{7097} (\bibinfo{year}{1987}), \urlprefix\url{https://link.aps.org/doi/10.1103/PhysRevB.35.7097}.

\bibitem[{\citenamefont{H{\"u}fner}(2013)}]{Hüfner2013}
\bibinfo{author}{\bibfnamefont{S.}~\bibnamefont{H{\"u}fner}}, \emph{\bibinfo{title}{Photoelectron Spectroscopy: Principles and Applications}}, Advanced Texts in Physics (\bibinfo{publisher}{Springer Berlin Heidelberg}, \bibinfo{year}{2013}), ISBN \bibinfo{isbn}{9783662092804}, \urlprefix\url{https://books.google.de/books?id=f6nvCAAAQBAJ}.

\bibitem[{\citenamefont{Choi et~al.}(1994)\citenamefont{Choi, Kneedler, and Kevan}}]{Choi1994}
\bibinfo{author}{\bibfnamefont{W.-K.} \bibnamefont{Choi}}, \bibinfo{author}{\bibfnamefont{E.}~\bibnamefont{Kneedler}}, \bibnamefont{and} \bibinfo{author}{\bibfnamefont{S.~D.} \bibnamefont{Kevan}}, \emph{\bibinfo{title}{Delocalization of the Fe 3d levels in the quasi-two-dimensional correlated insulator ${\mathrm{FePS}}_{3}$}}, \bibinfo{journal}{Phys. Rev. B} \textbf{\bibinfo{volume}{50}}, \bibinfo{pages}{15276} (\bibinfo{year}{1994}), \urlprefix\url{https://link.aps.org/doi/10.1103/PhysRevB.50.15276}.

\bibitem[{\citenamefont{Jin et~al.}(2022)\citenamefont{Jin, Yan, Kremer, Voloshina, and Dedkov}}]{Jin2022}
\bibinfo{author}{\bibfnamefont{Y.}~\bibnamefont{Jin}}, \bibinfo{author}{\bibfnamefont{M.}~\bibnamefont{Yan}}, \bibinfo{author}{\bibfnamefont{T.}~\bibnamefont{Kremer}}, \bibinfo{author}{\bibfnamefont{E.}~\bibnamefont{Voloshina}}, \bibnamefont{and} \bibinfo{author}{\bibfnamefont{Y.}~\bibnamefont{Dedkov}}, \emph{\bibinfo{title}{Mott–Hubbard insulating state for the layered van der Waals FePX$_3$ (X: S, Se) as revealed by NEXAFS and resonant photoelectron spectroscopy}}, \bibinfo{journal}{Sci. Rep.} \textbf{\bibinfo{volume}{12}}, \bibinfo{pages}{735} (\bibinfo{year}{2022}), ISSN \bibinfo{issn}{2045-2322}, \urlprefix\url{http://dx.doi.org/10.1038/s41598-021-04557-1}.

\bibitem[{\citenamefont{Budniak et~al.}(2022)\citenamefont{Budniak, Zelewski, Birowska, Woźniak, Bendikov, Kauffmann, Amouyal, Kudrawiec, and Lifshitz}}]{Budniak2022}
\bibinfo{author}{\bibfnamefont{A.~K.} \bibnamefont{Budniak}}, \bibinfo{author}{\bibfnamefont{S.~J.} \bibnamefont{Zelewski}}, \bibinfo{author}{\bibfnamefont{M.}~\bibnamefont{Birowska}}, \bibinfo{author}{\bibfnamefont{T.}~\bibnamefont{Woźniak}}, \bibinfo{author}{\bibfnamefont{T.}~\bibnamefont{Bendikov}}, \bibinfo{author}{\bibfnamefont{Y.}~\bibnamefont{Kauffmann}}, \bibinfo{author}{\bibfnamefont{Y.}~\bibnamefont{Amouyal}}, \bibinfo{author}{\bibfnamefont{R.}~\bibnamefont{Kudrawiec}}, \bibnamefont{and} \bibinfo{author}{\bibfnamefont{E.}~\bibnamefont{Lifshitz}}, \emph{\bibinfo{title}{Spectroscopy and Structural Investigation of Iron Phosphorus Trisulfide—FePS3}}, \bibinfo{journal}{Adv. Opt. Mater.} p. \bibinfo{pages}{2102489} (\bibinfo{year}{2022}), \urlprefix\url{https://onlinelibrary.wiley.com/doi/abs/10.1002/adom.202102489}.

\bibitem[{\citenamefont{Yang et~al.}(2024)\citenamefont{Yang, Ning, Zhou, Lu, Ma, Liu, Pu, and Wu}}]{Yang2024}
\bibinfo{author}{\bibfnamefont{K.}~\bibnamefont{Yang}}, \bibinfo{author}{\bibfnamefont{Y.}~\bibnamefont{Ning}}, \bibinfo{author}{\bibfnamefont{Y.}~\bibnamefont{Zhou}}, \bibinfo{author}{\bibfnamefont{D.}~\bibnamefont{Lu}}, \bibinfo{author}{\bibfnamefont{Y.}~\bibnamefont{Ma}}, \bibinfo{author}{\bibfnamefont{L.}~\bibnamefont{Liu}}, \bibinfo{author}{\bibfnamefont{S.}~\bibnamefont{Pu}}, \bibnamefont{and} \bibinfo{author}{\bibfnamefont{H.}~\bibnamefont{Wu}}, \emph{\bibinfo{title}{Understanding the Ising zigzag antiferromagnetism of FePS$_3$ and FePSe$_3$ monolayers}}, \bibinfo{journal}{Phys. Rev, B} \textbf{\bibinfo{volume}{110}}, \bibinfo{pages}{024427} (\bibinfo{year}{2024}), ISSN \bibinfo{issn}{2469-9969}, \urlprefix\url{http://dx.doi.org/10.1103/PhysRevB.110.024427}.

\end{thebibliography}
\end{document}


\title{Supplementary Information: Identifying band structure changes of FePS$_3$ across the antiferromagnetic phase transition}
\author{Benjamin Pestka}\affiliation{\AC}
\author{Jeff Strasdas}\affiliation{\AC}
\author{Gustav Bihlmayer}\affiliation{\FZJtheo}
\author{Adam K. Budniak}\affiliation{\Haifa}
\author{Marcus Liebmann}\affiliation{\AC}
\author{Niklas Leuth}\affiliation{\AC}
\author{Honey Boban}\affiliation{\FZJ}
\author{Vitaliy Feyer}\affiliation{\FZJ}
\author{Iulia Cojocariu}\affiliation{\FZJ}
\author{Daniel Baranowski}\affiliation{\FZJ}
\author{Simone Mearini}\affiliation{\FZJ}
\author{Yaron Amouyal}\affiliation{\HaifaMat}
\author{Lutz Waldecker}\affiliation{\ACa}
\author{Bernd Beschoten}\affiliation{\ACa}
\author{Christoph Stampfer}\affiliation{\ACa}
\author{Lukasz Plucinski}\affiliation{\FZJ}
\author{Efrat Lifshitz}\affiliation{\Haifa} 
\author{Peter Kratzer}\affiliation{\Dui}
\author{Markus Morgenstern}\affiliation{\AC}
\date{\today} 

\maketitle
\noindent {{Corresponding author: } 
M.~Morgenstern, email: mmorgens@physik.rwth-aachen.de } 
\tableofcontents

%

\section{Experimental details}
\label{sec:exp_det}

\subsection{Crystal growth}
\label{sub:growth} 
 FePS$_3$ crystals were synthesized in a quartz ampoule via chemical vapor transport (CVT) according to our previous recipe \cite{Budniak2022}. About 1 gram of the element mixture consisting of metal iron powder (from Sigma-Aldrich), red phosphorus (from Riedel-de Haën), and elemental sulfur (from Sigma-Aldrich) with Fe : P : S = 1 : 1 : 3.15 atomic ratio was grounded in an agate mortar. The 5 \% excess sulfur with respect to the 1 : 1 : 3 stoichiometric ratio served as a transport agent and increased the reaction yield. The sealed ampoule, evacuated to $<3.5\cdot10^{-5}$\,Torr by a turbomolecular pump, was placed inside a previously heated 3-zone furnace that was calibrated such that the mixture of elements (the substrate zone) was at $850^{\circ}$C and the deposition zone at $790^{\circ}$C. After a week, the furnace was turned off and the sample was left to cool down inside. Then, the ampoule was opened, and only recrystallized FePS$_3$ was collected and used for experiments.
\subsection{Flake preparation}
\label{sub:flake}

\begin{figure}[thb]
\centering
\includegraphics[width=0.485\textwidth]{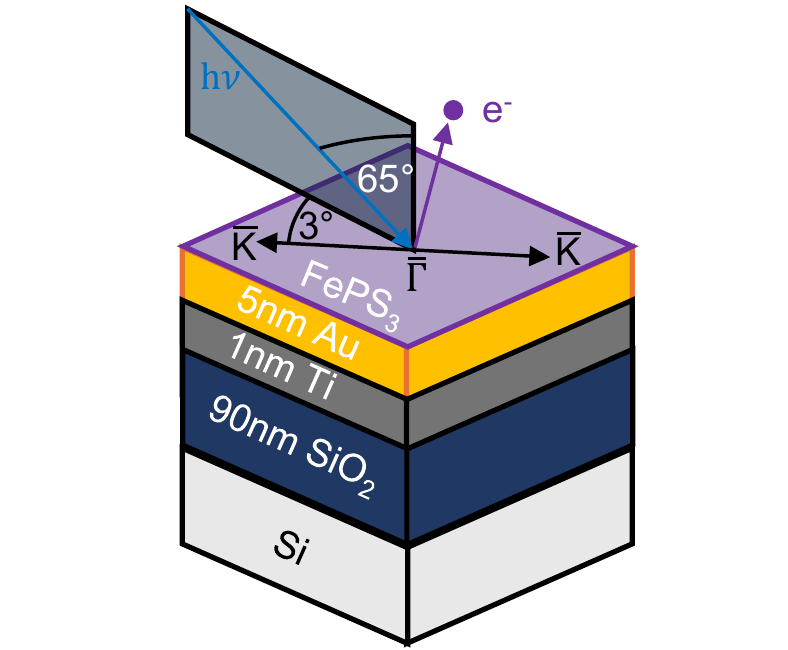}
\vspace{-0.4cm}
\caption{
{\bf Sample and photon beam.} Sketch of the sample structure with marked crystallographic directions of the $\rm FePS_3$ flake, 
the photon beam geometry and an exemplary emitted photoelectron. The $\Vec{E}$ field polarization direction is in the plane of incidence.
}
\label{Fig_S1}
\end{figure}

The FePS$_{\rm 3}$ flakes were exfoliated with an optimized scotch tape method on a Au/Ti ($5$\,nm/$1$\,nm) film on top of Si/SiO$_{\rm 2}$  (Figure~\ref{Fig_S1}) \cite{Strasdas2023}. 
The Si/SiO$_2$ (thickness SiO$_{\rm 2} = 90$\,nm) enables optical contrast of the thin FePS$_3$ flakes. Ti and Au were deposited at $370$\,K in ultra-high vacuum (UHV) which also improves the adhesion between  the flakes and the Au substrate \cite{Huang2020b,Velicky2018}. Before exfoliation, the substrate was exposed to plasma ashing using O$_2$ at $300$\,K with a power of $50$\,W and a flux of $100$\,sccm for $25$\,s. Within $10$~min afterwards, the substrate was placed on a hotplate at $60^\circ$\,C , where the flakes were exfoliated using an adhesive tape (Nitto ELP BT-150E-KL). The whole process is conducted in a cleanroom. The flake thickness was determined afterwards via atomic force microscopy (AFMi) in tapping mode (Figure~1, main text) using a commercial Bruker system. The stoichiometry of FePS$_3$ was confirmed by X-ray photoelectron spectroscopy (XPS) after the ARPES measurements  \cite{Budniak2022}. Apparent minor contributions of O and C are probably related to remaining residues of glue similar to the case of MnPS$_3$ \cite{Strasdas2023}.
%
\subsection{ARPES and XPS}
\label{sub:ARPES_XPS}
For XPS and µ-ARPES, we used the NanoESCA beamline of Elettra, the Italian synchrotron radiation facility, employing a FOCUS NanoESCA photoemission electron microscope (PEEM) that operated in $k$-space mapping mode at a base pressure $p < 5\cdot 10^{-11}$\,mbar \cite{Schneider2012}. The photoelectrons were collected from a spot-size with a diameter of $5-10$\,µm. The total energy resolution of beamline and analyser was $50$\,meV. For all measurements, we used p-polarized light with an incident angle of $65^{\circ}$ from the surface normal and $3^{\circ}$ azimuth angle relative to the $\overline{\rm K\Gamma K}$ direction (Figure~\ref{Fig_S1}) as determined by the $k_\parallel$ symmetry of the ARPES data (section~S1\ref{sub:BZdet}). Before the ARPES measurements, the sample was annealed in UHV at $200^\circ$\,C for $2$\,h.

\subsection{Raman spectroscopy}
\label{sub:Raman}
For Raman spectroscopy, we used a partly home-built low-temperature Raman setup. The incident laser (wavelength: $532$\,nm, intensity: $100$\,µW, linearly polarized) is focused onto the sample by a $100\times$ objective with a numerical aperture of $0.82$ leading to a spot diameter of $505 \pm 10$\,nm. 
The reﬂected and the scattered light are guided via a single-mode optical ﬁber to a spectrometer with a grating of $1200$\,lines/mm without polarization analysis \cite{Goldsche2018}.

\subsection{ARPES: Fermi level and work function}
\label{sub:Fermi_workfunciton}
\begin{figure}[htb]
\centering
\includegraphics[width=0.45\textwidth]{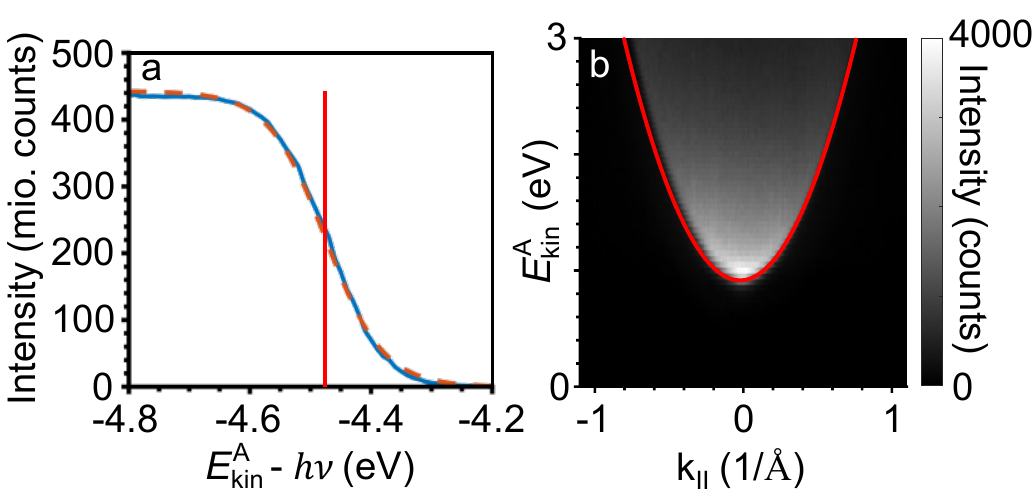}
\vspace{-0.4cm}
\caption{{\bf Determination  of Fermi energy and work function.} (a) ARPES intensity of the Au next to the FePS$_3$ flake of Figure~1d, main text. For each photoelectron energy $E_{\rm kin}^{\rm A}$,  
the intensity is integrated across the whole measured $k_\parallel$-space, $h\nu=58$\,eV, dashed orange line: fit by the Fermi distribution function, red line: resulting Fermi energy $E_{\rm F} =-4.475\pm 0.025$\,eV. (b) Intensity of secondary electron emission. 
The red line marks a fit using the free electron parabola 
to the onset of the secondary electron emission determined by the maximum of the second derivative of $I(E_{\rm kin}^{\rm A})$ at each $k_\parallel$. The apex of the parabola determines the work-function of $\rm FePS_3$ $\phi = 5.35$\,eV (Equation~\ref{eq:S2}).
}
\label{Fig_S4}
\end{figure}

The Fermi level $E_{\rm F}$ is determined by  ARPES on the gold surface next to the FePS$_3$ flake using a fit by the Fermi distribution function with temperature $T = 550$\,K that is dominated by the energy resolution of the detector (Figure~\ref{Fig_S4}a). The precision of the fit is 25\,meV. 

The work function of the analyzer $\phi^{\rm A}$ is directly related to the measured kinetic energy $E_\text{kin, F}^\text{A}$ and the used photon energy $h\nu$ by
\begin{equation}
 \label{eq:S1}
 E_\text{kin,F}^\text{A}+\phi^\text{A}=h\nu.
\end{equation}
This results in $\phi^{\rm A} =  4.475 \pm 0.025$\,eV (Figure~\ref{Fig_S4}a). To determine the work function $\phi$ of FePS$_3$, we  measure the angular dependence of the onset energy $E_{\rm kin, min}(k_{\rm \parallel})$ of the secondary photoelectrons (Figure~\ref{Fig_S4}b).
$E_{\rm kin, min}(k_{\rm \parallel})$ is  determined for each $k_{\rm \parallel}$ as $I''(E) = 0$, i.\,e. as the maximum slope of $I(E)$.
The resulting inflection points are then fitted by the parabola of free electrons in vacuum for $k_z = 0/{\rm \AA}$:
\begin{equation}
\label{eq:S2}
 E_\text{kin}^\text{A}(k_{\rm \parallel}) = \frac{\hbar^2}{2m} {k_{\rm \parallel}}^2 + \phi - \phi^\text{A}
\end{equation}
 ($m$: free electron mass) \cite{Sobota2021}. 
The minimum of the parabola is found at
 $E_\text{kin}^\text{A}(k_{\rm \parallel} = 
 0) = 0.875\,$eV leading to the FePS$_3$ work function  $\phi  = 5.35$\,eV.
\subsection{ARPES: Brillouin zone orientation and symmetry}
\label{sub:BZdet}

\begin{figure}[htb]
\centering
\includegraphics[width=0.45\textwidth]{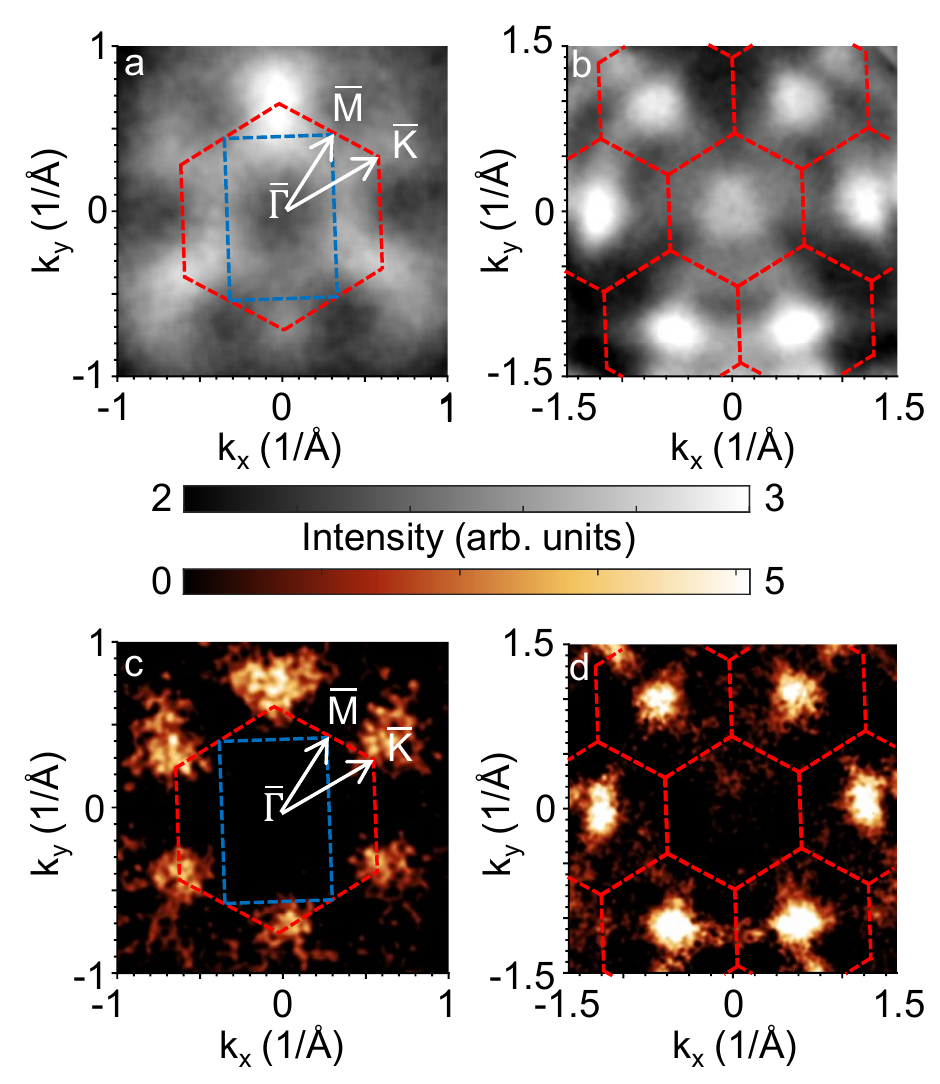}
\vspace{-0.4cm}
\caption{{\bf 
Brillouin zone determination.} (a) $I(k_x,k_y)$ plot at $E-E_{\rm F} = -7.7$\,eV, $90$\,K. The deduced geometric Brillouin zone (BZ) is marked in red with the position of $\overline{\Gamma}$, $\overline{\rm M}$ and $\overline{\rm K}$ indicated. The corresponding rectangular magnetic BZ is shown by blue-dashed lines. (b) $I(k_x,k_y)$ at $E-E_{\rm F} = -3,96$\,eV, 90\,K, for a larger $(k_x,k_y)$ area. First and second geometric BZs are marked in red. (c)-(d) Curvature plots $C(k_x,k_y)$ corresponding to a and b, respectively.
}
\label{Fig_S5}
\end{figure}

The $\overline{\Gamma}$ point in the central BZ is at the apex of the fit parabola to the onset of secondary electrons (Figure~\ref{eq:S5}b) that is determined with a precision of $\pm 0.006$\,\AA.
Moreover, the curvature of the parabola enables to attribute the angular distribution of photoelectrons to $k_\parallel$ values with a precision of $\pm 1$\,\%. 

The known atomic distances of the FePS$_3$ crystal \cite{Budniak2022,Kurosawa1983,Ouvrard1985}, 
result in a reciprocal unit cell with lengths of $|\overline{\Gamma{\rm M}}|=0.611$/\AA\,, $|\overline{\Gamma\rm K}|=0.705/$\AA\, and 
$|\overline{\Gamma \Gamma}|=1.222$/\AA. At the corresponding distances from the central $\overline{\Gamma}$, we indeed find hexagonally symmetric features in both,
the $I(k_x,k_y)$ (Figure~\ref{Fig_S5}a--b) and the $C(k_x,k_y)$ plots (Figure~\ref{Fig_S5}c--d). Such features also appear at various other energies. They are used to orient the two-dimensional BZs projections within the $(k_x, k_y)$ plane (red dashed lines in Figure~\ref{Fig_S5}) with a precision of $\pm 0.5^\circ$.

We do not observe any features with symmetries that are related to the rectangular magnetic Brillouin zone of the zigzag AFM configuration (blue dashed lines in Figure~\ref{Fig_S5}a, c) at any energy, neither in $I(k_x,k_y)$ nor in $C(k_x,k_y)$ plots. This is most likely due to the relatively small change of the wave functions by the magnetic order.


\subsection{ARPES: Background subtraction}
\label{sub:BG_subtractoin}
\begin{figure*}[htb!]
\centering
\includegraphics[width=0.9\textwidth]{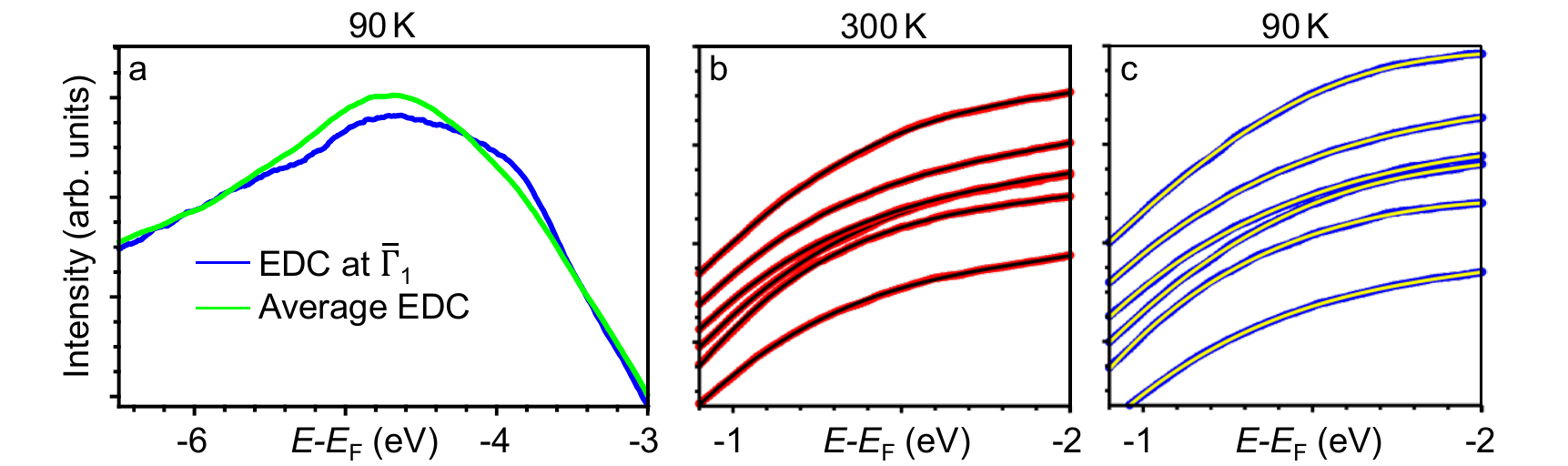}
\vspace{-0.4cm}
\caption{{\bf Background subtraction, $h\nu = 58$\,eV.} (a)  Blue: EDC at $\overline{\Gamma}_1$, green: average EDC of all $\boldsymbol{k}_\parallel$ along $\rm \overline{M\Gamma}_1\overline{M}$ within $\pm 0.5$/\AA\, around $\overline{\Gamma}_1$. 
(b) EDCs at various $k_{\rm \parallel}$ around the central $\overline{\Gamma}$ mutually offset for the sake of visibility, red dots: raw data, black lines: fourth order polynomial fits, $300$\,K. 
(c) same as b at $90$\,K, blue dots: raw data, yellow lines: fourth order polynomial fits.}
\label{Fig_S6}
\end{figure*}

In the main text, we used smoothing, normalization, and background subtraction of the raw data to increase the visibility of the changing features across $T_{\rm N}$.

Firstly, the data is smoothed by averaging across $\pm 40$\,meV in energy and $\pm 0.0291$/\AA\, in two perpendicular $\boldsymbol{k}_\parallel$ directions ($\pm 4$ pixel in each direction). Afterwards, each energy distribution curve (EDC) $I(E)$ is divided by its own integral for the purpose of normalization.\\
The subsequent background subtraction varied for the three different features discussed in the main text.
A linear background is subtracted to deduce the band shift of the P 3p$_z$/S 3p$_z$ band (Figure~3e--f, main text). 
To determine the energy shift of the Fe 3d$_{z^2}$ related rhombus (Figure~4, main text), we averaged all EDCs along the direction ${\rm \overline{M\Gamma}}_n{\rm \overline{M}}$ within $\pm0.5$/\AA\, around $\overline{\Gamma}_n$ 
and subtracted it from the EDCs at each $\boldsymbol{k}_\parallel$ in that region (Figure~\ref{Fig_S6}a). This largely removed the otherwise dominating flat Fe 3d band (peak in the green curve of Figure~\ref{Fig_S6}a), such that the band positions of the rhombus appear more clearly (Figure~4e, main text).

For the upper valence band that shows the relatively subtle separation in two bunches of bands above $T_{\rm N}$ (Figure~5a--b, main text), each EDC is fitted individually by a fourth order polynomial in the energy range from $E-E_{\rm F} = -0.9$\,eV to $-2$\,eV (Figure~\ref{Fig_S6}b--c). The fit is then subtracted revealing the reduced intensity between the two bunches of bands rather directly (Figure~5c--d, main text).


\subsection{ARPES: Curvature}
\label{sub:Curvature}

For better visibility of band structure features, we partially displayed the curvature $C(E,k_{\rm \parallel})$  of the raw ARPES intensity $I(E,k_{\rm \parallel})$ \cite{Peng2020}.  This emphasizes electronic bands by removing the smooth background from ineleastic scattering and by amplifying local maxima and minima of  $I(E,k_{\rm \parallel})$. 
We determine $C(E,k_{\rm \parallel})$  after Gaussian smoothing of the raw data in, both, energy ($\sigma_E = 75$\,meV) and $k_\parallel$ ($\sigma_{k_\parallel}=0.018\,$/\AA) via

\begin{equation}
 C(E,k_\parallel) = \frac{-I''(E,k_\parallel)}{(C_0 + I'(E,k_\parallel)^2)^{3/2}}\,\hspace{3mm}{\rm with}\,\hspace{3mm} C_0= a_0 |I'|_\text{max}^2.
 \label{eq:S1}
\end{equation}
 $I'$ and $I''$ are the first and second partial derivative  of the smoothed ARPES intensity $I(E,k_\parallel)$ with respect to energy, $|I'|_\text{max}$ is the maximum absolute value of the first derivative, while $a_0$ is a free parameter used to adjust the relative intensity of different bands. We have chosen $a_0=0.1$. 
For Fig.~2d, main text, an additional Gaussian smoothing is applied  after calculating the curvature with $\sigma_E=230$\,meV and $\sigma_k=0.068$\,/\AA in order to further reduce contrast over the displayed large energy scale. 

\begin{figure*}[htb]
\centering
\includegraphics[width=0.9\textwidth]{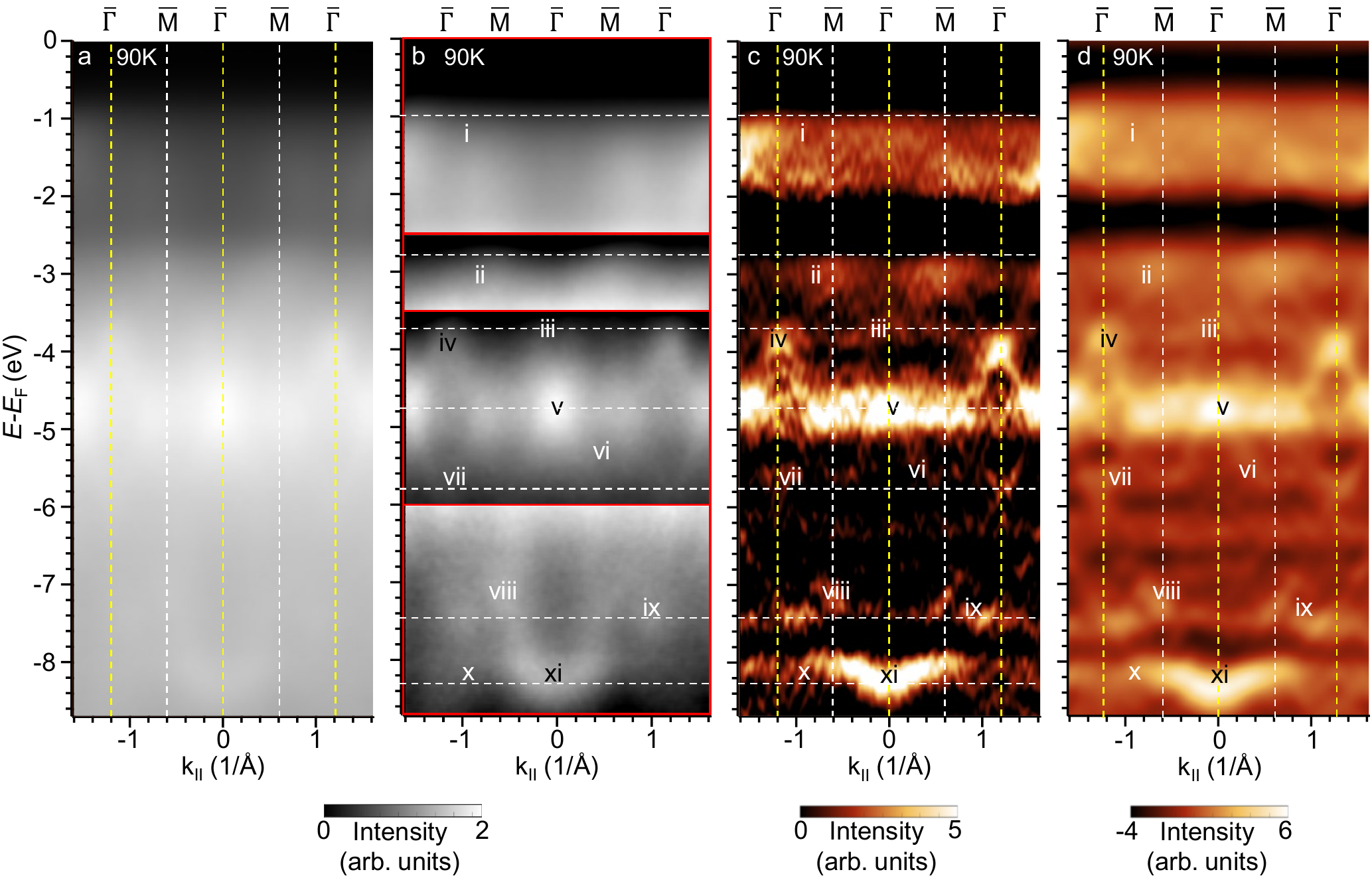}
\vspace{-0.4cm}
\caption{{\bf Raw data vs. curvature.} (a) Raw $I(E,k_{\rm \parallel})$ greyscale plot in ${\rm \overline{M\Gamma M}}$ direction after smoothing and normalizing each EDC to its integral value, dashed lines: centers and boundaries of the BZs, $90$\,K, $h\nu=58$\,eV. (b) Same data as a with different contrasts for different energy regions as marked by the red rectangles, white dashed lines: guides to the eye, features marked by i--xi are also visible in c.
(c) Curvature $C(E,k_{\rm \parallel})$ deduced from the data in a (Equation~1, main text). Same positions of horizontal lines and marks i--xi as in b, and of vertical lines as in a. (d) Same as c with additional Gaussian smoothing after curvature determination using $\sigma_E = 240$\,meV and $\sigma_k = 0.068$\,/\AA\, and also displaying negative curvature.  
}
\label{Fig_S6a}
\end{figure*}

To avoid artifacts by the curvature determination, caused e.\,g. by noise, it is mandatory to compare each feature with the raw data as shown exemplarily in Figure~\ref{Fig_S6a}. The raw data are smoothed and  each EDC is divided by its integral for normalization (section~1\ref{sub:BG_subtractoin}).
To enhance visibility of the correspondingly treated $I(E,k_\parallel)$ (Figure~\ref{Fig_S6a}a), the same data is also displayed at different contrasts for different energy regions
(red rectangles in Figure~\ref{Fig_S6a}b). The features discussed in the main text and a few additional features are marked by i--xi for comparison with the curvature plots (Figures~\ref{Fig_S6a}c-d). Two curvature plots are displayed, ones without and once with the additional smoothing after curvature determination. Moreover, Figure~\ref{Fig_S6a}c does only display the positive values of $C(E,k_{\rm \parallel})$ corresponding to peaks in the raw $I(E,k_{\rm \parallel})$ data, while Figure~\ref{Fig_S6a}d also shows negative values.

Multiple features are easily recognizable in both images (ii, iii, iv, v, vii, viii, ix, xi), while others require careful zooming (vi,x). Feature i shows the rising edge at high energy only in Figure~\ref{Fig_S6a}b, but appears distinct from the neighboring low energy bands in Figure~\ref{Fig_S6a}a, such that the separated bunch of bands in the curvature (Figure~\ref{Fig_S6a}c) is confirmed. A corresponding plateau-like region is also visible in the raw-data EDCs at $\overline{\Gamma}$ (Figure~2e, main text). Importantly, all features discussed in the main text are corroborated by comparison with the raw data. It is also apparent that the more steep, nearly vertical dispersions are less pronounced in the curvature, as e.\,g. the side facets of the rhombus in the 2$^{\rm nd}$ BZ, most strikingly on the right hand side which is due to the curvature determination along energy. 

Notice, moreover, that features in the curvature plot of Fig.~\ref{Fig_S6a}d that are not labeled appear at negative $C(E,k_{\rm \parallel})$, e.g. at $E-E_{\rm F}\in [-6\,{\rm eV},-7{\rm eV}]$, and are, hence, areas of reduced dip curvature in $I(E,k_{\rm \parallel})$.



\section{Computational Details}
\label{sec:Computational_Details}
\subsection{DFT+U calculations}
\label{sub:DFT+U}
The volume and shape of the magnetic unit cell of bulk FePS$_3$, as well as the positions of the atoms, were optimized by plane-wave based DFT+U calculations utilizing the VASP code \cite{Kresse1993,Kresse1996}. 
These were spin-polarized calculations with collinear, oppositely oriented spins at the two Fe positions.  The projector-augmented wave (PAW) method \cite{Blochl1994,Kresse1999}  was used to represent the orbitals of the ion cores. Specifically for Fe, the configuration $3p^6 \, 3d^7 \, 4s^1$ was used for the valence orbitals, while  
P and S were treated with 5 and 6 valence electrons, respectively. The cut-off for the plane-wave basis was set to $E_{cut} = 300$eV, and a $9 \times 5 \times 9$ {\it k}-point mesh was used to sample the Brillouin zone. 
Since FePS$_3$ is a layered material held together by van-der-Waals forces, the 
Grimme D2 pairwise interactions \cite{Grimme2006}
were added to the energy and forces. 
The calculations employed the 
PBE-GGA functional \cite{Perdew1996}; electronic correlation in the $3d$ shell of Fe was enhanced with the help of the DFT+U approach in the spherically-symmetric variant proposed by  
Dudarev \cite{Dudarev1998} with $U_{\rm eff} = U-J$ and $J=1.0$\,eV in all cases.  
As a double-counting correction scheme for the DFT+U calculations, the so-called fully-localized limit for the Fe $3d$ orbitals was used.

While bulk FePS$_3$ is described in the literature as a monoclinic crystal, i.e. with lattice vectors $a$ and $b$ perpendicular to each other, we prefer to use a hexagonal cell in the $a-b$ plane with an angle of $120^{\circ}$ between these vectors. The $b$-vector is chosen almost twice as long as the $a$-vector which allows us to describe the magnetic unit cell containing 4 Fe atoms. After optimization, $|a| = 5.923$\,{\AA} and  $|b| = 11.821$\,{\AA} are obtained. The $a$ and $b$ vectors span the plane of one FePS$_3$ layer. The stacking of layers is described by the $c$-vector of length $6.701$\,{\AA} that forms an angle of $98.5^{\circ}$ with the $a$ and $b$ vectors. 
In this stacking, a P atom of one layer sits above a S atom of the layer below. 
The interlayer distance between two adjacent S layers of different FePS$_3$ units amounts to $3.35$\,{\AA}. 
After relaxation, the distances of the Fe-Fe pairs are obtained as $3.307$, $3.421$ and $3.509$\,{\AA}. The longest Fe-Fe distance occurs between Fe atoms with parallel spin orientation. The relaxed bond length of the P dimers is found as $2.20$\,{\AA}, larger that the $2.02$\,{\AA} as reported previously from the analysis of neutron diffraction data \cite{Lanon2016}. All S atoms at the same side of the FePS$_3$ layer are located at the same height. In other words, there is no buckling within the S layer.\\

\begin{figure*}[htb]
\centering
\includegraphics[width=0.9\textwidth]{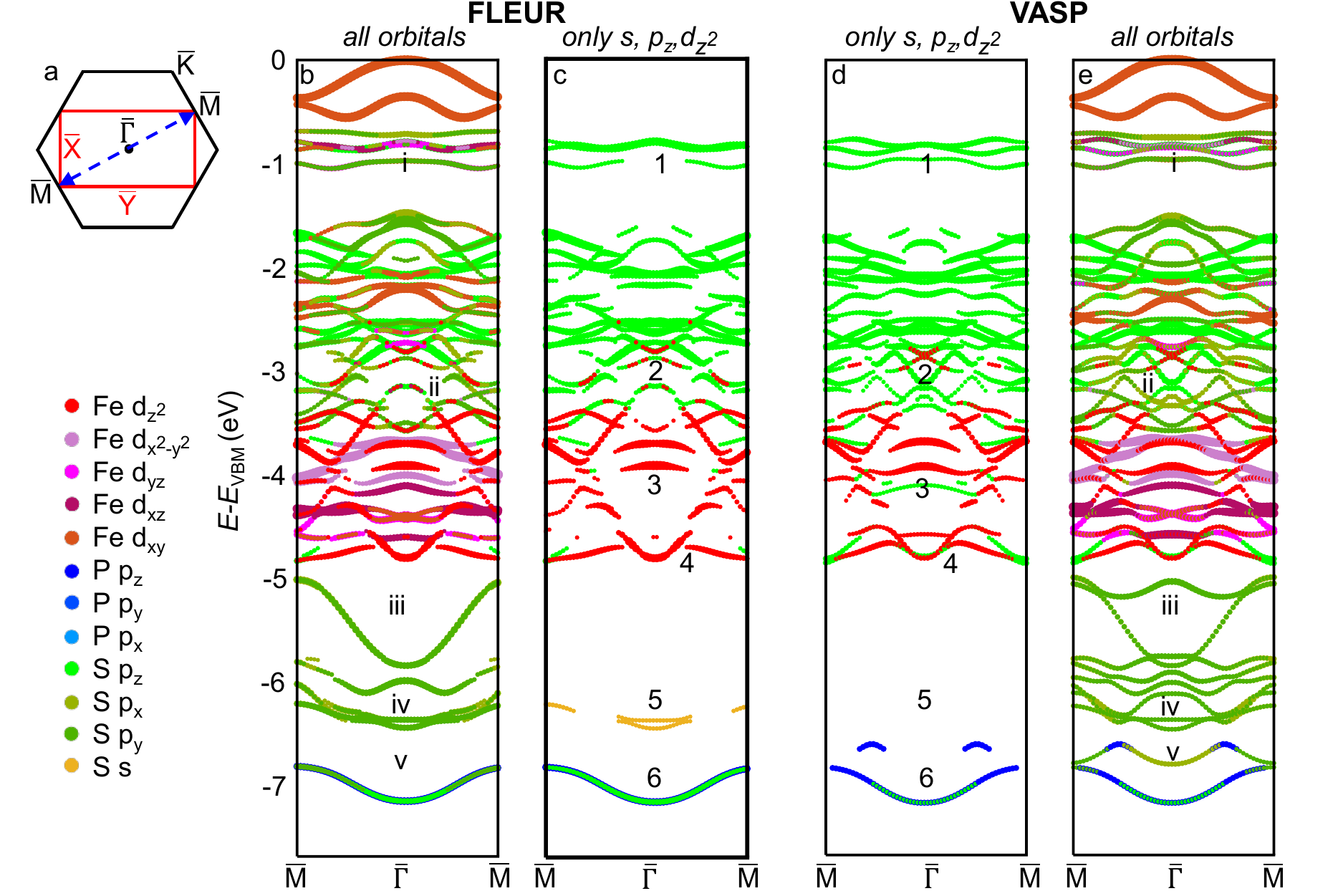}
\vspace{-0.4cm}
\caption{{\bf Comparison of band structures calculated by FLEUR (unfolded) and VASP (folded).} (a) Surface projected Brillouin zone (BZ) of the atomic arrangement (black) and of the AFM zigzag magnetic arrangement (red) of FePS$_3$ with marked high symmetry points and $k_\parallel$ direction of the plots in b--e (dashed blue arrow).  (b) Band structure of the AFM zigzag configuration that is unfolded from the magnetic to the atomic BZ \cite{Bui:23}, DFT+U (FLEUR), $U_{\rm eff} =  1.2$\,eV, $k_{\rm z}=-0.07$/\AA, all orbital contributions of each state are displayed, if larger than 10\,\%, the diameter of the symbol is proportional to the contribution, 100\,\% contribution corresponds to the symbol diameter in the legend on the left. Labels i--v are for comparison with e. (c) Same as b, but displaying only the contributions of s, p$_z$ and d$_{z^2}$-type orbitals as required by the simplified ARPES selection rules \cite{Moser2017}. Symbols 1--6 are for comparison with d. (d) Band structure within the magnetic BZ (without unfolding), DFT+U (VASP), $U_{\rm eff} = 1.2$\,eV, $k_{\rm z}=0$/\AA, only s, p$_z$ and d$_{z^2}$-type orbitals. (e) Same as d with all orbitals.
}
\label{Fig_S15}
\end{figure*}

For the distinct calculations with the full-potential linearized augmented planewave method, we employed the FLEUR code~\cite{fleurCode} using the geometry that have been optimized by the VASP calculations described above. Again, the generalized gradient approximation in the form of Ref.~\cite{Perdew1996} and DFT+$U$ with $U=1.4$~eV and $J=0.2$~eV was used. The muffin-tin radii were $1.33$~{\AA} for Fe and $0.99$~{\AA} for P and S. The basis set cutoff was $7.6$~{\AA}$^{-1}$. An unfolding procedure as described in Ref.~\cite{Bui:23} was used to plot the electronic bands obtained from the $(2 \times 1)$ calculation in a $(1 \times 1)$ Brillouin zone.

\subsection{Comparing unfolded band structure (FLEUR) and folded one (VASP)}
\label{sec:compare_FLEUR_VASP}

Figure~\ref{Fig_S15} compares the band structure after unfolding to the geometric BZ as calculated by FLEUR (Figure~\ref{Fig_S15}b, c) with the band structure within the magnetic BZ as calculated by VASP (Figure~\ref{Fig_S15}d, e). Both calculations use the same atomic geometry resulting from the atomic relaxation within the VASP calculation and employ the zigzag-type AFM order.
The values of $U_{\rm eff}=1.2$\,eV are also identical, while the $k_z$ values  slightly differ to an extent of minimal influence on the bands  (Figure~\ref{Fig_S12}). We display the $\overline{\rm M\Gamma M}$ direction that is identical in the atomic and in the magnetic BZ (Figure~\ref{Fig_S15}a). Nevertheless, bands from the surrounding atomic BZ are folded into the magnetic BZ, eventually also appearing along the $\overline{\rm M\Gamma M}$ direction. Both, the full band structure (Figure~\ref{Fig_S15}b, e) and the state contributions from selected orbitals as deduced from the simplified selection rules \cite{Moser2017} (Figure~\ref{Fig_S15}c, d) are displayed.

Additional bands are indeed visible for the non-unfolded band structure
(areas marked by v, iv, iii, 3, ii, 2, 1) that we attribute to backfolding of bands from the atomic into the magnetic BZ. Such bands  are typically not visible in ARPES, since the photon induced excitation projects to the same free-electron states with and without magnetic structure, while the backfolded states are barely changed by the magnetic structure. Hence, their projection to final state $\boldsymbol{k}$ values remains largely unchanged by the magnetic transition and the related BZ size reduction.  

\begin{figure*}[htb]
\centering
\includegraphics[width=0.9\textwidth]{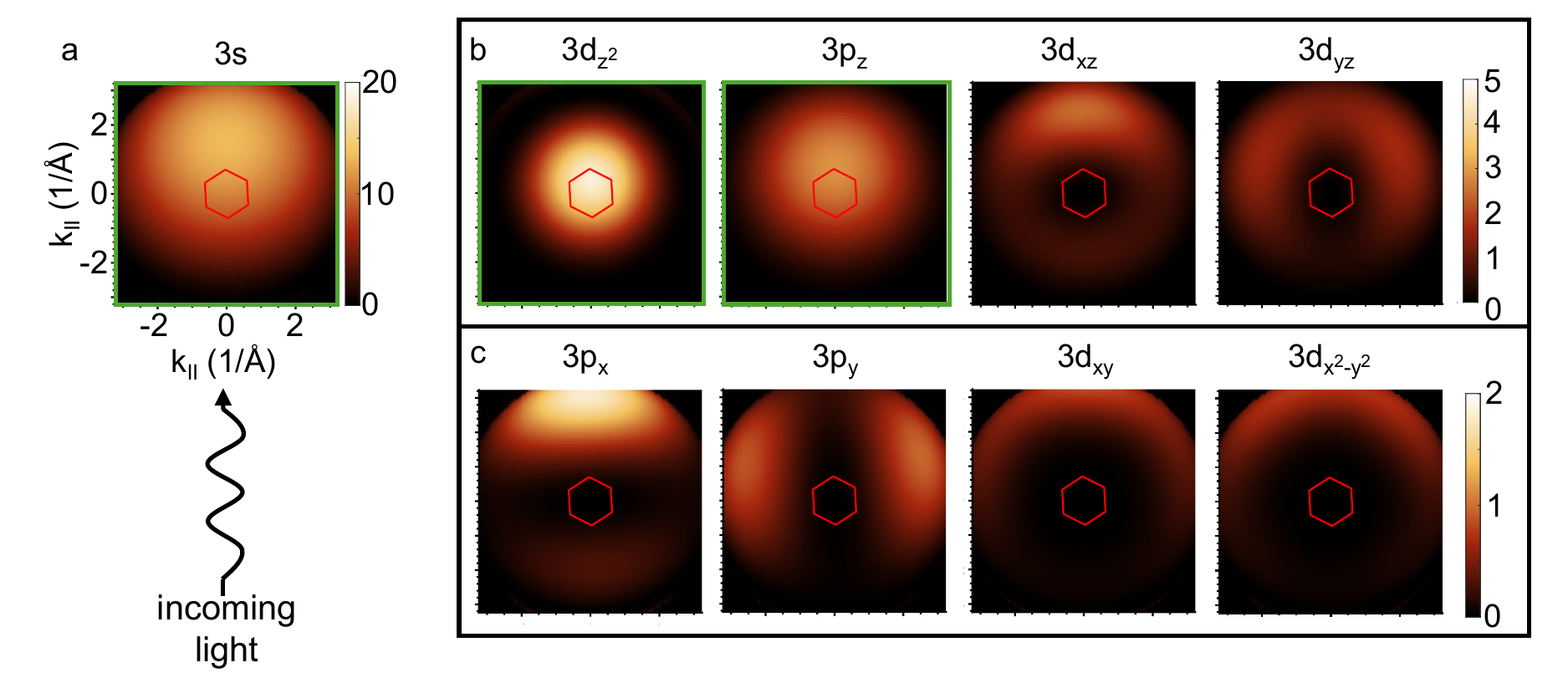}
\vspace{-0.4cm}
\caption{{\bf Matrix elements for different initial orbitals according to \cite{Moser2017}.} Calculated $\boldsymbol{k}_{{\rm f}\parallel}$ distribution of the matrix element 
$|M_{\boldsymbol{k}_{\rm f}, \ket{nlm}}|$ 
for excitations from the initial orbital $\ket{nlm}$, 
as marked on top within the ($x,y,z$) base, to a final plane wave with wave vector $\boldsymbol{k}_{\rm f}$, color scales in a, b and c are distinct, $k_\parallel$ ranges are identical in all plots, c displays only in-plane orbitals. The first atomic Brillouin zone of $\rm FePS_3$ is marked as a red hexagon. The incident p-polarized light impinges from the bottom (marked below a) at an angle of $65^{\circ}$ with respect to the surface normal.
Preferentially visible orbitals within the first BZ are framed in green. 
}
\label{Fig_S7}
\end{figure*}

Moreover, the bands of the two calculations are partially slightly offset in energy with respect to each other by $\sim 30$\,meV
(e.g. features $2$, $6$), benchmarking the precision of the calculations. 
Also the orbital contributions of the bands are partially distinct. For example, feature 6 has a stronger S 3p$_z$ contribution in the FLEUR calculation. The partial disappearance of the S 3p$_z$ part in the VASP data is caused by our strict threshold of 10\,\%. We crosschecked that the differences in orbital contributions are always in the $\pm 5$\,\% range. The same is correct for the S 3s contribution ($\sim 15$\,\%) to the band marked 5 in Figure~\ref{Fig_S15}c that is not apparent in the VASP calculation, since having only $7$\,\% contribution. The differences are likely caused by the cut-off radii for the projection in the FLEUR code due to the muffin-tin potentials (section~2\ref{sub:DFT+U}).  
Importantly, these minor differences barely influence the conclusions in the main text and the two calculations enable us to identify bands in the magnetic BZ that are barely visible by ARPES.

\subsection{Simplified selection rules for ARPES}
\label{sub:Selection}
The visibility of occupied electronic bands in ARPES depends on the matrix elements of their photoelectron excitation into final states that propagate into vacuum \cite{Hüfner2013}. 
A simple approach assumes the sudden dipole approximation and plane waves  as final vacuum states with exponential decay into the crystal \cite{Moser2017}.
We further simplify this model by ignoring the decay of the final state into the crystal (decay lengths $\lambda \rightarrow\infty$) and the negligible wavelength of the photon (photon wave vector $\boldsymbol{k}_{h\nu}=0$).

The matrix element M$_{\textbf{\textit{k}}_{\rm f} {\textbf{\textit{k}}}}$ between initial state at wave vector $\textbf{\textit{k}}$ and final state in vacuum with wave vector $\boldsymbol{k}_{\rm f}$ then reads (based on Equation~(20) in \cite{Moser2017})

\begin{equation}
\begin{split}
M_{\textbf{\textit{k}}_{\rm f}; 
\boldsymbol{k},\kappa} \propto  \textrm{i}\,\textbf{\textit{$E$}} \cdot \textbf{\textit{k}}_{\rm f}\braket{\textbf{\textit{k}}_{\rm f}\vert \textbf{0},\kappa} \cdot\delta(\boldsymbol{k}-\boldsymbol{k}_{\rm{f}}+ \boldsymbol{K})
\end{split}
\label{eq:S_matrixelements}
\end{equation}
Here, $\boldsymbol{E}$ 
is the electric polarization vector of the incoming light, 
$\ket{\boldsymbol{0},\kappa}$ is the atomically periodic part of the initial electronic state (Wannier state) for the band with index $\kappa$ taken for the unit cell at position $\boldsymbol{R}=\boldsymbol{0}$, 
and $\boldsymbol{K}$ is a reciprocal lattice vector.
The final state $\bra{...}$ is described as a single plane wave. 
The polarization vector $\boldsymbol{E}$ 
is adapted to the experiment, where the light impinges with an angle $\alpha = 65^{\circ}$ to the surface-normal and is polarized in the plane of incidence (p-polarization). As initial Wannier states, we use atomic orbitals in the $(x,y,z)$ base.

Figure~\ref{Fig_S7} shows the results for all relevant orbitals in the probed energy regime. The first atomic BZ is marked as a red hexagon. It is obvious that the 3s and 3d$_{z^2}$ states should have the strongest visibility in the first BZ followed by the 3p$_z$ states, while all other states have negligible visibility due to an intensity node in the BZ center. Within the surrounding BZs, that are also accessible in the experiment (Figure~\ref{Fig_S5}b, d), there might be additional contributions from 3p$_x$, 3p$_y$, 3d$_{yz}$, and 3d$_{xz}$ with directional dependence, but 3s, 3p$_z$ and 3d$_{z^2}$ contributions are still strongly dominating. 
This is the reason to use only states with significant contribution of 3s, 3p$_z$ or 3d$_{z^2}$ orbitals for comparison with the experiments.
Note, however, that the model is severely simplified by ignoring any type of orbital structure of the final states within the crystal by assuming them as pure plane waves.

%

\subsection{Adjusting $U_{\rm eff}$}
\label{sec:U_select}

\begin{figure*}[htb]
\centering
\includegraphics[width=1.0\textwidth]{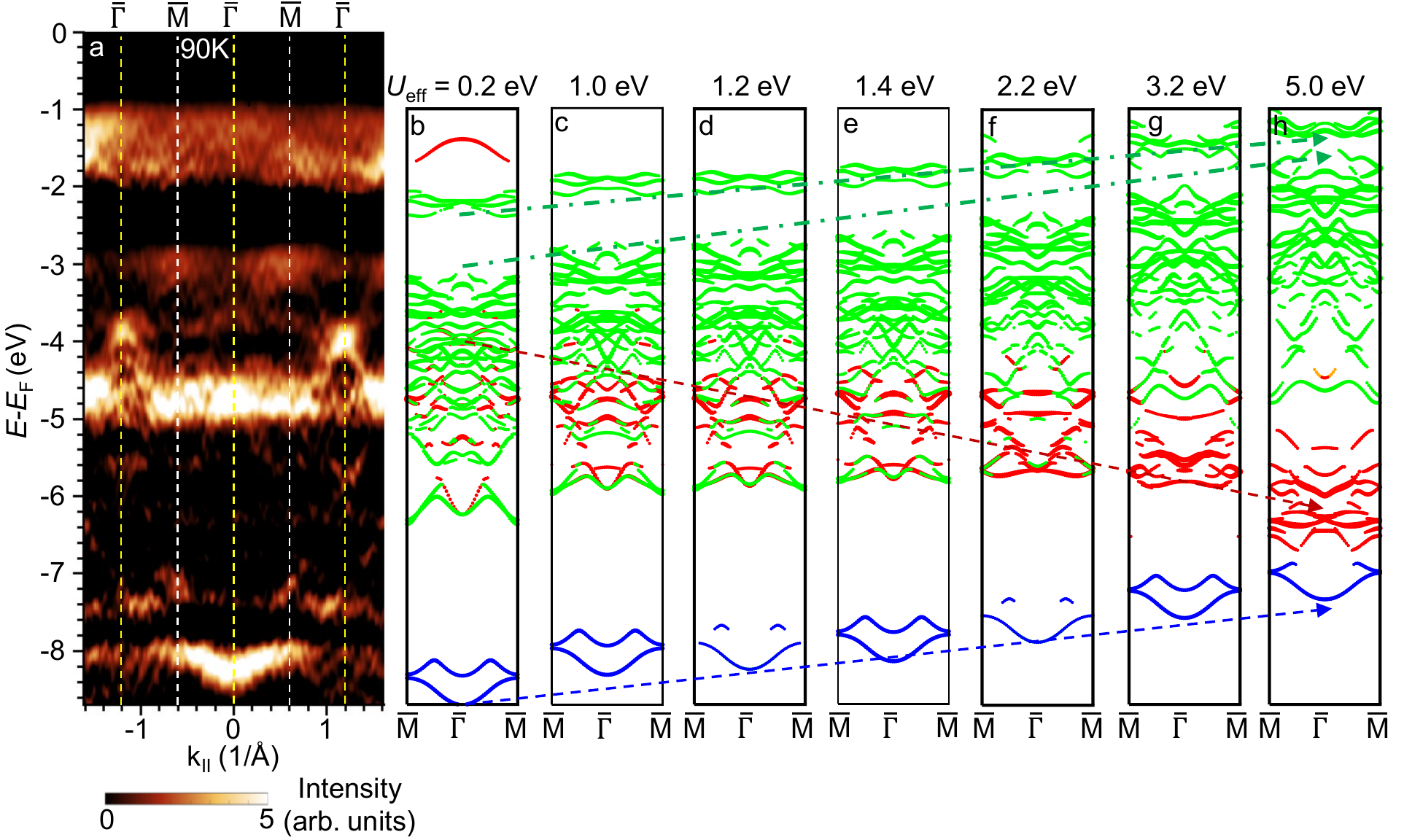}
\vspace{-0.4cm}
\caption{{\bf ARPES curvature vs. DFT+U band structure at different $U_{\rm eff}$.} 
(a) ARPES curvature, $90$\,K, $h\nu=58$\,eV. (b)--(h) Band structure from DFT+U calculation (VASP) at various $U_{\rm eff}$ as marked, $k_z = 0$/\AA. Only states with more 10\,\% contribution from $\rm d_{z^2}$, $\rm p_z$ or s orbitals are displayed, same color code as in Figure~\ref{Fig_S15}. The energy scale is identical to a, while the valence band maximum of the calculated band structures is always rigidly shifted to  $E-E_{\rm F}=-1$\,eV of a. 
Arrows mark continuous changes of band features with $U_{\rm eff}$: green: gap closing, red: downwards energy shift and orbital character change, blue: upwards energy shift.
}
\label{Fig_S8_b}
\end{figure*}
For the detailed comparison between DFT+U calculations and ARPES results, the onsite-energy $U_{\rm eff}$ has to be adapted. 
Figure \ref{Fig_S8_b} shows ARPES results at $k_z\approx 0/$\AA\,
(section~2\ref{sub:kzhnu}) in comparison with the calculated band structure at $k_z\approx 0/$\AA\, and various $U_{\rm eff}$ displaying the selected orbitals only (section~2\ref{sub:Selection}). The valence band maximum in the calculations is always aligned with $E-E_{\rm F}=-1$\,eV of the experiment.

Several trends appear with decreasing $U_{\rm eff}$ :
\begin{enumerate}
    \item A gap is continuously closing between two blocks of S 3p$_z$ bands (green arrows). Such gap is apparent in the ARPES data pointing to low $U_{\rm eff}$.
    \item A block of Fe 3d$_{z^2}$ bands moves downwards in energy and loses its S 3p$_z$ bands contribution, i.\,e. Fe 3d$_{z^2}$ bands and S 3p$_z$ bands separate  (red arrow). The resonance at the Fe 3p-3d transition that we observe  at $E-E_{\rm F}\approx -(4-6)$\,eV restricts the possible selection of $U_{\rm eff}$ markedly.
    \item The parabolic P 3p$_z$ band moves upwards (blue arrow).
\end{enumerate}

Comparing to the experiment, only the band structures at $U_{\rm eff}=1.0-1.4$\,eV match reasonably. At higher energy, the gap in the S bands is closing too strongly and the Fe 3d$_{z^2}$ bands are moving too much downwards. At lower $U_{\rm eff}$, the Fe 3d$_{z^2}$ contributions are too low to explain the observed resonance around $E-E_{\rm F}\approx -(4-6)$\,eV. The more precise selection is a bit arbitrary such that we used $U_{\rm eff}=1.2$\,eV as the central value of the three options.

\newpage
\subsection{Stretched DFT+U data}
\label{sub:stretched}

\begin{figure}[htb]
\centering
\includegraphics[width=0.43\textwidth]{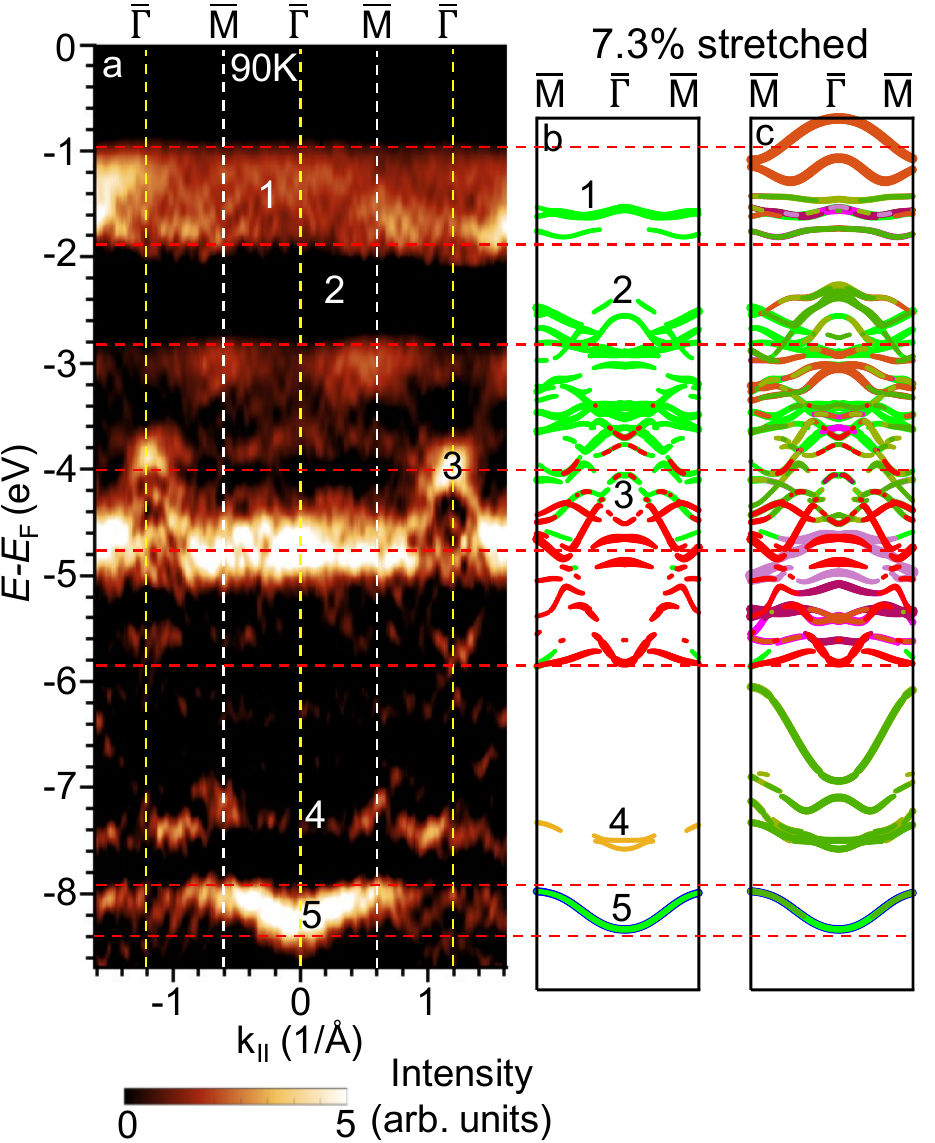}
\vspace{-0.4cm}
\caption{{\bf ARPES curvature vs. stretched DFT+U band structure} (a) ARPES curvature, $h\nu = 58$\,eV, $90$\,K, $\rm \overline{M\Gamma M}$-direction, 1-5 mark features referred to similar features in b. (b) Unfolded DFT+U band structure (FLEUR) with selected  s, $\rm p_z$ and $\rm d_{z^2}$ orbital character ($>10$\,\%), $U_{\rm eff} = 1.2$\,eV, $k_z=-0.07$/\AA, (c) same as b, but with all orbitals, same color code as in Figure~\ref{Fig_S15}.}
\label{Fig_S12a}
\end{figure}

The comparison between the ARPES data (Figure~\ref{Fig_S8_b}a) and the 
calculated band structure using the selected $U_{\rm eff}=1.2$\,eV  (Figure~\ref{Fig_S8_b}d) does still not match ideally, i.\,e. the lowest energy P 3p$_z$ band (blue) is too high in energy in the calculation and the upper S 3p$_z$ band is too low. This can be partly cured by stretching the DFT+U band structure slightly (Figure~\ref{Fig_S12a}b--c) in line with \cite{Nitschke2023}. We aligned the upper valence band of the calculation with $E-E_{\rm F} = -0.7$\,eV of the experiment and stretched the calculated band structure by 7.3\,\% providing favorable agreement of features 3, 4, and 5 in Figure~\ref{Fig_S12a}a--b, both in shape as well as in energetic position, but still less good agreement for the band gap size between feature 1 and 2, being less in the calculations.
For feature 3, both the rhombus and the flat bands (Fe 3d$_{z^2}$ character) are well aligned with the experimental data. The origin of the improvement by stretching is unknown, but might be related to correlations that are not captured by the DFT+U calculations or final state effects related to the photohole in the ARPES experiment.

\subsection{Relating $k_z$ to $h\nu$}
\label{sub:kzhnu}

\begin{figure*}[htb]
\centering
\includegraphics[width=1.0\textwidth]{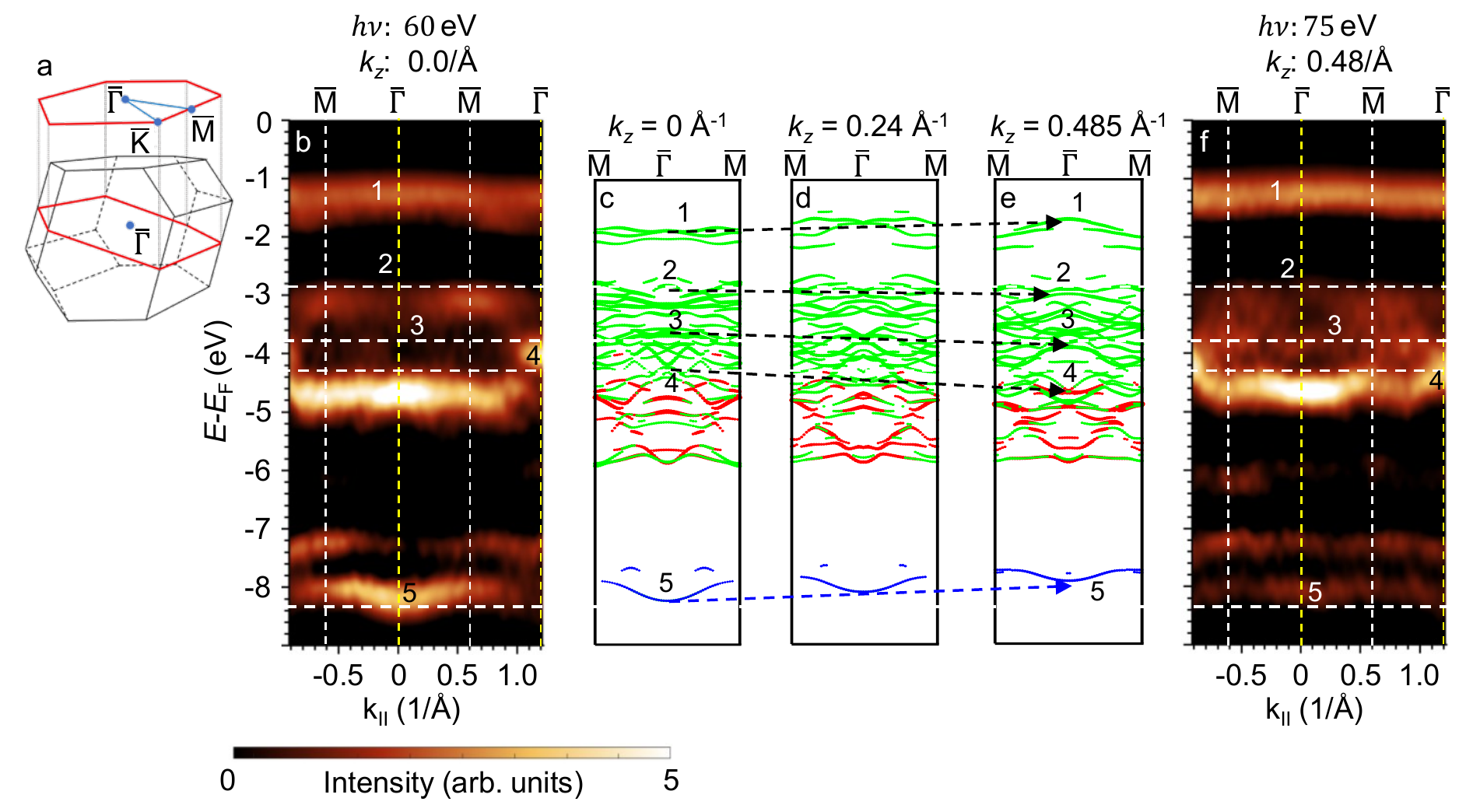}
\vspace{-0.4cm}
\caption{{\bf ARPES data vs. calculated band structure at various $k_z$.} 
(a) Brillouin zone of the bulk (black) and the monolayer (red, top) with corresponding projection (red in 3D BZ), high symmetry points and direction of ARPES data are marked. (b) ARPES curvature, $\rm \overline{M\Gamma M}$ direction, $h\nu=60$\,eV, $90$\,K, $k_z$ is deduced from the model of a free electron parabola for final states (see text),
horizontal lines are guides to the eye, numbers 1-5 mark features for comparison with b--e. (c)--(e) Band structures of DFT+U calculations (VASP), $U_{\rm eff} = 1.2$\,eV,  $k_z$ marked on top, only bands with s, $\rm p_z$ or $\rm d_{z^2}$ orbital character ($>10$\,\%) are displayed, same color code as in Figure~\ref{Fig_S15}, arrows: dispersion of the labeled features with $k_z$. (f) Same as b, but at $h\nu=75$\,eV.
}
\label{Fig_S12}
\end{figure*}

Since the DFT+U calculations do not reproduce the 
measured band structure in all details, it is difficult to adapt $k_z$ of the calculations precisely to the photon energy $h\nu$. We found a few general trends that can be used for a rough adaption. Figure~\ref{Fig_S12} shows the band structure obtained by DFT+U calculations ($U_{\rm eff}=1.2$\,eV) at various $k_z$ in comparison with ARPES data at two photon energies. The photon energies are selected to roughly cover  half of the BZ in $k_z$ direction (reciprocal lattice vector $G_\perp = 0.9652/$\AA) as described below. 
Some features that change with $k_z$ are marked in Figure~\ref{Fig_S12}c-e. The lower P 3p$_z$ band (5) is moving upwards and gets flatter with increasing $k_z$, the Fe 3d$_{z^2}$ rhombus (4)  gets compressed in energy, the bands in the lower block of S 3p$_z$ character (2,3) move slightly downwards and change their detailed in-plane dispersions, and the upper S 3p$_z$ bands move upwards and extend in energy range.  
Some similar differences are observed between the two ARPES curvature plots, namely the lowest band (5) is at higher energy and flatter for $h\nu=75$\,eV than at $h\nu=60$\,eV, the upper half of the rhombus (4) is less steep at $h\nu=75$\,eV, and the upper band (1) is slightly higher in energy at $h\nu=75$\,eV. On the other hand, the changes in the block of S 3p$_z$ bands (2,3) are distinct from the calculations and the increased width of the upper S 3p$_z$ bands is also not observed.

To assign $k_z$ values to the experimental data, we choose the model of a free electron parabola
of final states shifted by an inner potential $V_0$ within the crystal reading \cite{Sobota2021}
\begin{equation}
\label{eq:S5}
E_{\rm final}(\textbf{\textit{k}}_{\rm f}) = \frac{\hbar^2\textbf{\textit k}_{\rm f}^2}{2m}-V_0,
\end{equation}
with the final state energy within the crystal $E_{\rm final}$. 
We assume that the ARPES data at $h\nu = 60$\,eV and $E-E_{\rm F}= -8.25$\,eV belong to $k_z=0$, such that the corresponding states at $\boldsymbol{k}_\parallel=\boldsymbol{0}$ have a wave vector identical to a reciprocal lattice vector $\boldsymbol{G}$. Using an in-plane component $\boldsymbol{G}_\parallel=\boldsymbol{0}$ and Equation~(\ref{eq:S2}), one obtains
\begin{equation}
V_0 = \frac{\hbar^2}{2m} (
n G_\perp)^2 -h\nu-(E-E_{\rm F}-\phi)
\label{eq:S6}
\end{equation}
with the perpendicular reciprocal lattice vector of ${\rm FePS_3}$ $G_\perp = 0.9652/$\AA \cite{Budniak2022,Kurosawa1983,Ouvrard1985}, an integer $n$, and the work function of FePS$_3$ $\phi=5.35$\,eV (section~1\ref{sub:Fermi_workfunciton}). 
This results in $V_0=11.1$\,eV for $n=4$ as first positive value and $V_0=43.3$\,eV for $n=5$. A typical inner potential is between a few eV and $20$\,eV \cite{zhang2022}, such that we opt for $V_0=11.1$\,eV. Relating $k_z$ to $h\nu$ then reads ($\theta$ polar angel of the photoelectrons with respect to the surface normal) \cite{Hüfner2013}
\begin{equation}
k_z = \frac{1}{\hbar}\sqrt{2m\,(h\nu+(E-E_{\rm F}-\phi)\,\cos^2(\theta)+V_0)}.
\label{eq:S7}
\end{equation}

We get $k_z\approx -0.07\,/$\AA\ at
$h\nu = 58$\,eV as used in the main text and
 $k_z\approx -0.48\,/$\AA\ at
$h\nu = 75$\,eV (Figure~\ref{Fig_S12}f).
The typical variation at fixed $h\nu$ is $\Delta k_z^{E}\approx0.2/$\AA\, in the displayed energy range and $\Delta k_z^{\theta}\approx0.18/$\AA\, in the displayed $k_\parallel$ range.


\section{Additional results}

\subsection{Photon energy dependence of ARPES data}
\label{sub:hnu_dependence}

\begin{figure*}[htb]
\centering
\includegraphics[width=1.0\textwidth]{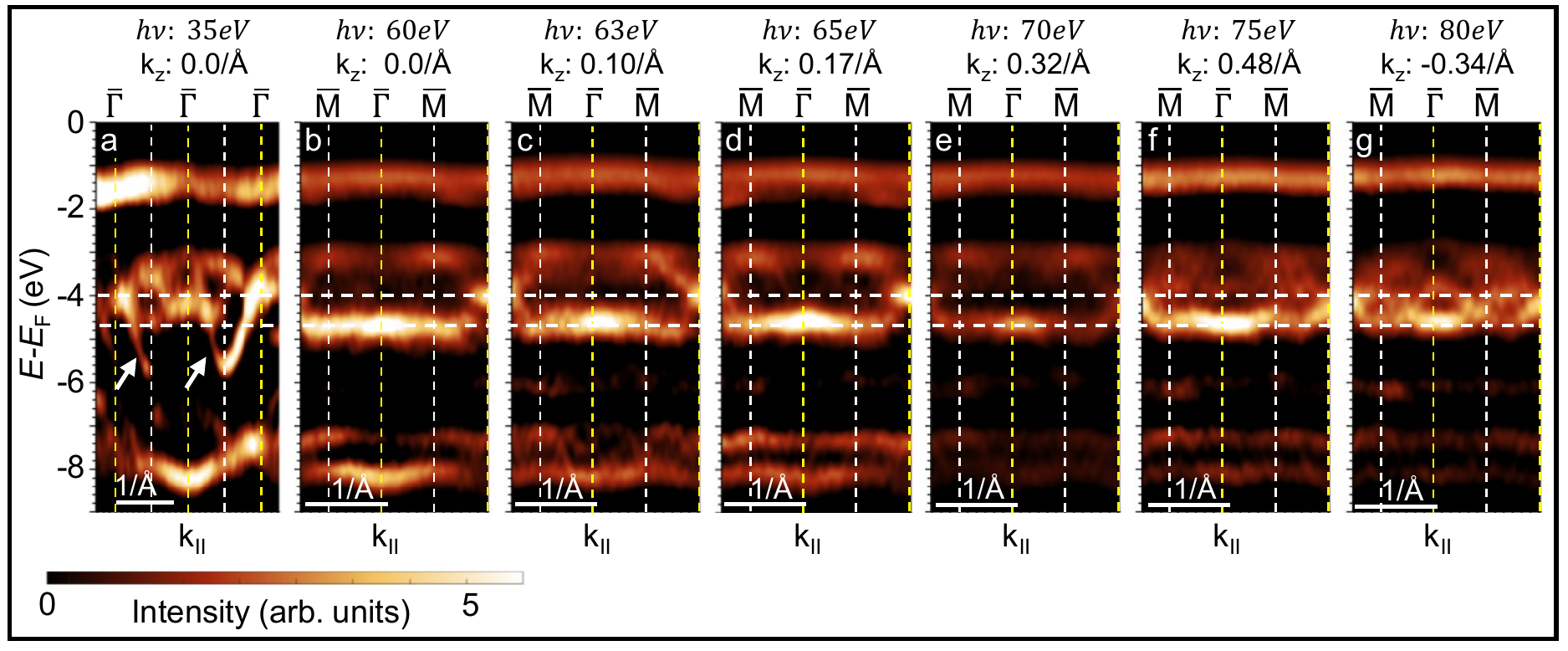}
\vspace{-0.4cm}
\caption{{\bf ARPES curvature for different photon energies,} (a)-(g) $\rm \overline{M\Gamma M}$ direction, $T = 300$\,K, $h\nu$ as marked on top, same color code for all images, $k_z$ is deduced for $E-E_{\rm F}= -8.25$\,eV using an inner potential $V_0=11.1$\,eV (section~2\ref{sub:kzhnu}).(a) The data at $h\nu=35$\,eV are recorded on a different flake (thickness: $16$ layers) with wider $k_\parallel$ range, arrows are discussed in  the text. 
}
\label{Fig_S20}
\end{figure*}

The photon energy dependence of the ARPES data depends on both, the change of $k_z$ and the sensitivity to different orbitals and band energies via final state effects (matrix elements) or initial state resonances. Figure~\ref{Fig_S20} shows a whole set of ARPES plots for different photon energies at 300\,K. Using the inner potential $V_0$ from section~
2\ref{sub:kzhnu}, we added $k_z$ values that covered more than half the BZ for $h\nu=60-80$\,eV.
The previously addressed changes with increasing $k_z$ (section~2\ref{sub:kzhnu}) are apparent as the flattening and upwards shift of the low energy P 3p$_z$ band, the slight upwards shift of the S 3p$_z$ band at $E-E_{\rm F}\approx -1.7$\,eV, and the less steep Fe 3d$_{z^2}$ rhombus. Additionally, some systematic changes of the bands at $E-E_{\rm F}\in [-3,-4]$\,eV
appear, namely more intensity in the range $E-E_{\rm F}\approx -4$\,eV at large $k_z$ accompanied by a reduced intensity at the central $\Gamma$ and 
$E-E_{\rm F}\approx -3$\,eV. 

As discussed in section~2\ref{sub:kzhnu}, only the 
change of dispersion of the P 3p$_z$ band is also apparent in the DFT+U band structures. This raises the suspicion that most of the changes are caused by the sensitivity such that distinct groups of bands appear with large intensity. This is corroborated by the low-energy ARPES plot ($h\nu=35$\,eV) that according to $V_0$ has the same $k_z=0$ as at $h\nu=60$\,eV, but looks markedly different. Only the upper S 3p$_z$ band is quite similar, while the P 3p$_z$ has an even stronger dispersion that might be related to uncertainties in the determination of $k_z$. More strikingly, the bands at $E-E_{\rm F}\approx -4$\,eV are completely different reproducing previous ARPES results at $h\nu= 21.2$\,eV, $21.7$\,eV and $40.8$\,eV \cite{Nitschke2023,Yan2023b,Koitzsch2023}. 

\begin{figure}[htb]
\centering
\includegraphics[width=0.45\textwidth]{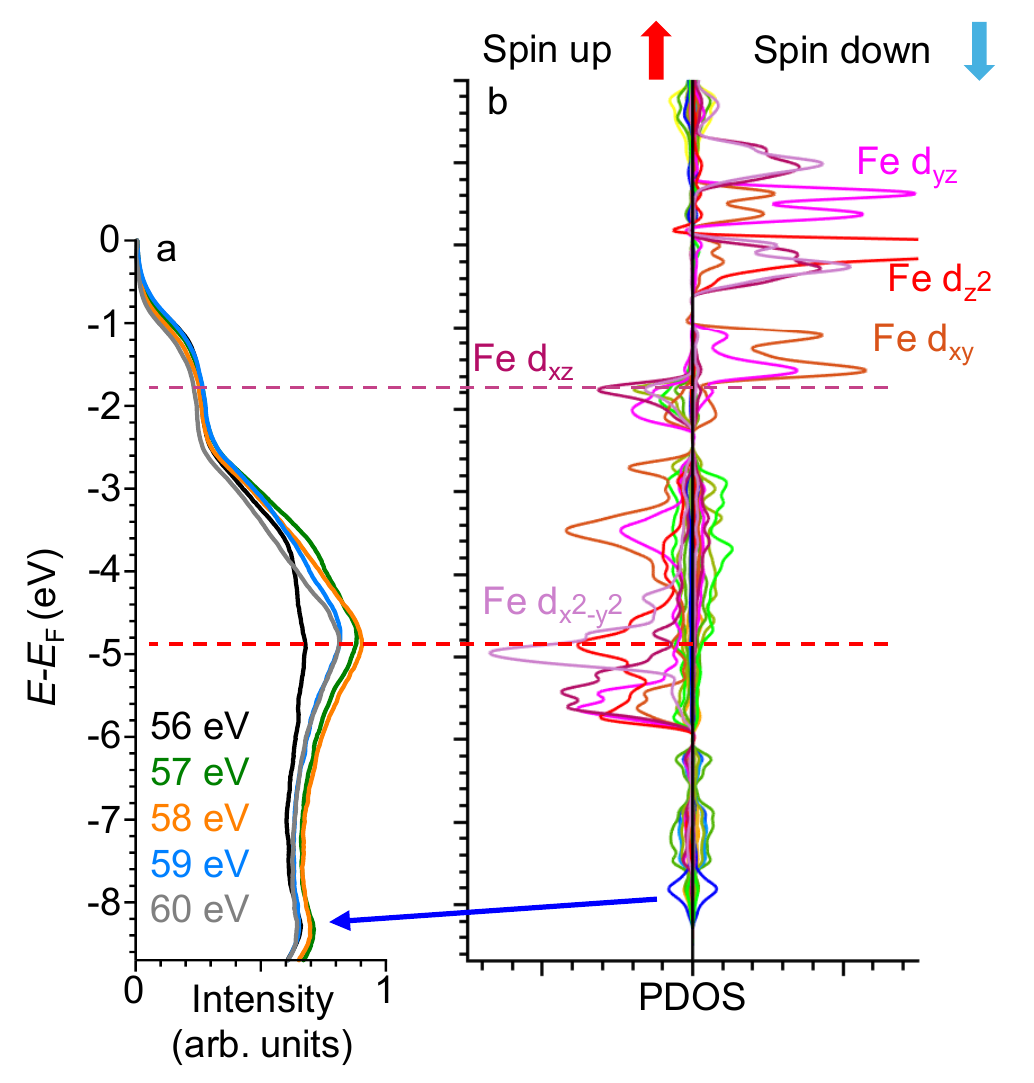}
\vspace{-0.4cm}
\caption{{\bf Resonant ARPES compared with orbital contributions of Fe 3d levels.} (a) Raw ARPES intensity at $\overline{\Gamma}$ for different $h\nu$ as color-coded, showing the resonant enhancement of Fe $\rm 3d_{z^2}$ band at $E-E_{\rm F} \sim 4.8$\,eV, same as Figure~2e, main text, ARPES intensity is normalized to the photon current at the entrance mesh of the analyser, horizontal lines mark the two main peaks. (b) Spin resolved density of states form DFT +U calculations (VASP), projected to different orbitals using the color-code of Figure~\ref{Fig_S15}, $U_{\rm eff} = 1.2$\,eV,$k_z=0$. Blue arrow marks the P $\rm 3p_z$ contribution. 
}
\label{Fig:P_Dos}
\end{figure}

A main difference to the data at large $h\nu$ is the absence of the flat Fe $\rm 3d_{z^2}$ band at $E-E_{\rm F}\sim-4.8$\,eV. We conjecture that this is a consequence of the Fe 3p$\rightarrow$3d resonance at $h\nu\approx 58$\,eV (Figure~\ref{Fig:P_Dos}a). However, the upper part of the Fe 3d$_{z^2}$ rhombus is still visible, which might be due to its accompanying S 3p contributions (section~3\ref{sec:orbital}). It, moreover, exhibits a different dispersion in the lower energy part, where it appears as a parabola around $\overline{\rm M}$ centered at $E-E_{\rm F}=-5.8$\,eV (white arrows in Figure~\ref{Fig_S20}a).
\clearpage

This might be identified with some of the bands in the stretched DFT+U data (Figure~\ref{Fig_S15}b), albeit these bands are less continuous. Finally, the bunch of S bands around $E-E_{\rm F}=-4$\,eV in the central BZ are distinct likely due to different band sensitivities. Indeed, these bands resemble the ones at $h\nu=75-80$\,eV, presumably at a distinct $k_z$.
This shows that sensitivities are an important ingredient for the ARPES data at various $h\nu$ due to the dense band sequences in FePS$_3$.

Another interesting aspect of the $h\nu$ dependence arises from the comparison  of the resonant enhancement of Fe 3d levels by the Fe 3p\,$\rightarrow$\,3d resonance (Figure~\ref{Fig:P_Dos}a) with the measurement of \cite{Jin2022}, Figure~5, that used the Fe 2p\,$\rightarrow$\,3d resonance at $h\nu\approx 708$\,eV averaging across $|\boldsymbol{k}_\parallel|<3.5/$\AA. While in our measurements only the flat Fe 3d$_{z^2}$ bands at
$E-E_{\rm F}\approx -5$\,eV get amplified, the other measurement also observes an amplification of the bands at 
$E-E_{\rm F}\approx -1.5$\,eV. This can be rationalized by the Fe $\rm 3d_{xz}$ and $\rm 3d_{yz}$ character of the latter bands (Figure~\ref{Fig:P_Dos}b). The orbitals are not accessible in the inner BZ at $h\nu\approx 58$\,eV according to the simplified matrix elements (Figure~\ref{Fig_S7}b), but can contribute at $h\nu\approx 700$\,eV due to the accordingly large averaged $\boldsymbol{k}_\parallel$ range covering 7 BZs.



\subsection{Complete $C(E,k_{\rm \parallel})$ plots below and above $T_{\rm N}$} 
\label{sub:T_dependence}

\begin{figure}[htb]
\centering
\includegraphics[width=0.45\textwidth]{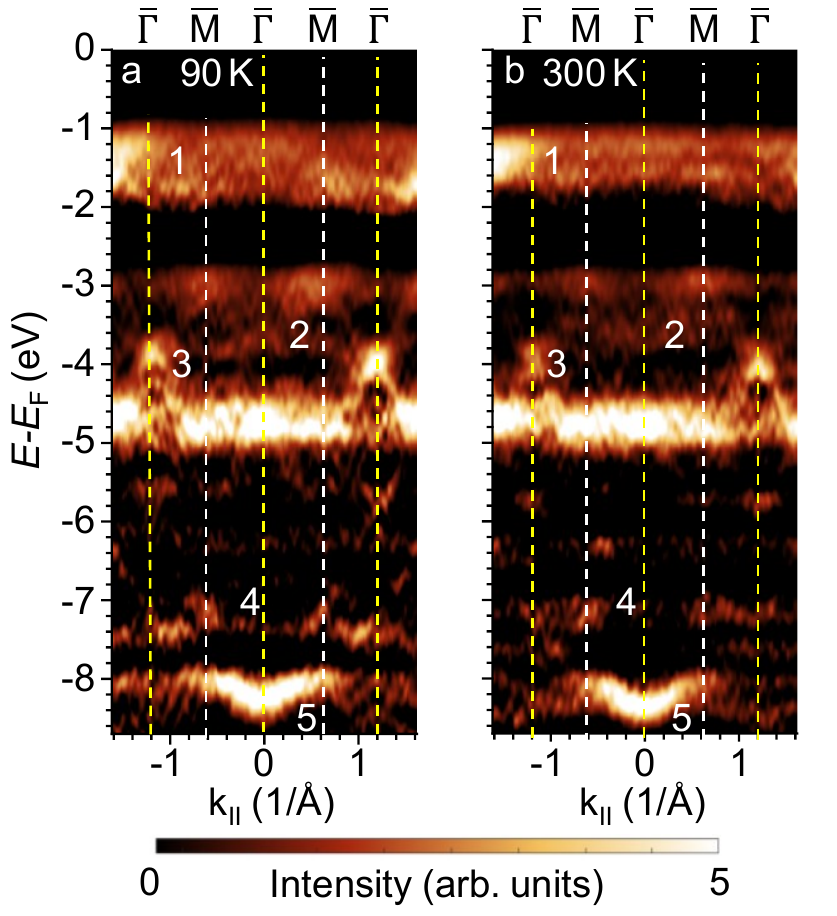}
\vspace{-0.4cm}
\caption{{\bf ARPES curvature at $90$\,K and $300$\,K}(a) ARPES curvature, $h\nu = 58$\,eV, $90$\,K, $\rm \overline{M\Gamma M}$ direction. 1-5 mark features for comparison with b (see text) (b) Same as a, but at $300$\,K.
}
\label{Fig_S19}
\end{figure}

The main text compares ARPES data 
above and below $T_{\rm N}$ in separate energy windows only. The ARPES curvature for the complete energy range 
is shown in Figure~\ref{Fig_S19}. The changes discussed in the main text are apparent, namely the line of reduced intensity at 300\,K in the upper S 3p$_z$ band (feature 1) and the downwards shift of the Fe 3d$_{z^2}$ rhombus (feature 3) as well as of the P 3p$_z$ parabola (feature 5). More subtle features in the energy range marked by 2 and 4 are not evident enough in the raw data and, hence, a possible task for future studies.   


\subsection{Energy shift of Fe 3d$_{z^2}$ rhombus} 
\label{sub:rhombus}

\begin{figure}[htb]
\centering
\includegraphics[width=0.41\textwidth]{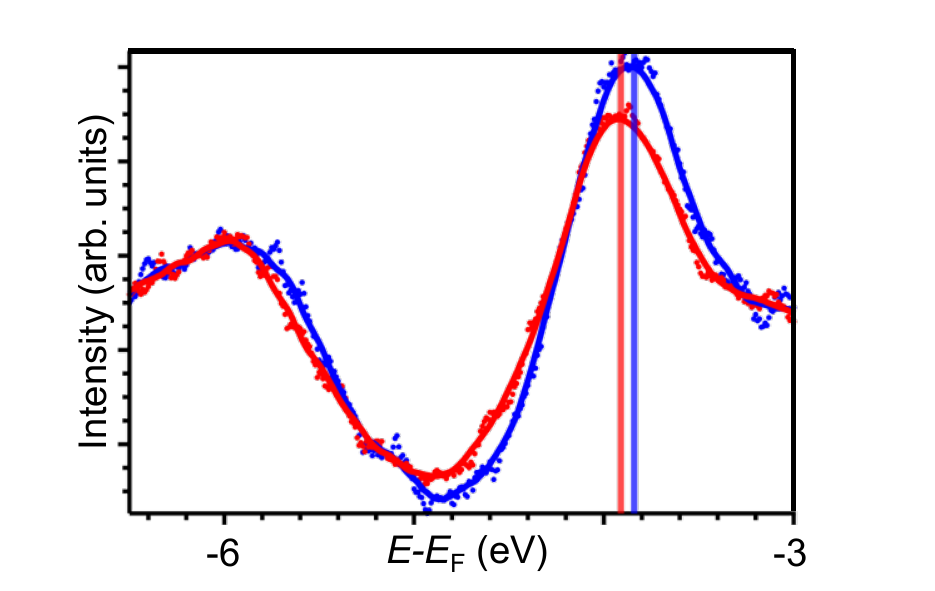}
\vspace{-0.4cm}
\caption{{\bf EDC at $\overline{\Gamma}_2$ across $T_{\rm N}$ (Fe 3d$_{z^2}$ rhombus).} Dots: EDCs at $\overline{\Gamma}_2$ after background subtraction as in Figure~4e, main text, red: $300$\,K, blue: $90$\,K, solid lines: EDCs after box smoothing for $\pm 125$\,meV, vertical lines: maxima positions of the smoothed data.
}
\label{Fig_S21}
\end{figure}

The energy shift of the Fe $\rm 3d_{z^2}$ rhombus between $300$\,K and $90$\,K is determined by comparing the EDCs at $\overline{\Gamma}$ in the second BZ after background subtraction as in Figure~4e, main text. After box smoothing the EDCs for $\pm 125$\,meV, we determined the maxima around $E-E_{\rm F} \approx -3.9$\,eV for five of the six $\overline{\Gamma}$'s in the second BZ, while for $\overline{\Gamma}_4$ it could not be determined precisely enough due to an increased single-to-noise ratio. 
As an example, Figure~\ref{Fig_S21} shows the two EDCs at $\overline{\Gamma}_2$ after background subtraction (dots) and after smoothing (solid line) with marked maxima (vertical lines) revealing a shift of $70\pm 30$\,meV. 
The average shift of the 5 EDCs at $\overline{\Gamma}_n$ amounts to  $74\pm32$\,meV. 

\subsection{$\overline{\rm M}\overline{\Gamma}\overline{\rm M}$ vs. $\overline{\rm K}\overline{\Gamma}\overline{\rm K}$ direction} 
\label{sub:K_Gamma_K}
\begin{figure}[htb]
\centering
\includegraphics[width=0.45\textwidth]{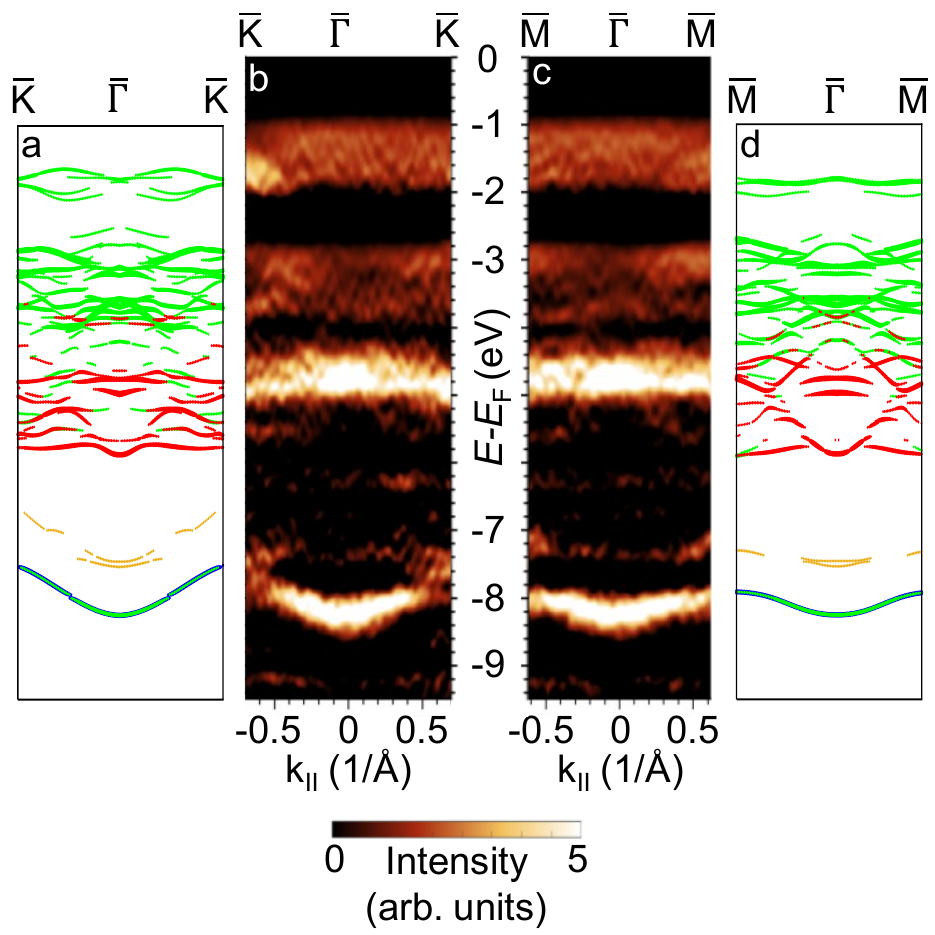}
\vspace{-0.4cm}
\caption{{\bf $\rm \overline{K\Gamma K}$ vs. $\rm \overline{M \Gamma M}$ direction.} (a) Band structure from DFT+U calculation (FLEUR), $\rm \overline{K\Gamma K}$ direction, $U_{\rm eff} = 1.2$\,eV, $k_z=-0.07$/\AA, only bands with orbital character of s-, $\rm p_z$- and $d_{z^2}$-type are shown, color code as in Figure~\ref{Fig_S15} (green: S $\rm 3p_z$, red: Fe $\rm 3d_{z^2}$, blue: P $\rm 3p_z$, yellow: S 3s). (b) ARPES curvature, $h\nu=58$\,eV, $90$\,K, $\rm \overline{K\Gamma K}$ direction.
(c) Same as b for $\rm \overline{M \Gamma M}$ direction. (d) Same as a for $\rm \overline{M \Gamma M}$ direction. The calculated band structures are shifted rigidly such that the valence band maximum is at $E-E_{\rm F}=-1$\,eV of the ARPES data. 
}
\label{Fig_S18}
\end{figure}

 Figure~\ref{Fig_S18} compares the $\rm \overline{K\Gamma K}$ and $\rm \overline{M \Gamma M}$ direction for the experimental and the DFT+U data. 
The difference between the two high-symmetry directions are small within the calculations. They cannot be resolved in the experiment except that 
the nearly parabolic P 3p$_z$ band exhibits less curvature in $\rm \overline{M \Gamma M}$  direction in both, calculation and experiment.

\begin{figure}[htb]
\centering
\includegraphics[width=0.45\textwidth]{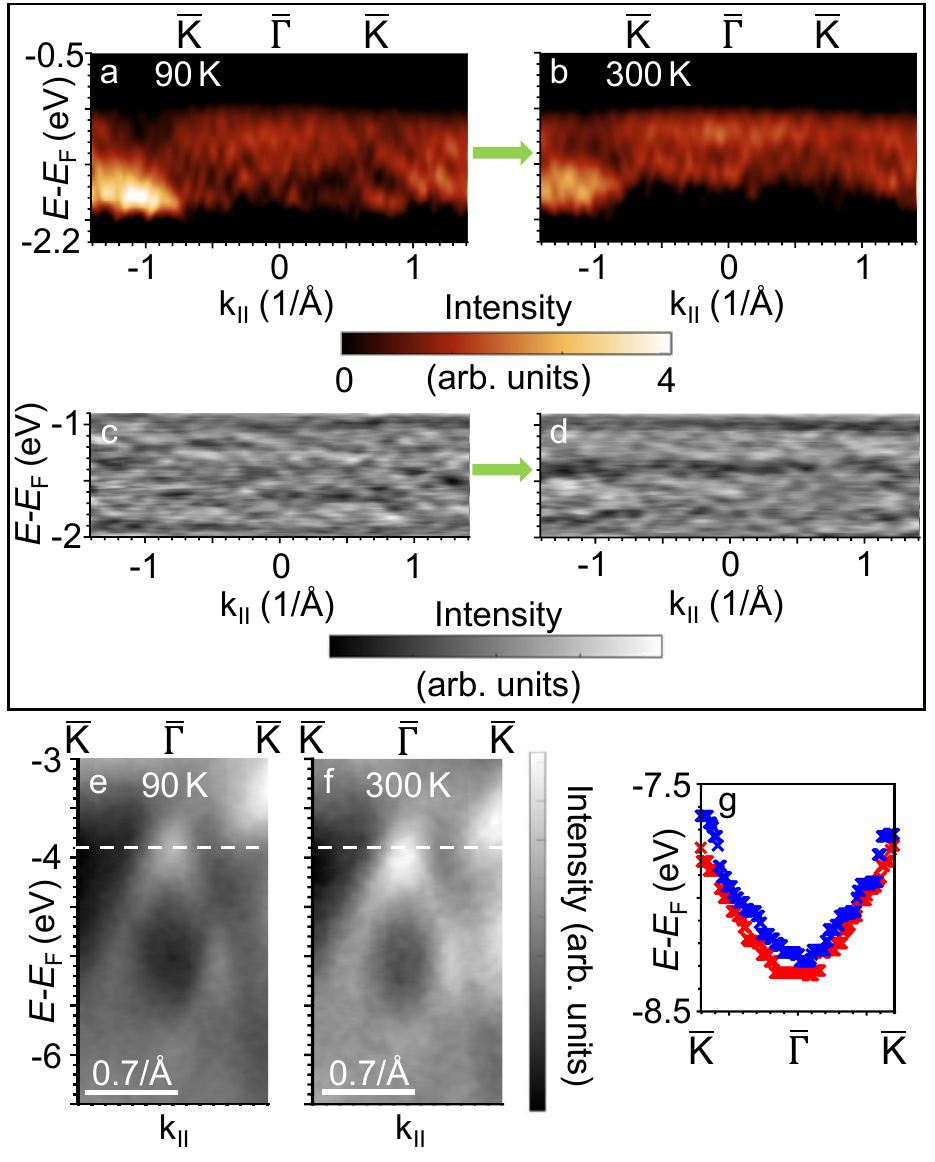}
\vspace{-0.4cm}
\caption{{\bf Changes across $T_{\rm N}$, $\rm \overline{K\Gamma K}$-direction.} (a),(b) ARPES curvature, $h\nu =58$\,eV,  $T$ as marked (upper S 3p$_z$ bands). (c),(d) Raw ARPES data at the same energy range as a,b with subtracted 4th order polynomial at each EDC, as in Figure~5c,d main text, 
$h\nu =58$\,eV, $T$ as marked, 
green arrow points to the line of reduced intensity at $300$\,K. (e),(f) Raw ARPES data around $\overline{\Gamma}_5$ after background subtraction as in Figure~4d, main text (Fe 3d$_{z^2}$ rhombus), $h\nu =58$\,eV, $T$ as marked, white line: guide to the eye. (g)  Intensity maxima of the P 3p$_z$ parabola like band deduced as in Figure~3, main text, blue: $90$\,K, red: $300$\,K.
}
\label{Fig_S22}
\end{figure}

The differences across the Néel temperature in $\rm \overline{K \Gamma K}$ direction are the same as in $\rm \overline{M \Gamma M}$ direction (Figure~\ref{Fig_S22}), namely a  splitting of the  upper S 3p$_{z}$ bands in two groups at $300$\,K  (Figure~\ref{Fig_S22}a--d), a downwards shift of the Fe 3d$_{z^2}$ rhombus by roughly 80\,meV at 300\,K (Figure~\ref{Fig_S22}e,f), and a downwards shift by $80\pm 40$\,meV of the P 3P$_z$ parabola-like band (Figure~\ref{Fig_S22}g).

\begin{figure}[htb]
\centering
\includegraphics[width=0.45\textwidth]{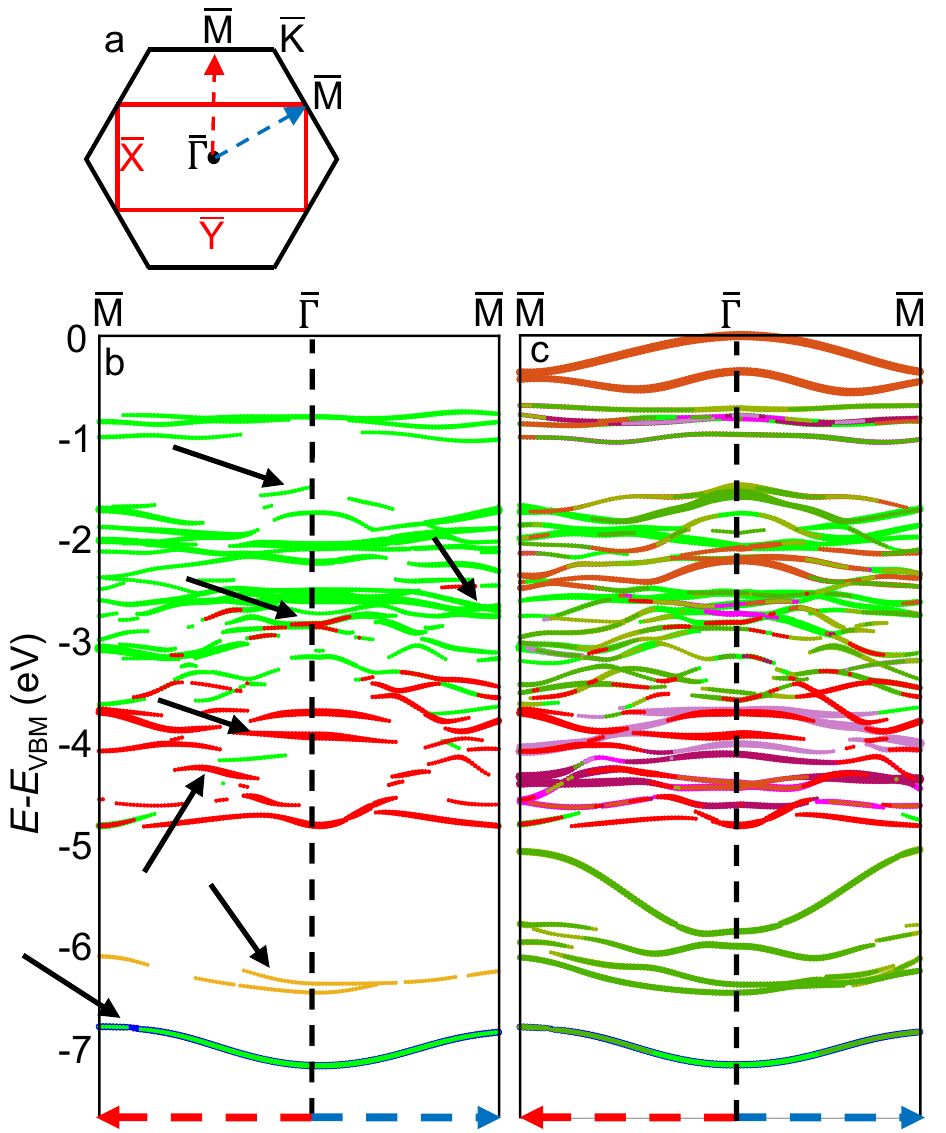}
\caption{{\bf Different $\rm \overline{M\Gamma M}$-directions.} (a) Geometric (black) and magnetic (red)  2D BZ with marked high symmetry points and two $\overline{\Gamma}\rm \overline{M}$ paths (dashed arrows), 
(b) Band structure from DFT+U calculations (FLEUR) after unfolding to the geometric BZ, $U_{\rm eff}=1.2$\,eV, $k_z = -0.07$/\AA. The two $\rm \overline{\Gamma M}$ paths from a are marked by arrows of according color. Black arrows point to distinctions between the two paths, symbol diameter is proportional to the orbital contribution, same color code as in Figure~\ref{Fig_S15}. (c) Same as b, but showing all orbital contributions above the 10\,\% threshold.}
\label{Fig_S23}
\end{figure}

The reduction to the rectangular magnetic BZ of the AFM zigzag arrangement leads to a distinction between the three $\rm \overline{ \Gamma M}$ directions of the geometric BZ (Figure~\ref{Fig_S23}a).
Two of them are unchanged, albeit getting additional bands via backfolding (Figure~\ref{Fig_S23}a), while the third $\rm \overline{ \Gamma M}$ direction gets cut into two pieces at $\rm \overline{Y}$  of the magnetic BZ (Figure~\ref{Fig_S23}a). For the $\rm \overline{ \Gamma K}$ direction, all three paths are cut by the magnetic BZ, but again two of them are identical, in that case the two shorter ones.   

However, the resulting changes in the calculated band structure are small, both for the selected (Figure~\ref{Fig_S23}b, arrows) and for all orbital contributions (Figure~\ref{Fig_S23}c). They could not be resolved in the experiment remaining as a future task.

\subsection{Orbital projections of DFT+U states}
\label{sec:orbital}

\begin{figure*}[htb]
\centering
\includegraphics[width=0.9\textwidth]{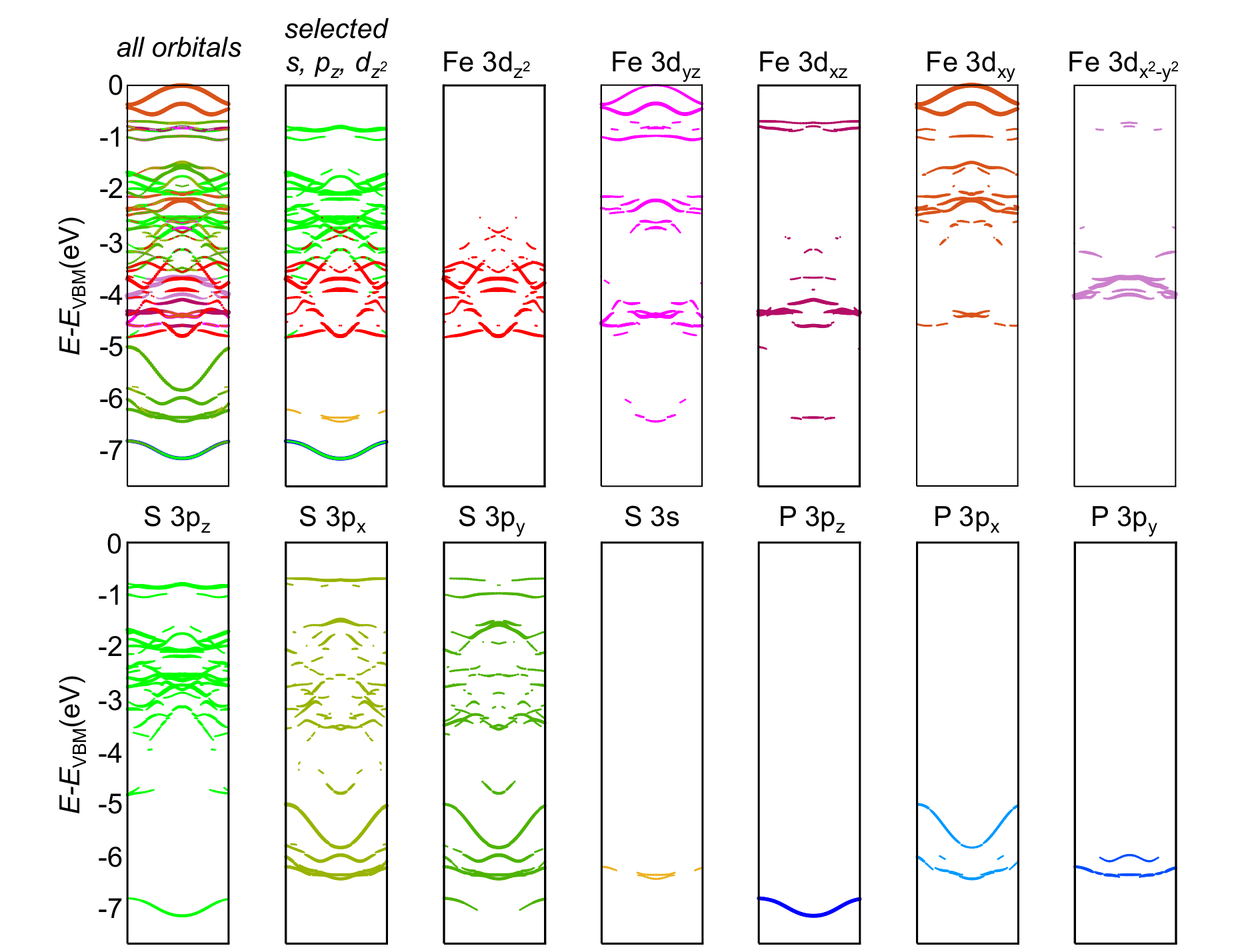}
\vspace{-0.4cm}
\caption{{\bf Orbital character of calculated states.} Band structure calculated by DFT+U (FLEUR), $U_{\rm eff}=1.2$\,eV, $k_z=-0.07$/\AA.  As marked on top, we display all orbital projections, the selected 3s, 3$\rm p_z$ and 3$\rm d_{z^2}$ orbitals according to the simplified selection rules \cite{Moser2017}, and each orbital contribution individually. Contributions are only shown, if larger than $10$\,\% of the state. The size of the symbol is proportional to the percentage of orbital contribution.
}
\label{Fig_S10}
\end{figure*}
Figure \ref{Fig_S10} shows the orbital contributions of each calculated state separately. Only contributions larger than $10$\,\% are displayed. For comparison, the total band structure and the selected orbitals from the simplified selection rules are added. This enables to highlight additional orbital contributions to the bands that change across $T_{\rm N}$. 
Firstly, the S 3p$_z$ bands at $E-E_{\rm VBM} \approx -1$\,eV exhibit significant contributions of all 3p orbitals as well as of the Fe 3d$_{xy}$, 3d$_{xz}$ and 3d$_{yz}$ orbitals. This indicates that these bands might be related to a superexchange process of the three Fe orbitals via the S atoms, that is then responsible for the rearrangement of the bands below $T_{\rm N}$. Secondly, the Fe 3d$_{z^2}$ rhombus has barely any contributions from other Fe 3d orbitals, but shows contributions of S 3p$_x$ across its whole extension, S 3p$_z$ at its top part and S 3p$_y$ at its bottom part. Hence, again S bands are involved. The origin of the energy shift remains unknown, but might be due to charge redistribution between the Fe orbitals. Finally, the lower P 3p$_z$ band does not show any contribution from Fe 3d orbitals, but of S 3p$_z$ and, to a smaller extent, of S 3p$_y$. It is tempting to assume that the known strong third nearest-neighbor exchange in FePS$_3$  is responsible for the energy shift of this band. It is believed that the third nearest neighbor exchange path requires a strong exchange coupling of the S 3p orbitals via the P atoms \cite{Lanon2016, Kim2021}, such that additional repelling charge appears at the P cores.


\subsection{Néel vs. zigzag arrangement (DFT+U)} 
\label{sub:Neel_vs_zig_zag}
\begin{figure*}[htb]
\centering
\includegraphics[width=0.9\textwidth]{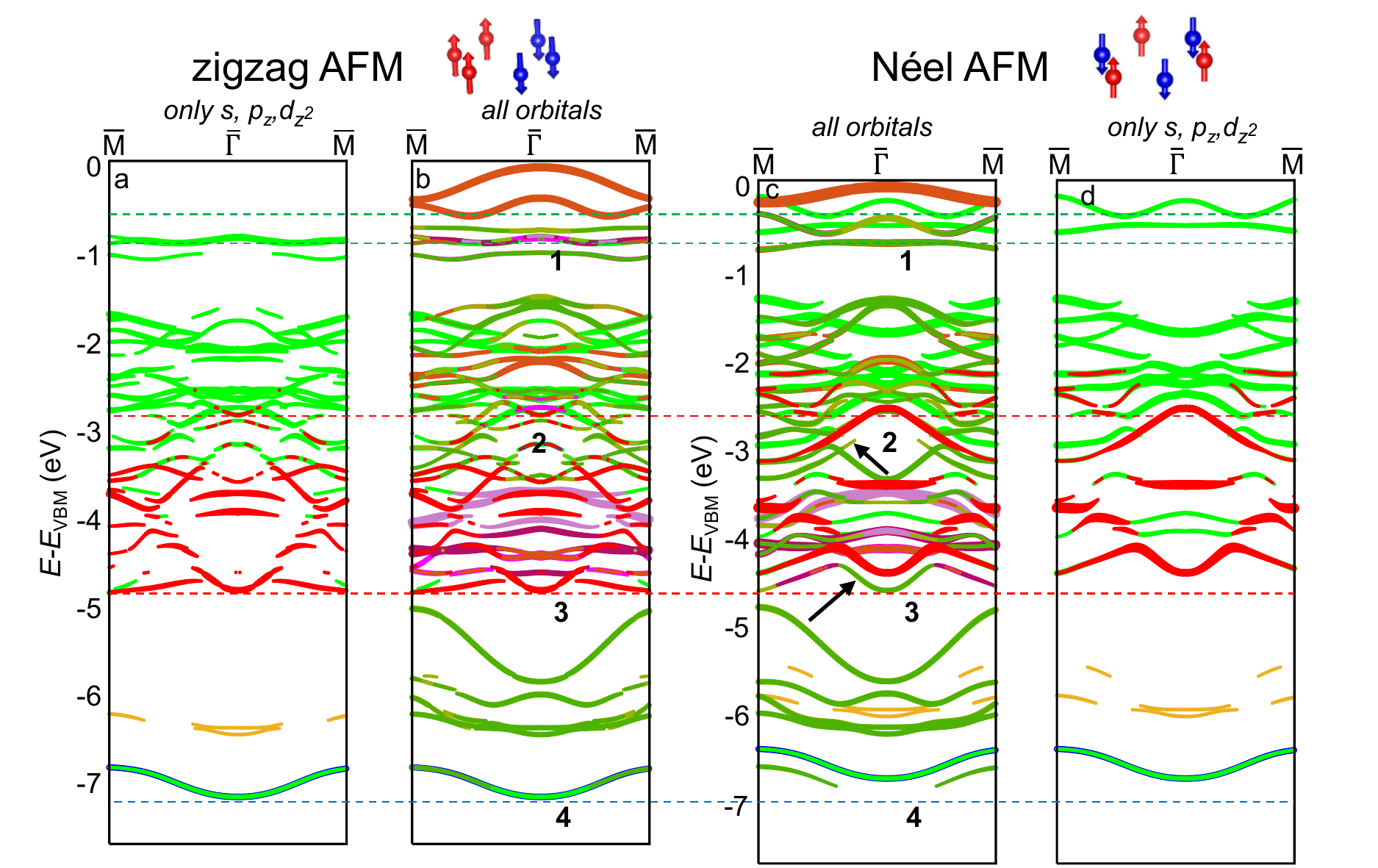}
\vspace{-0.4cm}
\caption{{\bf Band structure of AFM zigzag vs. Néel structure.} 
(a) Band structure from DFT+U (FLEUR) for the antiferromagnetic zigzag structure (model on top of b) after unfolding to the geometric BZ, $U_{\rm eff}=1.2$\,eV, $k_z=-0.07/$\AA\, only s, p$_z$ and $d_{z^2}$ orbitals are shown, color-code as in Figure~\ref{Fig_S15}. (b) Same as a, but for all  orbitals, numbers 1-4 mark features compared to c (c) Same as b, but for the antiferromagnetic Néel structure (model on top of d), arrows show band splittings with respect to b as discussed in the text.  (d) Same as a, but for the antiferromagnetic Néel structure.
}
\label{Fig_S13}
\end{figure*}
To relate the identified changes in the experimental band structure across $T_{\rm N}$ (main text) to the DFT+U calculations, we do not calculate the paramagnetic arrangement, but change the zigzag AFM arrangement of the ground state to the Néel structure with alternating spin orientations  (Figure~\ref{Fig_S13}). We use the same atomic arrangement as after relaxation of the zigzag ground state.
This emulates the paramagnetic (PM) state in the sense that the spins see a different spin environment than in the zigzag phase as they do in the PM state consisting of disordered Fe spin direction. Hence, we expect that the same bands react on the change of magnetic configuration and on the AFM-PM phase transition.  

For energy alignment, we do not use the valence band maximum, but match the energy of bands with a nearly unchanged dispersion such as the S 3p$_y$ bands at $E-E_{\rm VBM}\in[-5,-6]$\,eV. This leads to a downwards shift of the valence band maximum by $240$\,meV with respect to the valence band of the zigzag arrangement as can be rationalized straightforwardly by the significant flattening of the upper Fe 3d band that determines the valence band maximum.  Unfortunately, this change is not accessible by ARPES, since the bands are caused by Fe 3d$_{xy}$ and 3d$_{yz}$ orbitals (Figure~\ref{Fig:P_Dos}).

Using the alignment, both, the P 3p$_z$ band (feature 4) and the Fe 3d$_{z^2}$ rhombus (features 2-3) show an upwards shift by 100-200\,meV in the Néel structure accompanied by a splitting from a lower energy S 3p$_y$ band.
This indicates that these bands indeed react relatively strongly on the magnetism by energy shifts as observed with similar strengths in the experiment. The upper 
S 3p$_z$ bands (feature 1) also change exhibiting a broader, more pronounced energy distribution and an upwards energy shift in the Néel structure. Consequently, changes on the energy scale that we observe in the experiment are in line with induced changes by magnetic order. 

Other changes in the calculations are not observed experimentally. Firstly, the upper Fe 3d$_{yz}$/3d$_{xy}$ bands change drastically witnessing its strong involvement in the exchange. Moreover, the Fe rhombus splits into the upwards moving 3d$_{z^2}$ part and the nearly unchanged S 3p$_y$ contributions (arrows in Figure~\ref{Fig_S13}c) implying a lost hybridization that might weaken the nearest-neighbor ferromagnetic superexchange. Both is related to bands that are not visible by our ARPES experiment.
Finally, the upper part of the Fe 3d$_{z^2}$ rhombus gets less steep in dispersion. This might be caused by the larger distance to the next Fe atom with parallel spin in the Néel structure as probably not relevant for the paramagnetic case.

\FloatBarrier